%% file: hls_hff.tex
\documentclass[useAMS,usenatbib]{mn2e}

\usepackage{graphicx}
\usepackage{multirow}

\title[Complete census of \textit{Herschel} sources within \textit{HFF}]{A complete census of \textit{Herschel}-detected\thanks{\textit{Herschel} is an ESA space observatory with science instruments provided by European-led Principal Investigator consortia and with important participation from NASA.} infrared sources within the \textit{HST} Frontier Fields}
\author[T.~D.~Rawle et al.]
{\parbox{\textwidth}{
T.~D.~Rawle,$^{1,2}$\thanks{E-mail: \texttt{tim.rawle@sciops.esa.int}}
B.~Altieri,$^{1}$
E.~Egami,$^{3}$
P.~G.~P\'{e}rez-Gonz\'{a}lez,$^{4}$
F.~Boone,$^{5}$
B.~Clement,$^{6}$
R.~J.~Ivison,$^{7,8}$
J.~Richard,$^{6}$
W.~Rujopakarn,$^{9,10}$
I.~Valtchanov,$^{1}$
G.~Walth,$^{3}$
B.~J.~Weiner,$^{3}$
A.~W.~Blain,$^{11}$
M.~Dessauges-Zavadsky,$^{12}$
J.-P.~Kneib,$^{13}$
D.~Lutz,$^{14}$
G. Rodighiero,$^{15}$
D.~Schaerer,$^{12,5}$
I.~Smail$^{16}$
}\vspace{0.4cm}\\
\parbox{\textwidth}{
$^{1}$European Space Astronomy Centre (ESA/ESAC), Science Operations Department, E-28691 Villanueva de la Ca\~{n}ada, Madrid, Spain\\
$^{2}$ESA / Space Telescope Science Institute (STScI), 3700 San Martin Drive, Baltimore, MD 21218, USA\\
$^{3}$Steward Observatory, University of Arizona, 933 N. Cherry Ave, Tucson, AZ 85721, USA\\
$^{4}$Departamento de Astrof\'isica, Facultad de CC. F\'isicas,Universidad Complutense de Madrid, E-28040 Madrid, Spain\\
$^{5}$Universit\'e de Toulouse, UPS-OMP, CNRS, IRAP, 9 Av. colonel Roche, BP 44346, 31028, Toulouse Cedex 4, France\\
$^{6}$Univ Lyon, Univ Lyon1, Ens de Lyon, CNRS, Centre de Recherche Astrophysique de Lyon UMR5574, F-69230, Saint-Genis-Laval, France\\
$^{7}$European Southern Observatory, Karl-Schwarzschild-Str. 2, D-85748 Garching bei M\"unchen, Germany\\
$^{8}$Institute for Astronomy, University of Edinburgh, Royal Observatory, Blackford Hill, Edinburgh EH9 3HJ, UK\\
$^{9}$Department of Physics, Faculty of Science, Chulalongkorn University, 254 Phayathai Road, Pathumwan, Bangkok 10330, Thailand\\
$^{10}$Kavli Institute for the Physics and Mathematics of the Universe (WPI),The University of Tokyo Institutes for Advanced Study, The University of Tokyo, Kashiwa, Chiba 277-8583, Japan\\
$^{11}$Department of Physics \& Astronomy, University of Leicester, University Road, Leicester LE1 7RH, UK\\
$^{12}$Observatoire de Gen\`{e}ve, Universit\'{e} de Gen\`{e}ve, 51 Ch. des Maillettes, 1290, Sauverny, Switzerland\\
$^{13}$Laboratoire d'Astrophysique EPFL, Observatoire de Sauverny, Versoix, 1290, Switzerland\\
$^{14}$Max-Planck-Institut f\"ur extraterrestrische Physik, Postfach 1312, Giessenbachstrasse 1, D-85741 Garching, Germany\\
$^{15}$Dipartimento di Fisica e Astronomia `Galileo Galilei', Universit\'a di Padova, Vicolo dell'Osservatorio, 3, I-35122 Padova, Italy\\
$^{16}$Centre for Extragalactic Astronomy, Department of Physics, Durham University, South Road, Durham DH1 3LE
}}
\begin{document}

\date{20 Feb 2015}

\pagerange{\pageref{firstpage}--\pageref{lastpage}} \pubyear{2015}

\maketitle

\label{firstpage}

\begin{abstract}
We present a complete census of all \textit{Herschel}-detected sources within the six massive lensing clusters of the \textit{HST} Frontier Fields (HFF). We provide a robust legacy catalogue of 263 sources with \textit{Herschel} fluxes, primarily based on imaging from the \textit{Herschel Lensing Survey} (HLS) and PEP/HerMES Key Programmes. We optimally combine \textit{Herschel}, \textit{Spitzer} and \textit{WISE} infrared (IR) photometry with data from \textit{HST}, VLA and ground-based observatories, identifying counterparts to gain source redshifts. For each \textit{Herschel}-detected source we also present magnification factor ($\mu$), intrinsic IR luminosity and characteristic dust temperature, providing a comprehensive view of dust-obscured star formation within the HFF. We demonstrate the utility of our catalogues through an exploratory overview of the magnified population, including more than 20 background sub-LIRGs unreachable by \textit{Herschel} without the assistance gravitational lensing.
\end{abstract}

\begin{keywords}
infrared: galaxies; submillimetre: galaxies; galaxies: star formation
\end{keywords}

\section{Introduction}
\label{sec:intro}

The \textit{Hubble Space Telescope} (\textit{HST}) Frontier Fields (HFF)\footnote{http://www.stsci.edu/hst/campaigns/frontier-fields/} is an ongoing programme (2013--2016) to obtain ultra-deep imaging of six intermediate-redshift galaxy clusters, using 840 orbits of Director's Discretionary Time (PI: Lotz; Lotz et al., in preparation). The primary goal is to exploit the gravitational lensing effect of these massive foreground structures to study background galaxies in the very early Universe. HFF employs both the Advanced Camera for Surveys (ACS) and the Wide Field Camera 3 (WFC3) to gain imaging throughout the rest-frame optical and near-infrared (NIR). Although the observations are intrinsically shallower than e.g. the \textit{Hubble Extremely Deep Field} (XDF; \citealt{ill13-6}), the additional effective depth yielded by the wide-scale cluster lensing provides a first glimpse of the galaxy population to be probed in detail by the \textit{James Webb Space Telescope} (\textit{JWST}).

HFF comprises six clusters chosen primarily for their predicted lensing strength, low zodiacal background and observability by large ground-based facilities (e.g. ALMA, VLT and/or those on Mauna Kea). The ACS/WFC3 parallel mode provides two distinct footprints per cluster. A central field is aligned with the highest-magnification cluster core, while the parallel location is constrained by guide star availability and selected to maximise image quality (ie avoiding bright stars). HFF includes two northern and four southern clusters, covering a range of Right Ascension and redshifts within $z\sim0.30-0.55$ (see Table \ref{tab:fields}).

Rest-frame optical and NIR imaging probes the unobscured evolved stellar component of extragalactic sources. An understanding of the processes governing star formation is required to answer fundamental questions concerning galaxy evolution. Much of the star-formation activity in the Universe is shrouded in dust \citep[see the review,][]{cas14-45}, giving an imprint of star formation characteristics on the recycled light in the far-infrared (FIR). The \textit{Herschel Space Observatory} \citep{pil10-1} was launched in 2009 with the specific goal of understanding galaxy dust and obscured star formation. By the mission end in 2013, \textit{Herschel} had obtained more than 25000~h of FIR data.

\begin{table*}
\caption{Summary of the HFF central (C) and parallel (P) footprints, including available imaging. Positions are the nominal \textit{HST} pointing and redshift ($z$) is for the target cluster. Coverage is tabulated for SPIRE (S), PACS (PPP $=$ 70+100+160~$\mu$m), MIPS 24~$\mu$m (M), IRAC (IIII $=$ channels 1--4), \textit{WISE} (W), \textit{HST} (H; * for complete HFF imaging at the time of writing), VLA (V). This table is representative only: partial coverage may result in fewer available bands at the footprint extremities.}
\label{tab:fields}
\begin{tabular}{lrrrrlcclclc}
\hline
\multicolumn{1}{l}{Cluster} & \multicolumn{1}{c}{$z$} & \multicolumn{1}{c}{} & \multicolumn{1}{c}{RA} & \multicolumn{1}{c}{Dec} & \multicolumn{7}{l}{Coverage...} \\
\hline
\multirow{2}{*}{A2744} & \multirow{2}{*}{0.308} & C & 00:14:21.2 & --30:23:50.1 & S & --PP & M & IIII & W & H* & --- \\
 & & P & 00:13:53.6 & --30:22:54.3 & S & --- & --- & --- & W & H* & --- \\
\hline
\multirow{2}{*}{A370} & \multirow{2}{*}{0.375} & C & 02:39:52.9 & --01:34:36.5 & S & --PP & M & IIII & W & H & V \\
 & & P & 02:40:13.4 & --01:37:32.8 & S & --- & M & II-- & W & --- & V \\
\hline
\multirow{2}{*}{MACSJ0416.1--2403 (M0416)} & \multirow{2}{*}{0.396} & C &  04:16:08.9 & --24:04:28.7 & S & --PP & --- & II-- & W & H* & V \\
 & & P & 04:16:33.1 & --24:06:48.7 & S & --- & --- & --- & W & H* & V \\
\hline
\multirow{2}{*}{MACSJ0717.5+3745 (M0717)} & \multirow{2}{*}{0.545} & C & 07:17:34.0 & +37:44:49.0 & S & --PP & M & IIII & W & H* & V \\
 & & P & 07:17:17.0 & +37:49:47.3 & S & --- & M  & IIII & W & H* & V \\
\hline
\multirow{2}{*}{MACSJ1149.5+2223 (M1149)} & \multirow{2}{*}{0.543} & C & 11:49:36.3 & +22:23:58.1 & S & PPP & --- & II-- & W & H & V \\
 & & P & 11:49:40.5 & +22:18:02.3 & S & --- & --- & II-- & W & H & V \\
\hline
\multirow{2}{*}{RXCJ2248.7--4431 (AS1063)} & \multirow{2}{*}{0.348} & C & 22:48:44.4 & --44:31:48.5 & S & PPP & M & IIII & W & H & --- \\
 & & P & 22:49:17.7 & --44:32:43.8 & S & --- & --- & I--- & W & --- & --- \\
\hline
\end{tabular}
\end{table*}

\textit{Herschel} large-area surveys of well-studied extragalactic fields have provided a new understanding of dust temperature \citep[e.g.][]{hwa10-75,mag14-86}, dust mass \citep{san14-30,row14-1017} and spectral energy distribution (SED) shape \citep{ber13-100} in IR/submillimetre galaxies. Building upon previous \textit{Spitzer}-based studies of star formation rate (SFR) evolution \citep[e.g.][]{per08-234}, the large \textit{Herschel} sample size of \citet{elb11-119} allowed the identification of an IR ``main sequence'', including local, normal star-forming spirals. In contrast, luminous star-forming galaxies lie above this sequence in a seemingly distinct ``star-burst'' phase. We have now begun quantifying and characterising the physical origin of this relationship \citep[e.g.][]{rod14-19,spe14-15,bet15-113,sch15-74}.

\textit{Herschel} blank fields have yielded a large sample of galaxies at $z<1$, while also glimpsing a few of the brightest star-forming galaxies at higher redshift \citep[e.g.][]{swi12-1066,wal12-93,rie13-329}. However, observations suffer from the intrinsic limit imposed by instrumental confusion noise \citep{ngu10-5}, effectively restricting them to galaxies brighter than a demi-ULIRG ($L_{\rm IR} > 5\times10^{11}$~L$_{\sun}$) at $z\ga1.5$, and ULIRGs beyond $z\sim2$.

Gravitational lensing by massive galaxy clusters can be of great benefit in the far-infrared \citep{sma97-5}, boosting background sources out of the confusion noise. There is an added advantage that foreground cluster galaxies are nearly transparent, contributing only a small fraction to the total observed flux. The \textit{Herschel} Lensing Survey (HLS; \citealt{ega10-12, ega-prep}) was the largest \textit{Herschel} programme dedicated to exploiting cluster lensing. The survey design was ably demonstrated by the early observation of the well-known $z=2.8$ lensed LIRG behind the Bullet cluster \citep{rex10-13}. Accounting for the magnification factor $\mu\sim75$, the intrinsic flux of the source at 250~$\mu$m is $<1$~mJy, significantly fainter than the nominal confusion limit of SPIRE ($5\,\sigma\sim28$~mJy; \citealt{ngu10-5}). For a sample of HLS lensed sources, \citet{skl14-149} explored multi-wavelength characteristics, concentrating on optical extinction and star-formation histories, while \cite{des15-50} investigated their molecular gas content and found evidence for a non-universal dust-to-gas ratio. \citet{com12-4} and \citet{raw14-59} highlighted another facet of gravitational lensing, as the magnification and conservation of surface brightness enabled observation of an HLS-discovered source at $z=5.2$ with sub-kpc resolution millimetre interferometry.

While the lensed, high redshift galaxies are the primary interest of the HFF programme, many cluster members are also in the observations. Generally, cluster galaxies are quiescent, with little or no dust to produce FIR emission. However, several \textit{Herschel} studies have each detected 10s of cluster members at intermediate redshift \citep{raw10-14,per10-40,cop11-680,raw12-29,raw12-106}. Recently, \citet{raw14-196} analysed the HFF cluster A2744, demonstrating that IR SFRs and optical morphologies can descriminate cluster formation processes such as group in-fall and cluster merging.

In this paper, we primarily provide a robust flux catalogue of \textbf{all} \textit{Herschel}-detected sources within the \textit{HST} Frontier Fields. We also present a `value-added' catalogue of optical counterparts on a best-effort basis, enabling the calculation of redshift-dependent and intrinsic physical properties such as magnification, IR luminosity and dust temperature. The observations are detailed in Section~\ref{sec:obs}, while Section~\ref{sec:phot} describes the compilation of the \textit{Herschel} flux catalogue, including photometric extraction and band-merging methodology. Section~\ref{sec:cps} explains the `value-added' catalogue, and derives intrinsic star formation properties. Section~\ref{sec:disc} presents an initial exploration of some of the most interesting IR-bright sources in HFF. Section~\ref{sec:conc} summarises the paper. We assume a standard cosmology with $H_{0} = 70$~km~s$^{-1}$~Mpc$^{-1}$,$\Omega_{\mathrm M} = 0.3$, $\Omega_{\mathrm \Lambda} = 0.7$.

\textit{Herschel} imaging, catalogues and source IR SEDs associated with this paper can be downloaded from the public flavour of the Rainbow Database.\footnote{https://rainbowx.fis.ucm.es}

\section{Observations and data reduction}
\label{sec:obs}

This section describes IR and ancillary data for the \textit{HST} Frontier Fields. Each cluster is observed in two distinct regions, referred to as the central and parallel footprints. The data availability for each footprint at the time of writing is summarised in Table \ref{tab:fields}.

\subsection{\textit{Herschel}}


\begin{table*}
\caption{\textit{Herschel} programmes and observation IDs (OBSIDs) for the Frontier Fields.}
\label{tab:obsids}
\begin{tabular}{lllll}
\hline
\multicolumn{1}{l}{Cluster} & \multicolumn{2}{l}{Programme} & \multicolumn{1}{l}{PACS OBSIDs} & \multicolumn{1}{l}{SPIRE OBSIDs} \\
\hline
A2744 & HLS & KPOT\_eegami\_1 & 1342188251--2 ($n=2$) & 1342188584 \\
A370 & PEP & KPGT\_dlutz\_1 & 1342223332--3 ($n=2$) & --  \\
& HerMES & KPGT\_soliver\_1 & -- & 1342201311--8, 1342248002--4 ($n=11$)  \\
M0416 & HLS & OT2\_eegami\_5 & 1342250291--2 ($n=2$) & 1342241122 \\
M0717 & HLS & KPOT\_eegami\_1 & 1342219416--7 ($n=2$) & 1342193012 \\
M1149 & HLS & KPOT\_eegami\_1 & 1342211797--8 ($n=2$) & 1342222841 \\
& GT & GT1\_dlutz\_4 & 1342221954--7 ($n=4$)  & 1342210511 \\
AS1063 & HLS & KPOT\_eegami\_1 & 1342188222--3 ($n=2$) & 1342188165 \\
& HLS & OT2\_trawle\_3 & 1342270947--8 ($n=2$) & -- \\
\hline
\end{tabular}
\end{table*}

The \textit{Herschel Space Observatory} included two broadband imagers, PACS \citep{pog10-2} and SPIRE \citep{gri10-3}. They probed $\lambda\sim70-500$~$\mu$m, which brackets the peak of the IR dust component out to $z\sim4$. The central HFF footprints were observed by both instruments, while parallel fields are covered by SPIRE only. Here we describe the programmatic origin, observing set-up and reduction for the \textit{Herschel} data, as summarised in Table \ref{tab:obsids}.

\subsubsection{PACS}
\label{sec:pacs}

The PACS instrument allowed an observer to choose between 70 or 100~$\mu$m bands in a blue channel, collected simultaneously with the red channel at 160~$\mu$m. For all six central Frontier Fields footprints, imaging is available within the PACS 100 and 160~$\mu$m bands. In addition, M1149 and AS1063 were also covered at 70~$\mu$m.

Five clusters were observed by PACS as part of the \textit{Herschel} Lensing Survey (HLS; \citealt{ega10-12, ega-prep}), which combines an Open-Time Key Programme (KPOT) and an Open-Time Cycle 2 (OT2) Programme (both PI: E. Egami). These (100, 160~$\mu$m) observations consist of two orthogonal scan maps, each comprising 18 repetitions of thirteen parallel 4--arcmin scan legs (total observing time per cluster, $t_{\rm obs}=4.4$~h). The sixth Frontier Field (A370) was part of the PACS Evolutionary Probe (PEP; \citealt{lut11-90}) Guaranteed-Time Key Programme (KPGT). The PEP cluster observing strategy was almost identical to HLS, but with 22 repetitions per orthogonal scan map ($t_{\rm obs}=5.2$~h).

A GT Cycle 1 (GT1) Programme re-observed M1149 in all three PACS bands (PI: D. Lutz), consisting of two cross-scans per blue band (70 and 100~$\mu$m; each $t_{\rm obs}=0.9$~h), comprising three repetitions of 20 parallel 3--arcmin scan legs in a point source map. This also gave 1.8~h of additional depth at 160~$\mu$m, concentric with the HLS observation, but over a smaller area.

An ancillary HLS OT2 Programme re-targeted AS1063 to gain 70~$\mu$m (PI: T. Rawle). The programme added a further two cross-scans with 27 repetitions each (4--arcmin scan legs), at 70 and 160~$\mu$m ($t_{\rm obs}=6.4$~h). This data is coincident with the HLS observations.

Finally, A2744 is located on the edge of the South Galactic Plane coverage of the wide-field PACS observations from H--ATLAS (PI: S. Eales; \citealt{eal10-499}). The survey was completed in parallel (PACS+SPIRE) mode and achieved a $5\,\sigma$ depth at 100~$\mu$m of $\sim$120~mJy \citep{iba10-38}. As this is very shallow compared to HLS we do not include the data in our co-added map (hence it is ignored in Table \ref{tab:obsids}). The H--ATLAS PACS data for the A2744 parallel field (not covered by HLS), contain no detections.

Regardless of origin, all PACS data were reduced homogeneously. Calibrated time-stream data (level one frames) were taken directly from the \textit{Herschel} archive and processed by \textsc{UniHipe} to produce fits files in the format required by \textsc{Unimap} \citep{pia12-3687,pia15-1471}. \textsc{Unimap} employs a Generalised Least Squares (GLS) method to produce the final maps (combining all OBSIDS), and includes a second timeline deglitcher, detection and correction for detector signal jumps, and an advanced drift removal via an Alternate Least Square method. This algorithm has several advantages compared to the naive map-maker \textsc{photProject} (the erstwhile standard for archive products), as it removes the need to filter 1/f noise drift in the PACS bolometers: 1) variations in background flux are conserved rather than smoothed; 2) source fluxes are not truncated, removing the need to mask known sources and allowing blind PSF--fit photometry without using correction functions; 3) all turnaround data can be used, including those observed at zero scan speed, which increases the depth at the edge of the map.

The final PACS images extend to a radius of $\sim$4--5~arcmin from each cluster core. Generally this only covers the central HFF footprint, although three galaxies on the inner edge of the parallel fields are also just within the PACS maps. The beam sizes have FWHM $=$ 5.2, 7.7 and 12~arcsec at 70, 100 and 160~$\mu$m respectively. The sensitivity of our prior-based flux catalogues (see Section \ref{sec:priors}) at the centre of the PACS maps are given in Table \ref{tab:depth}.

\subsubsection{SPIRE}

SPIRE operated at 250, 350 and 500~$\mu$m simultaneously, and achieved confusion-limited depth in a very short time. Four HFF clusters were part of the HLS OTKP programme, consisting of 20 repetitions in large scan map mode, each with two 4-arcmin scans and cross-scans (per cluster, $t_{\rm obs}=1.7$~h). SPIRE coverage of M0416 is from HLS OT2, and was achieved via a 10-repetition small scan map (1 scan and one cross-scan of 4~arcmin length; $t_{\rm obs}=0.4$~h). The resulting map is a little smaller and shallower than the original HLS observations. A370 was included in the KPGT Herschel Multi-tiered Extragalactic Survey (HerMES; \citealt{oli12-1614}), and the data combines eight small scan maps (6 repetitions each) and three large scan maps (1 repetition, 7 scans and 7 cross-scans of 38~arcmin length). The overall SPIRE observing time for A370 was $t_{\rm obs}=3.5$~h, but due to the wider footprint this corresponds to a similar central depth as the HLS maps. H--ATLAS parallel mode SPIRE observations ($5\,\sigma$ at 250~$\mu$m $\sim33$~mJy; \citealt{cle10-8}) also cover A2744, but are ignored as the shallow data contribute nothing extra to the confusion-limited HLS.

For all six clusters, images were produced via the standard reduction pipeline in \textsc{Hipe} \citep{ott10-139} v12 (v12.2 calibration product) with median baseline removal and destriper. Including all turnaround data, the final SPIRE images extend to a cluster centric radius of $\sim$10~arcmin (HLS OTKP), $\sim$7~arcmin (M0416) and $\sim$30~arcmin (A370). In all clusters except M0416, both central and parallel footprints are covered. For M0416, one third of the parallel field falls outside of the SPIRE image. The beam sizes are large (18, 25, 36~arcsec, respectively) and all three bands are confusion limited (5$\,\sigma_{\rm conf}$ $\approx$ 28, 32, 33~mJy; \citealt{ngu10-5}). However, this confusion limit assumes that the local source density is unknown, but with a prior knowledge of source distribution from shorter wavelength data (see Section \ref{sec:priors}), we reach a significantly lower flux limit in the central fields (see Table \ref{tab:depth}).

\begin{table}
\caption{Median 5$\,\sigma$ depth (in mJy) of the prior-based \textit{Herschel} catalogues within the central HFF footprints.}
\label{tab:depth}
\begin{tabular}{lrrrrrr}
\hline
\multicolumn{1}{c}{Cluster} & \multicolumn{1}{c}{PACS} & \multicolumn{1}{c}{PACS} & \multicolumn{1}{c}{PACS} & \multicolumn{1}{c}{SPIRE} & \multicolumn{1}{c}{SPIRE} & \multicolumn{1}{c}{SPIRE} \\
& \multicolumn{1}{c}{70} & \multicolumn{1}{c}{100} & \multicolumn{1}{c}{160} & \multicolumn{1}{c}{250} & \multicolumn{1}{c}{350} & \multicolumn{1}{c}{500} \\
\hline
A2744 & -- & 4.5 & 8.8 & 13.9 & 14.9 & 13.0 \\
A370 & -- & 4.1 & 8.3 & 13.9 & 17.2 & 14.0 \\
M0416 & -- & 4.3 & 8.0 & 17.5 & 14.6 & 15.5 \\
M0717 & -- & 4.5 & 8.5 & 18.6 & 17.9 & 17.7 \\
M1149 & 7.3 & 4.5 & 8.6 & 14.9 & 15.2 & 17.8 \\
AS1063 & 3.1 & 4.7 & 8.2 & 16.2 & 17.1 & 17.5 \\
\hline
\end{tabular}
\end{table}

\subsection{Infrared ancillary data}

The prior-based photometry method described in Section \ref{sec:priors} depends heavily on observations from the \textit{Spitzer Space Telescope}. The IR SED fitting procedures in Section \ref{sec:irsed} use all available IR photometry $\lambda\ga3$~$\mu$m. Here we briefly describe IR imaging that originates from facilities other than \textit{Herschel}.

\subsubsection{\textit{Spitzer}}

{\it Spitzer} data for HFF were extracted from the {\it Spitzer} Heritage archive.\footnote{http://irsa.ipac.caltech.edu/applications/Spitzer/SHA} For IRAC, we started from the reduced, flux-and-WCS calibrated images provided in the archive as `corrected Basic Calibrated Data' (cBCD) and mosaic them together using the procedure developed by \citet{hua04-44}. This procedure includes pointing refinement, distortion correction, drizzling to a scale half of the original ($\sim$0.6~arcsec~pixel$^{-1}$) and correction of detector artefacts (most noticeably, mux-bleeding). In the case of MIPS, we started from the post-BCD products downloaded from the archive and used \textsc{Mopex} v1.8 for the flat fielding, jailbar removal and mosaicking. We used a pixel scale of 1.2~arcsec~pixel$^{-1}$.

All six cluster centres are covered by imaging in the 3.6 and 4.5~$\mu$m IRAC bands (beam size of 1.7~arcsec), with median 5$\,\sigma$ sensitivities of $\sim$1.7 and $\sim$1.6~$\mu$Jy respectively. Four clusters are also observed using the IRAC 5.8 and 8~$\mu$m channels (beam size of 1.7, 1.9~arcsec), yielding a median 5$\,\sigma$ depth of $\sim$5.2~$\mu$Jy, and by MIPS 24~$\mu$m (6~arcsec beam diameter) with a median 5$\,\sigma$ sensitivity of $\sim$90~$\mu$Jy. The parallel fields unfortunately have a less uniform \textit{Spitzer} coverage, with the details described in Table \ref{tab:fields}.

The heterogeneity results in no attempt by this study to derive stellar masses from the NIR. We defer this aspect to future analyses, which will be able to employ the deeper, homogeneous IRAC observations currently underway.

\subsubsection{\textit{WISE}}

The four-band, all-sky \textit{WISE} mission covers 3.4, 4.6, 12, 22~$\mu$m \citep{wri10-1868}. Imaging and catalogues for all of the HFF footprints were obtained from the NASA/IPAC Infrared Science Archive (IRSA).\footnote{http://irsa.ipac.caltech.edu/Missions/wise.html} Sensitivity in each band varies across the sky, but is typically 0.25, 0.35, 3.0, 18.0~mJy (5$\,\sigma$), with beam sizes 6.1, 6.4, 6.5, 12~arcsec (PSF FWHM). Source blending within the 12~arcsec beam at 22~$\mu$m is a major issue, and we generally disregard fluxes in this band for SED fitting.

\subsection{Counterpart ancillary data}

Knowledge of the source redshift is a pre-requisite for the derivation of magnification and intrinsic physical properties from IR SEDs. Redshifts are generally obtained from optical data, so we identify a candidate optical counterpart for each IR source on a best-effort basis (Section \ref{sec:cps}). Here we briefly describe the imaging and spectroscopy relevant to that process.

\subsubsection{\textit{HST} imaging}
\label{sec:hstimg}

The fully completed \textit{HST} Frontier Fields programme will comprise seven-band observations from ACS (F435W, F606W and F814W) and WFC3 (F105W, F125W, F140W, F160W). The wider field ACS images ($\sim3.4\times3.4$~arcmin) define our area of interest. Smaller ($\sim2.0\times2.3$~arcmin) WFC3 footprints are concentric with the ACS coverage, and are useful for probing the redder sources often associated with \textit{Herschel} flux.

At the time of writing, HFF observations for A2744, M0416 and M0717 are complete, while M1149, A370 and AS1063 are due to be concluded by Cycle 22 (2015--16). For counterpart identification, we use the best available \textit{HST} imaging taken directly from the Hubble Legacy Archive.\footnote{http://hla.stsci.edu/} We defer to a future paper \textit{HST}-derived homogeneous photometric redshifts, stellar masses and morphologies, as this will be possible only when the full HFF observing program is complete. Section \ref{sec:photzs} details the heterogeneous photometric redshifts that we do employ in this study.

\subsubsection{VLA}
\label{sec:vla}

A useful stepping stone in counterpart identification is broadband radio interferometry, which probes the same star formation dominated sources as \textit{Herschel} with arcsec-scale spatial resolution.

M0416, M0717 and M1149, are covered by public Jansky VLA 3~GHz ($\sim$10~cm) imaging data (programme ID: VLA/13B-038, PI: M. Aravena). The resulting maps easily cover HFF central and parallel footprints, with restored beam sizes (FWHM) of 2.5$\times$1.9, 1.8$\times$1.0 and 1.8$\times$1.0~arcsec respectively and a $5\,\sigma$ sensitivity limit of $\sim$6~$\mu$Jy beam$^{-1}$. For A370, we use the VLA 1.4~GHz ($\sim$21~cm) source catalogue presented in \citet{wol12-2}, which covers both HFF footprints within a 40$\times$40~arcmin field. The map has a synthesised beam of 1.8$\times$1.6~arcsec, and a 5$\,\sigma$ noise of $\sim$30~$\mu$Jy beam$^{-1}$ for the HFF centre.

\subsubsection{Spectroscopy}
\label{sec:speczs}

Many of the spectroscopic redshifts used within this study are from large public compilations: \citet{owe11-27} for A2744 (11 redshifts), \citet{wol12-2} for A370 (12), \citet{ebe14-21} for M0416, M0717 and M1149 (23 in total) and \citet{gom12-79} for AS1063 (7). Several further publications provide individual redshifts, as noted in the catalogue tables.

We also present 33 previously unpublished redshifts:

\paragraph{HST grism}

Our catalogue includes 19 spectroscopic redshifts from our own reduction of the publicly available \textit{HST} data of the Grism Lens-Amplified Survey from Space (GLASS; \citealt{sch14-36,wan15-29}), covering the central $\sim$2$\times$2~arcmin of each cluster. The \textit{HST} grism data were reduced using \textsc{aXe} \citep{kum09-59}, and multiple visits were drizzled together using \textsc{MultiDrizzle}, with the routine \textsc{tweakshifts} determining offsets.

\paragraph{LBT/MODS}

Optical specta for 13 objects in M0717 (2 masks) and M1149 (1 mask) were obtained using the 6$\times$6~arcmin FoV LBT/MODS on 19 January 2013 and 8 April 2013. MODS was used in the grating mode with resolutions of 1850 (blue channel) and 2300 (red), pixel scales 0.120~arcsec~pixel$^{-1}$ and 0.123~arcsec~pixel$^{-1}$, and a combined wavelength range of 3200--10500~\AA. The average seeing was between 0.57--0.97~arcsec. Data reduction used \textsc{modsCCDred} and \textsc{modsIDL},\footnote{http://www.astronomy.ohio-state.edu/MODS/Software} with sky-subtraction based on the \citet{kel03-688} 2D B-splines algorithm, reducing residuals from skylines and minimizing noise in the spectra.

\paragraph{Magellan/IMACS}

One spectrum is derived from optical observations of A370 obtained using the Magellan/IMACS f/4 camera on 5 September 2013. The 150 lines/mm grating provides a dispersion of 1.453~\AA~pixel$^{-1}$ and covers a wavelength range of 3650--9740~\AA.  IMACS f/4 has a 15.4$\times$15.4~arcmin FoV with a pixel scale of 0.111~arcmin~pixel$^{-1}$. The average seeing during the observation was 0.68~arcsec. We employed the \textsc{cosmos} data reduction package \citep{dre11-288}, which includes sky-subtraction based on the \citet{kel03-688} algorithm.

\subsubsection{Photometric redshifts}
\label{sec:photzs}

Four of the Frontier Fields (M0416, M0717, M1149, AS1063) are included in the Cluster Lensing And Supernova Survey with Hubble (CLASH; \citealt{pos12-25}). Optical counterparts in these clusters without a spectroscopic redshift rely on well-constrained CLASH photometric redshifts, based on 16-band \textit{HST} photometry or 5--8 band 8~m-class ground-based imaging \citep{ume14-163}. We direct the reader to the above papers or data archive\footnote{https://archive.stsci.edu/prepds/clash/} for further details.

We do not attempt to produce our own optical photometric redshift estimates for the remaining clusters as the complete merging of all optical/NIR ground-based observations will be presented in future papers. However, for counterparts without either a spectroscopic or CLASH photometric redshift, we use an ``IR photometric redshift" estimate, described in Section \ref{sec:irz}.

\section{Merged IR flux catalogue}
\label{sec:phot}

In this section we describe the band-merged photometry methodology, culminating in the presentation of our \textit{Herschel} flux catalogue for HFF.

Band-merged catalogues in the wavelength range covered by \textit{WISE}, \textit{Spitzer} and \textit{Herschel} were produced within the Rainbow Cosmological Surveys Database framework \citep{per08-234}. Photometric extraction follows the method described in \citet{per10-15}. The same procedure was followed for each cluster.

\begin{figure*}
\centering
\includegraphics[width=85mm]{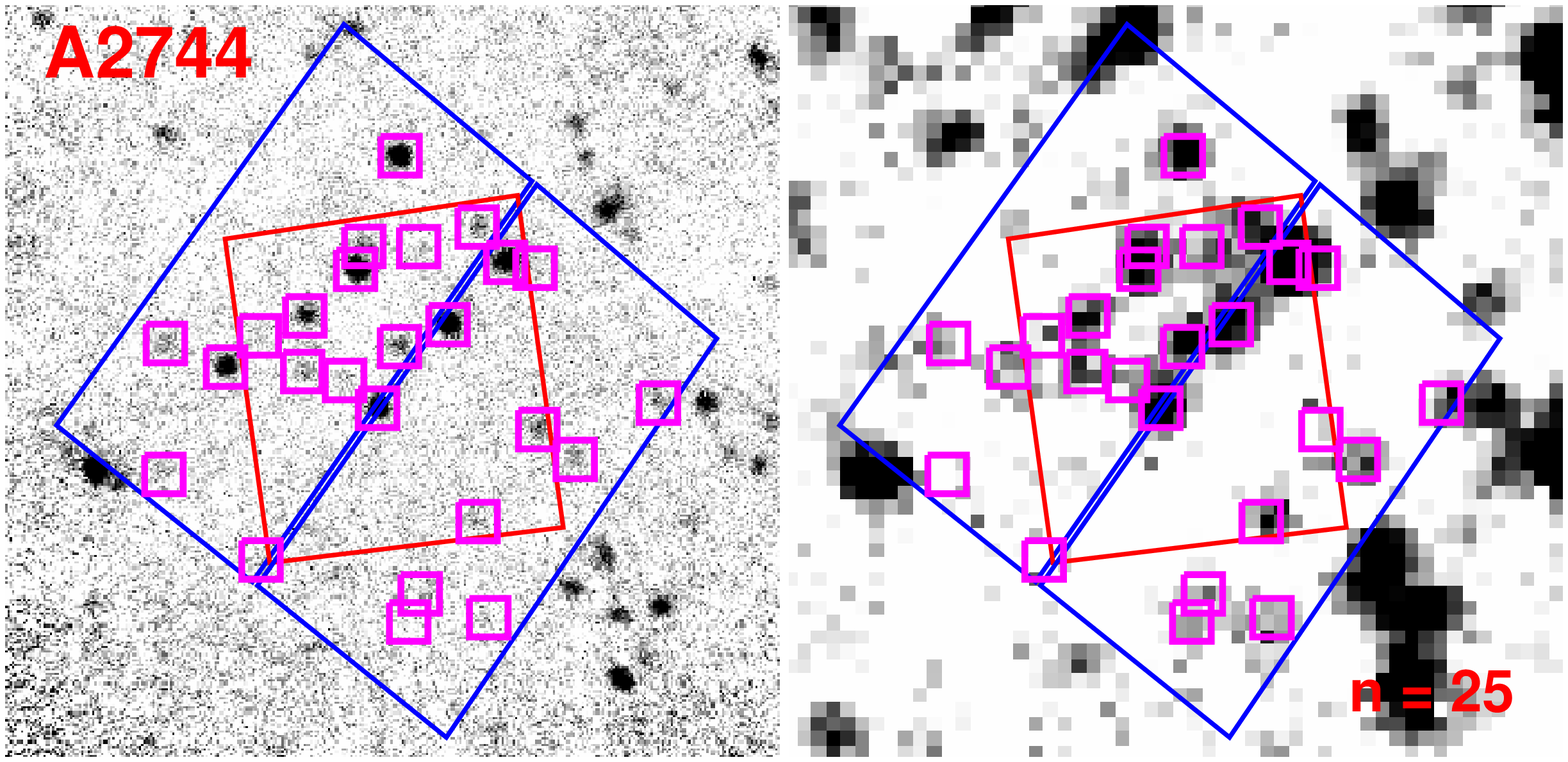}
\hspace{4mm}
\includegraphics[width=85mm]{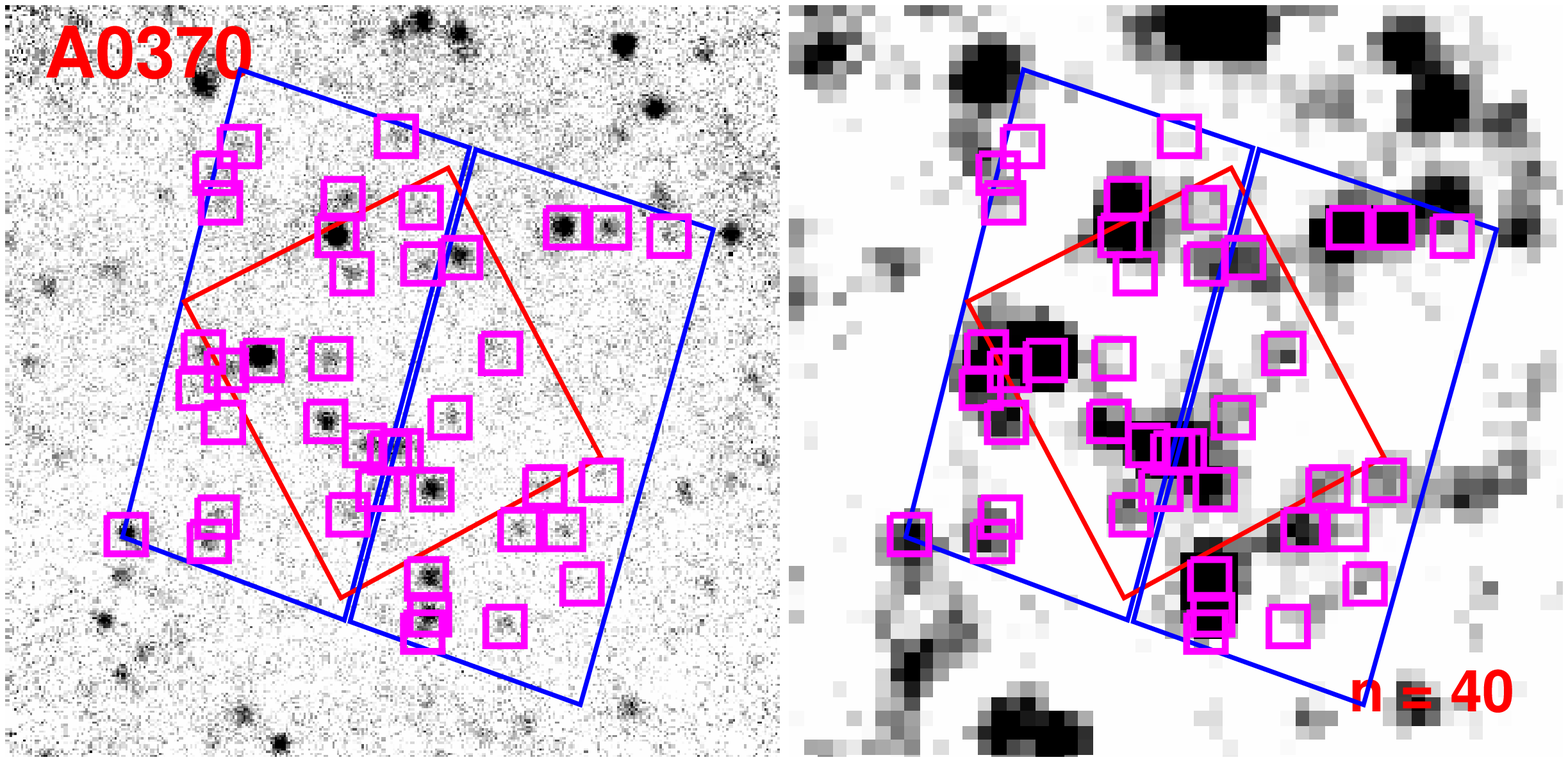}\\
\includegraphics[width=85mm]{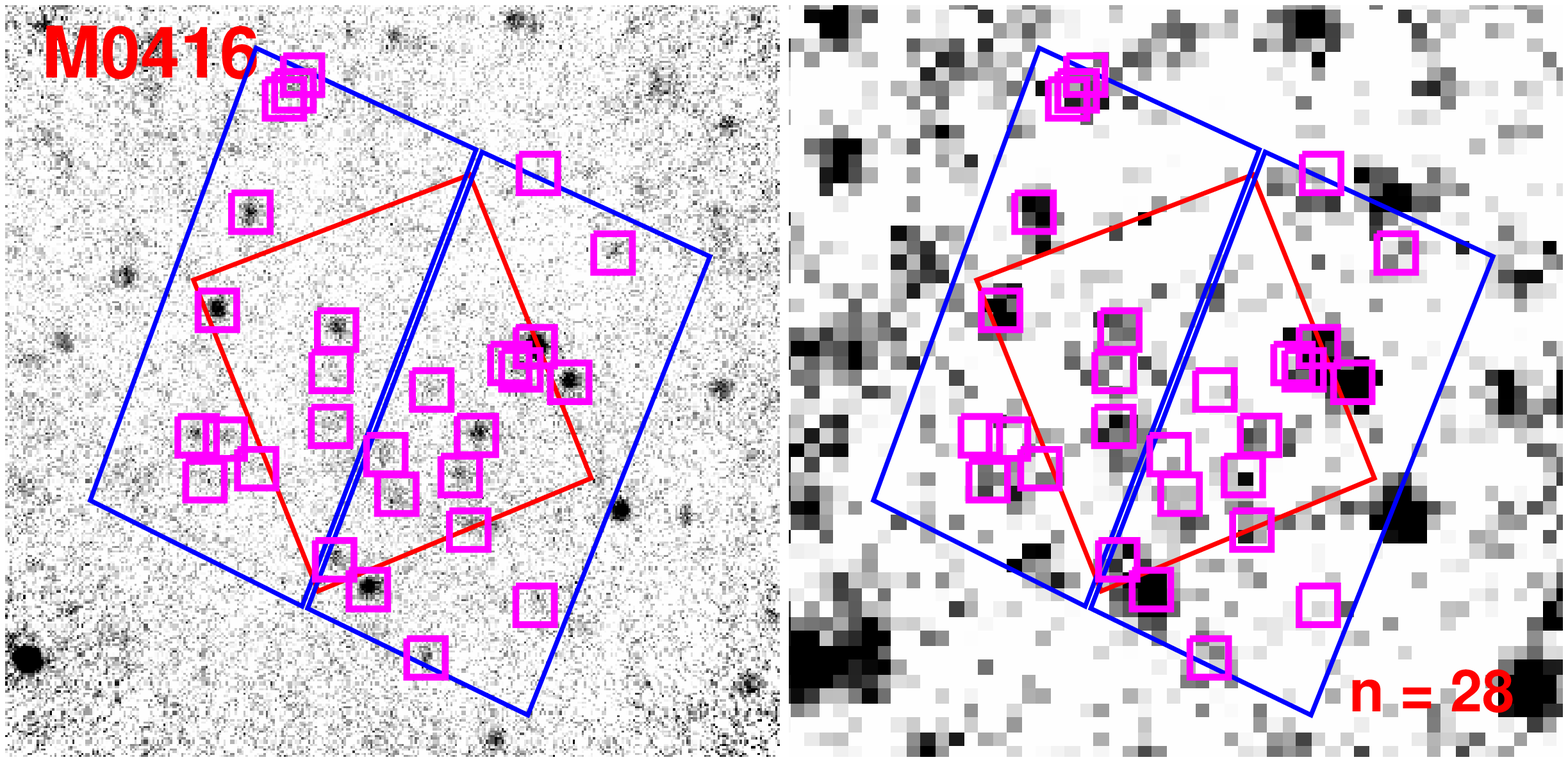}
\hspace{4mm}
\includegraphics[width=85mm]{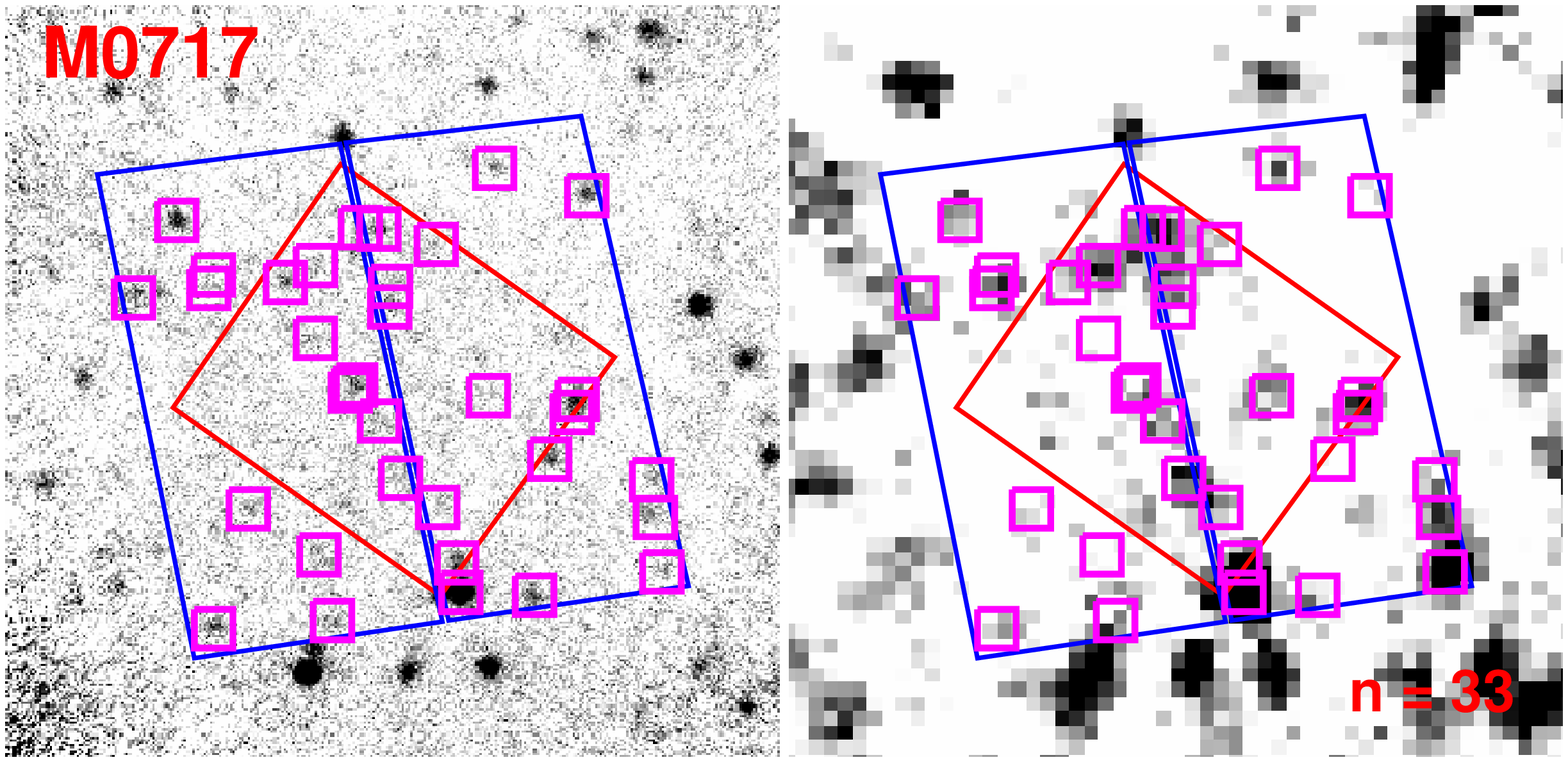}\\
\includegraphics[width=85mm]{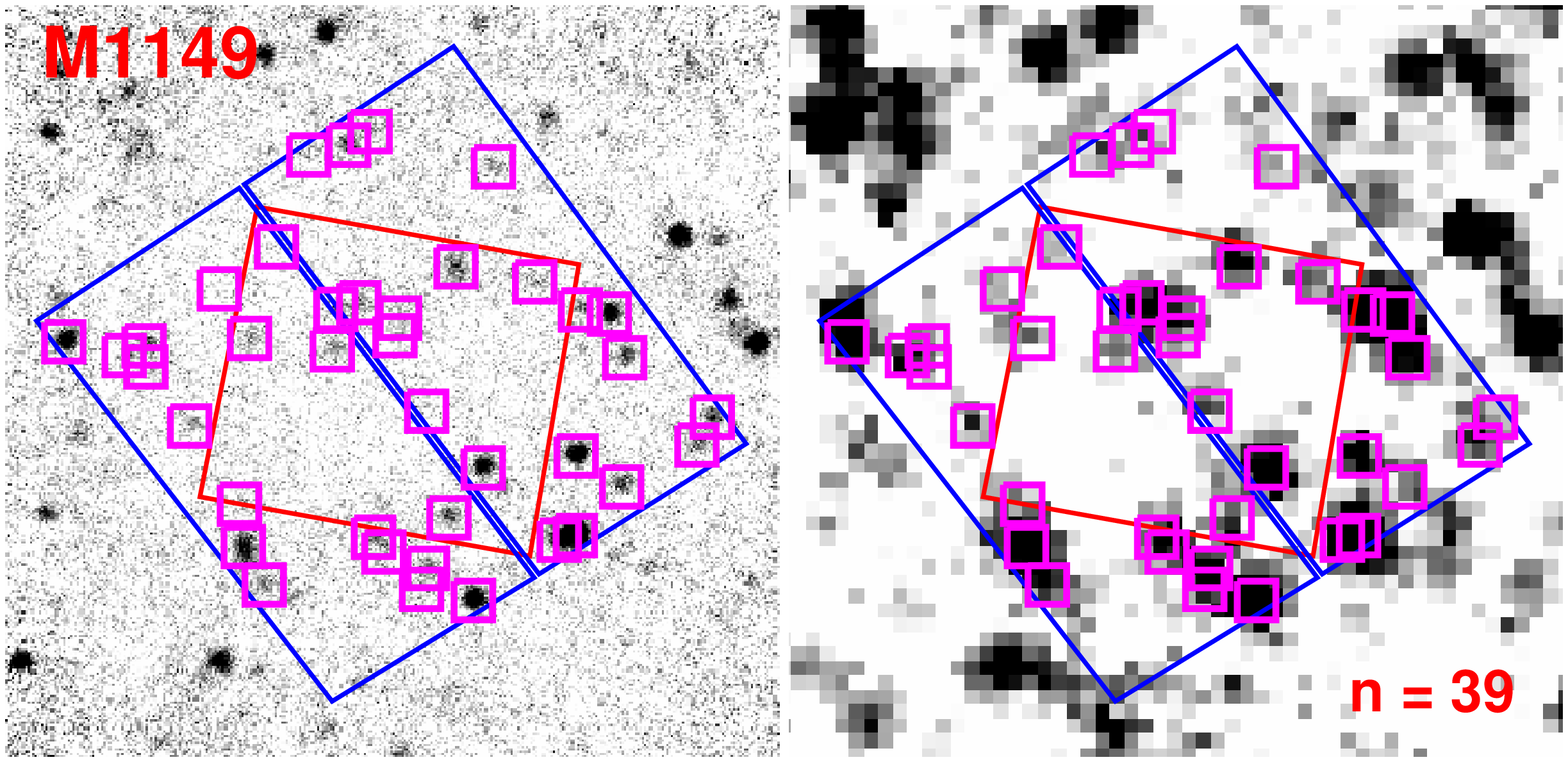}
\hspace{4mm}
\includegraphics[width=85mm]{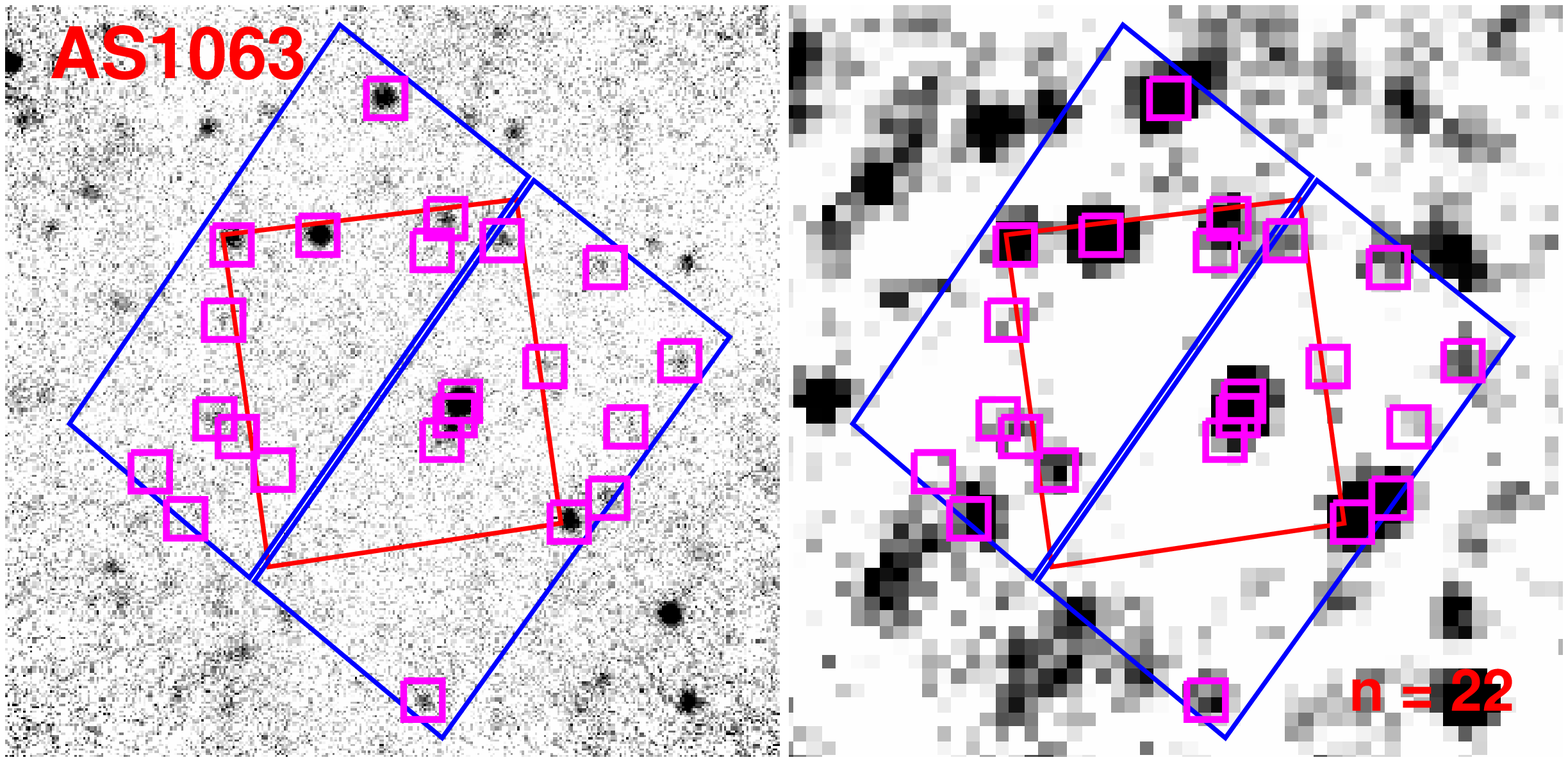}
\caption{Herschel imaging for the HFF central regions: PACS 100~$\mu$m (\textit{left of pair}) and SPIRE 250~$\mu$m (\textit{right of pair}). HFF footprints are marked in red (WFC3) and blue (ACS). \textit{Herschel}-detected sources are highlighted by magenta squares.}
\label{fig:fov_cent}
\end{figure*}

\subsection{Image alignment}

\textit{Herschel} maps were aligned to \textit{Spitzer} astrometry using a global x--y offset based on a stack of the brightest few tens of \textit{Herschel} point sources. In reality, the offset was only calculated for PACS 100 and SPIRE 250~$\mu$m, as the well-known \textit{Herschel} optical models ensure that offsets between bands of the same instrument are well calibrated. Hence we applied the same offset for both PACS bands and a second one for all three SPIRE bands. At 500~$\mu$m, this assumption is crucial, as the large beam size and small number of sources within the \textit{Spitzer}-observed regions render a direct offset characterisation very uncertain. Generally, the offsets required to align pipeline \textit{Herschel} images to the \textit{Spitzer} astrometry are small, with a mean of 1.1 and 1.6~arcsec for PACS and SPIRE respectively (c.f. FWHM $>$5~arcsec).

We also tested the application of a more elaborate transformation to the \textit{Herschel} imaging, including stretch and rotation as well as offset. However, the final maps were negligibly different, the degree-of-freedom was worryingly high for the number of available reference points, and the band-to-band transferal of the transformation was significantly more complicated. Therefore we decided to adopt the simpler x--y-offset transform.

\subsection{Multi-band photometry}
\label{sec:priors}

Photometric extraction in the \textit{Herschel} bands rely on prior catalogues from the bluer, and higher-spatial resolution, \textit{Spitzer} IRAC and MIPS imaging. Therefore, we begun by creating 3$\,\sigma$ direct-detection catalogues for each \textit{Spitzer} band in turn, using an iterative aperture photometry technique. Each pair of \textit{Spitzer} catalogues were merged via a simple counterpart algorithm, using a maximum search radius of the largest PSF FWHM in the pair, to produce a master IRAC/MIPS catalogue. The IRSA \textit{WISE} catalogue was trivially linked to the IRAC/MIPS catalogue by finding the nearest IRAC--\textit{WISE} source pairs.

Photometry for each \textit{Herschel} image used a simultaneous PSF-fitting algorithm, allocating FIR flux in each band to shorter-wavelength priors. While this method is very good at de-blending close, but ultimately resolved complexes of \textit{Herschel} flux, the large number of priors from deep IRAC imaging runs the risk of incorrectly de-blending \textit{Herschel} point sources into several constituent parts. To avoid the latter, we replaced close (sub-\textit{Herschel} PSF FWHM) groups of \textit{Spitzer} sources with a single `pseudo-source', corresponding to the position of the brightest source in that group. For each group, the number (`multiplicity') and identity of all sources is propagated through the scripts. The choice of position for a pseudo source is ultimately not significant, as we also allowed small offsets within the simultaneous PSF fitting mechanism. Furthermore, although we start with the location of the brightest grouped source for the prior-based fit, that does not influence which IRAC source we ultimately select as the counterpart (see below, and Section \ref{sec:cps}).

We began with PACS 100~$\mu$m, as the band with the best combination of spatial resolution, sensitivity and coverage for all clusters. We adapted the \textit{Spitzer} prior catalogue by grouping sources within 5~arcsec diameters of each other and then proceeded with the simultaneous PSF fitting, allowing for small offsets necessary due to the effect of different pixel sizes. Fluxes were derived via aperture photometry on the fitted PSF, allowing us to use the same aperture and corrections as for a direct detection. The resulting catalogue automatically includes prior information for each PACS source, and where a `pseudo-source' is allocated PACS flux, all potential counterparts are recorded. For sources detected by PACS but too faint for, or beyond the coverage of, \textit{Spitzer}, we also produced a list of direct-detections based on the residual image, ie the remaining 100~$\mu$m sources after all the fitted flux has been removed. Flux was measured using aperture photometry and the direct detections were added to the PACS 100~$\mu$m photometry catalogue and flagged.

The full procedure was then repeated for 70 and 160~$\mu$m, using the PSF FWHM to define the \textit{Spitzer} prior catalogue grouping diameter. After completing the simultaneous fitting and direct detection stages, we merged the independent PACS catalogues. First we matched common IRAC/MIPS priors and psuedo-sources, and then used a search radius of 4~arcsec to match the remainder. For PACS sources undetected in one or two other PACS band, we force photometric measurement in the residual images.

For the SPIRE bands, the same iterative procedure was applied, varying the prior catalogue grouping scale to correspond with the SPIRE PSF FWHM. The only other difference is the inclusion of the direct-detection PACS sources in the prior catalogue (in effect a merged \textit{Spitzer}/PACS prior list). Once a catalogue was produced for each of the three SPIRE bands independently (including the direct detection step on the residual maps), we proceeded with the SPIRE-band merging, following the example of the PACS bands, with forced photometry in the cases of single-band non-detections.

The prior-based methodology helps to de-blend confused \textit{Herschel} sources, and ensures a straight-forward combination of the inter-connected IRAC, WISE, MIPS, PACS and SPIRE catalogues. Furthermore, multiple possible priors from the grouping procedure can often be reduced to a single dominant prior by accounting for the location of intermediary MIPS or PACS flux and/or the inferred SED. For example, most SPIRE sources have a single PACS counterpart, which in turn corresponds to a single MIPS source, which itself has one plausible IRAC counterpart (based on flux or SED shape). So although several IRAC sources may have been included within the grouping for a SPIRE source, the intermediate bands suggest a unique counterpart.

\subsection{HFF \textit{Herschel} flux catalogue}
\label{sec:fluxcat}

\begin{table*}
\caption{Observed PACS and SPIRE fluxes for \textit{Herschel}-detected sources within the \textit{HST} Frontier Fields.}
\label{tab:cat1}
\begin{tabular}{lrrrrrrr}
\hline
\multicolumn{1}{c}{ID (ref)$^1$} & \multicolumn{1}{c}{Field$^2$} & \multicolumn{1}{c}{$S_{70}$} & \multicolumn{1}{c}{$S_{100}$} & \multicolumn{1}{c}{$S_{160}$} & \multicolumn{1}{c}{$S_{250}$} & \multicolumn{1}{c}{$S_{350}$} & \multicolumn{1}{c}{$S_{500}$} \\
 & & \multicolumn{1}{c}{mJy} & \multicolumn{1}{c}{mJy} & \multicolumn{1}{c}{mJy} & \multicolumn{1}{c}{mJy} & \multicolumn{1}{c}{mJy} & \multicolumn{1}{c}{mJy} \\
\hline
HLSJ001412.7--302359 (P) & A2744 C & -- & 5.2 $\pm$ 0.9 & 10.5 $\pm$ 1.8 & 13.1 $\pm$ 2.9 & 6.8 $\pm$ 3.0 & -- \\
HLSJ001415.3--302423 (P) & A2744 C & -- & 5.6 $\pm$ 0.8 & 10.2 $\pm$ 2.1 & 14.8 $\pm$ 2.8 & -- & -- \\
HLSJ001416.7--302304 (P) & A2744 C & -- & 2.5 $\pm$ 0.6 & 7.5 $\pm$ 1.3 & -- & -- & -- \\
HLSJ001416.5--302410 (P) & A2744 C & -- & 7.2 $\pm$ 0.9 & 11.8 $\pm$ 1.6 & 9.8 $\pm$ 5.5 & 8.4 $\pm$ 5.9 & 6.3 $\pm$ 9.2 \\
HLSJ001418.5--302246 (P) & A2744 C & -- & 7.6 $\pm$ 0.9 & 12.4 $\pm$ 1.6 & 12.6 $\pm$ 3.3 & 6.3 $\pm$ 2.8 & -- \\
HLSJ001417.6--302301 (P) & A2744 C & -- & 25.5 $\pm$ 2.1 & 62.4 $\pm$ 5.1 & 66.1 $\pm$ 5.2 & 47.5 $\pm$ 4.8 & 21.5 $\pm$ 4.8 \\
HLSJ001418.5--302448 (P) & A2744 C & -- & 4.5 $\pm$ 0.9 & 11.9 $\pm$ 1.6 & 18.5 $\pm$ 2.6 & 14.2 $\pm$ 2.9 & 8.1 $\pm$ 2.6 \\
HLSJ001418.0--302529 (S) & A2744 C & -- & -- & -- & 7.7 $\pm$ 2.3 & 12.1 $\pm$ 3.4 & 14.2 $\pm$ 3.3 \\
... & ... & ... & ... & ... & ... & ... & ... \\
\hline
\\
\end{tabular}
\\
\raggedright
 \textit{[Note: The full table is published in the electronic version of the paper. A portion is shown here for illustration.]}\\
$^1$ \textit{Herschel} ID derived from PACS (P) or SPIRE (S) catalogue position\\
$^2$ C=central region; P=parallel\\
\end{table*}

\begin{figure}
\centering
\includegraphics[width=41.5mm]{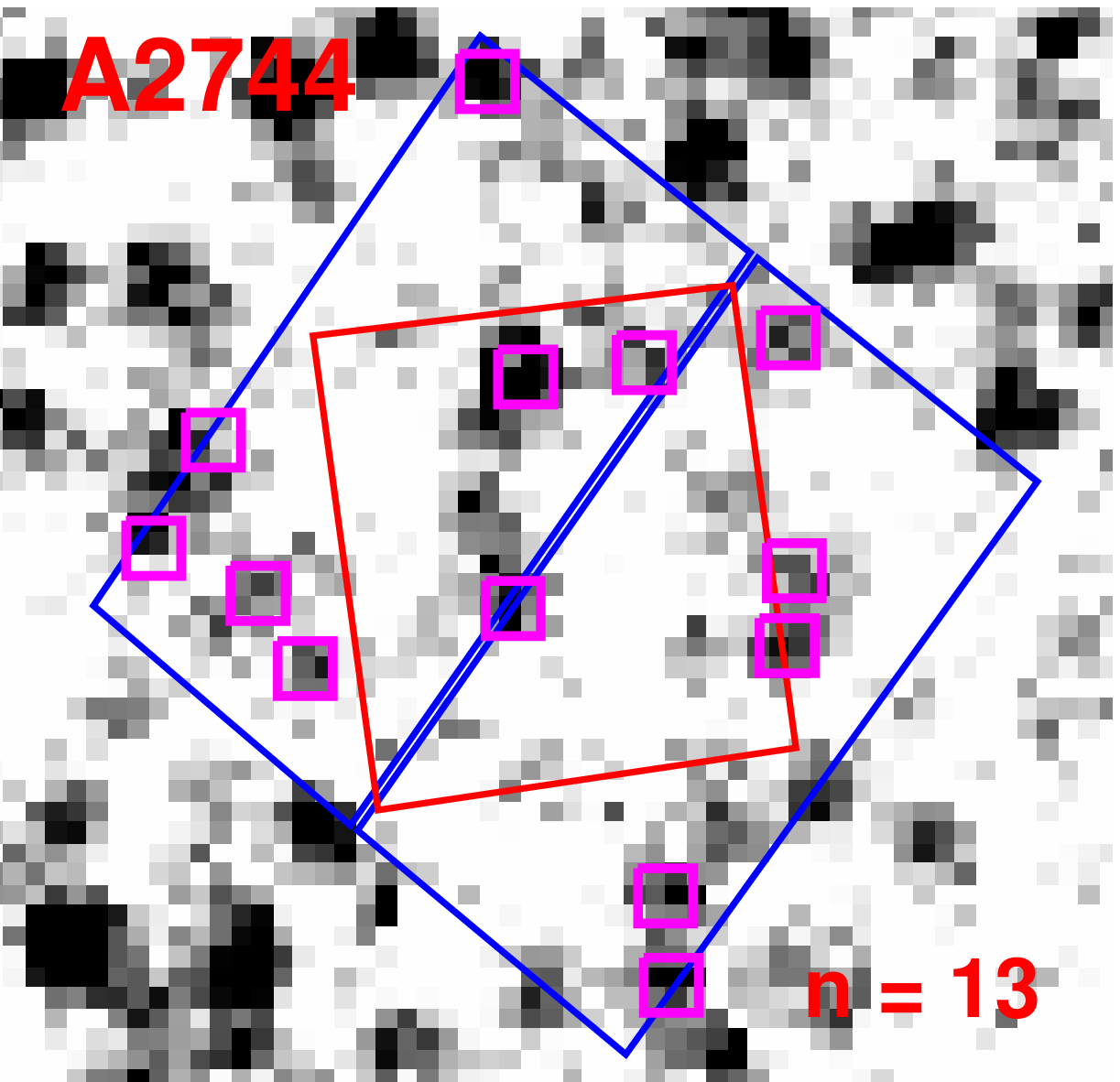}
\includegraphics[width=41.5mm]{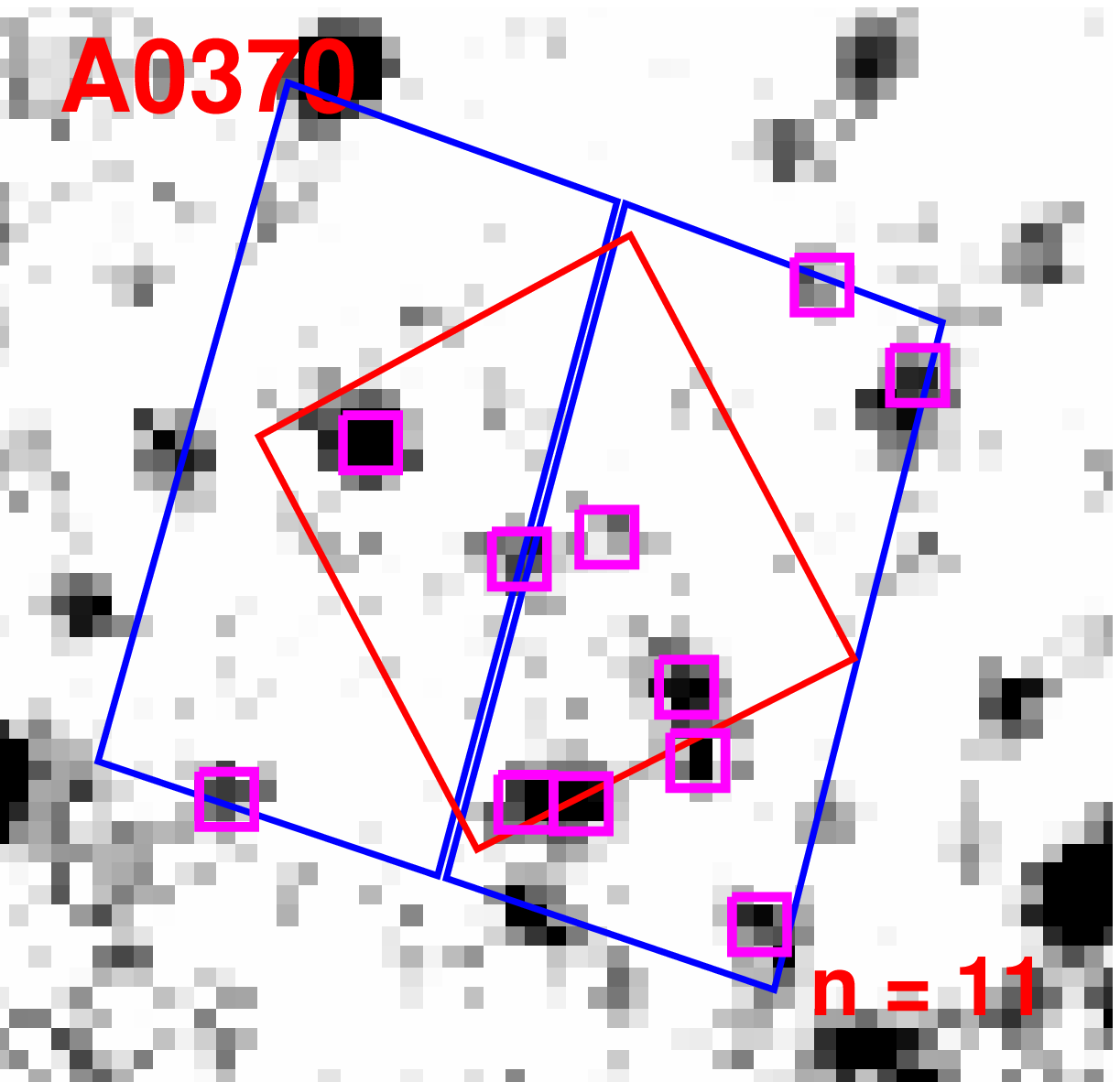}\\
\includegraphics[width=41.5mm]{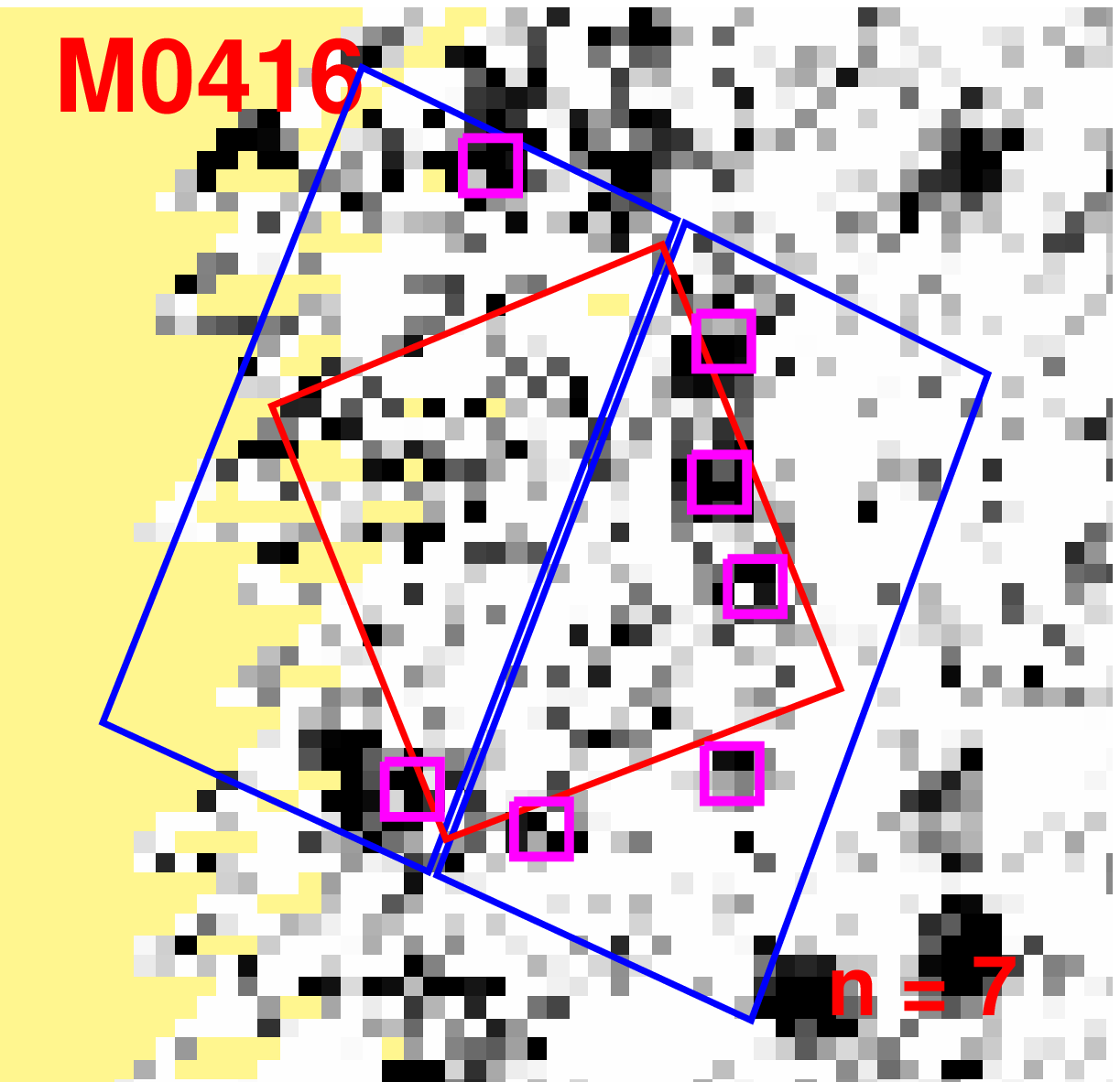}
\includegraphics[width=41.5mm]{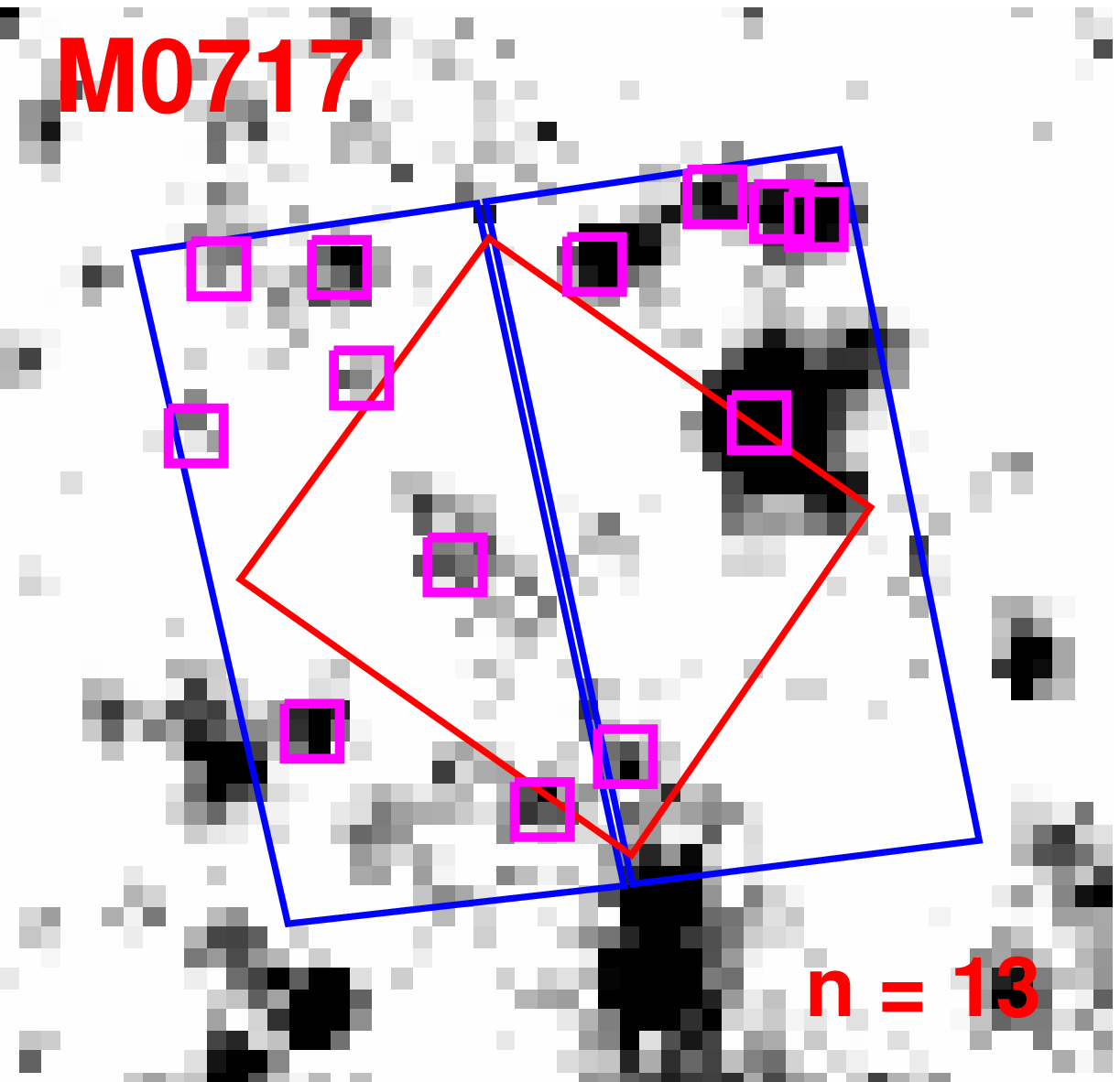}\\
\includegraphics[width=41.5mm]{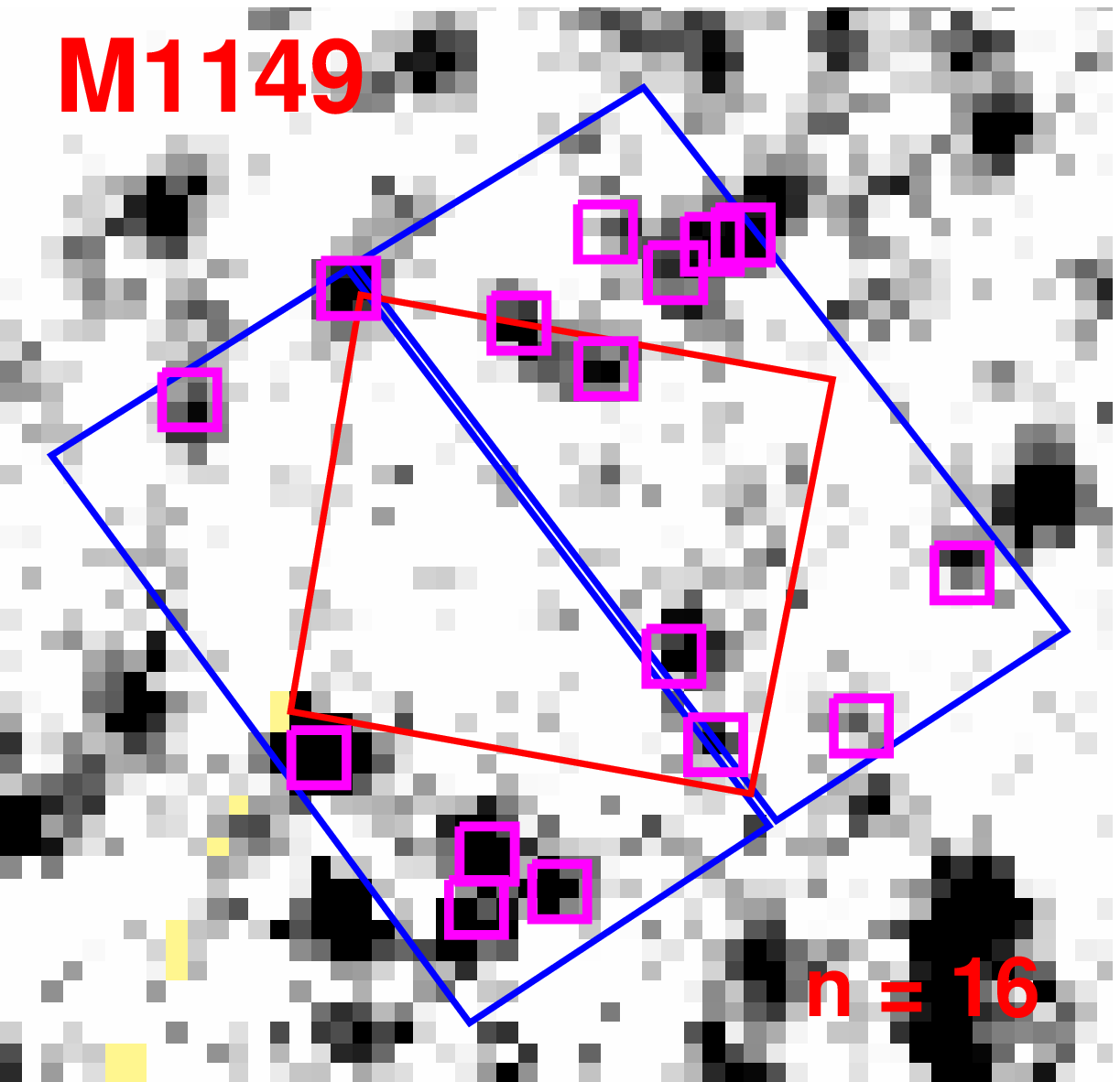}
\includegraphics[width=41.5mm]{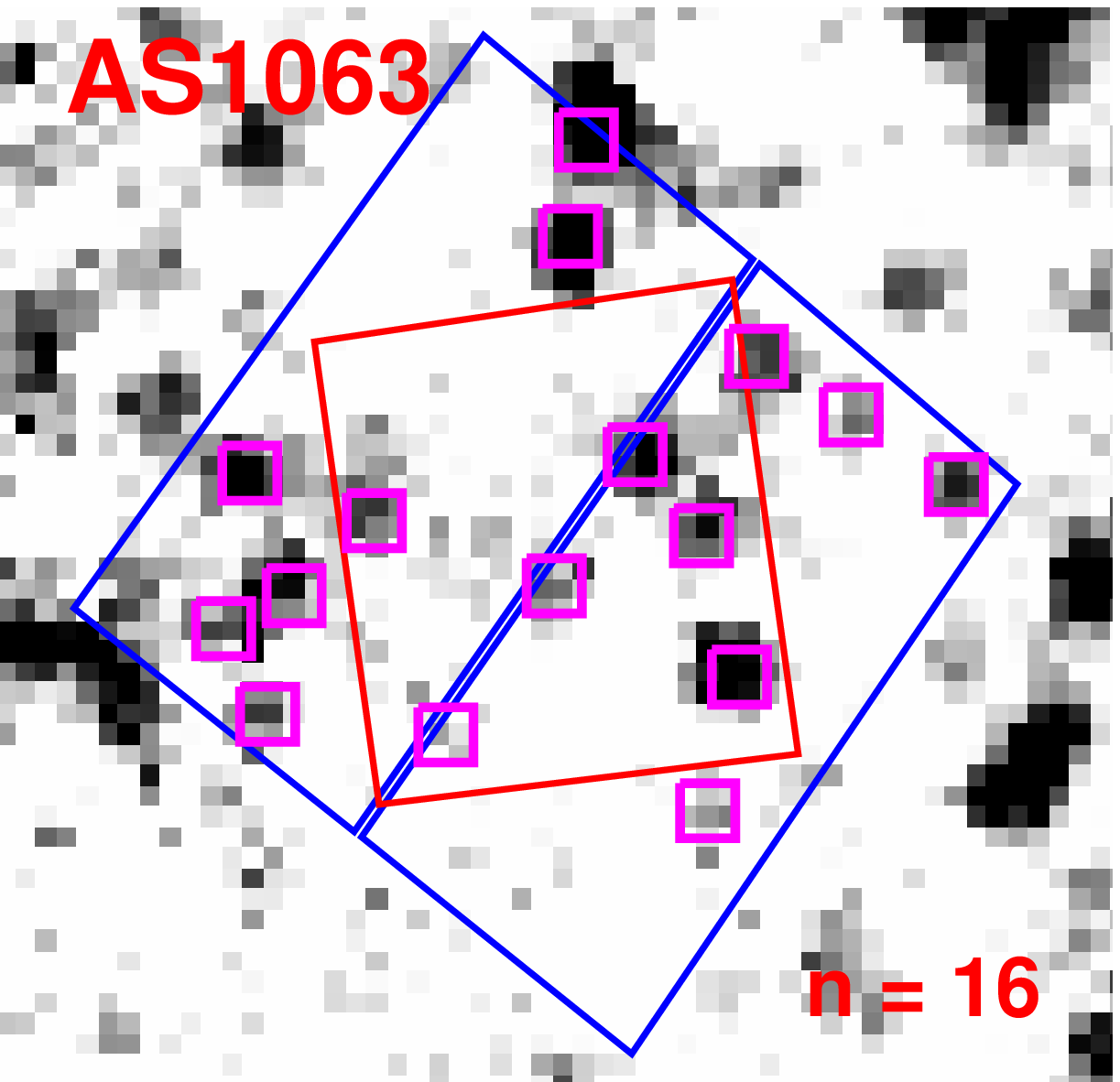}
\caption{Herschel SPIRE 250~$\mu$m imaging for the HFF parallel fields (markings as in Figure \ref{fig:fov_cent}). Regions beyond the SPIRE coverage are shaded yellow.}
\label{fig:fov_par}
\end{figure}

Our primary goal is to create a \textit{Herschel} catalogue within the HFF footprints, so every source beyond the ACS image boundaries\footnote{http://www.stsci.edu/hst/campaigns/frontier-fields/HST-Survey} is ignored.

The only other selection criterion concerns the definition of ``\textit{Herschel}-detected'': any source having at least two $>$4$\,\sigma$-significance \textit{Herschel} fluxes, regardless of filter or detection by any of the ancillary facilities. We stress that by simultaneously fitting to priors, we will account for and remove fainter ($<$4$\,\sigma$) \textit{Herschel} source flux from the fluxes of brighter galaxies, even if they themselves are not included in the final catalogue. However, we caution that we may be bias against the faintest, reddest IR sources which exhibit a dust continuum peak at SPIRE 500~$\mu$m or beyond and are extremely faint in bluer \textit{Herschel} bands.

The final catalogue contains 263 \textit{Herschel}-detected sources, 187 within the central regions and a further 76 within the parallel footprints. The difference in source density (the two footprints have equal area) can be trivially linked to the lack of \textit{Spitzer} and PACS coverage for large parts of the parallel fields, meaning that detection of a \textit{Herschel} source relies on SPIRE alone. Figures \ref{fig:fov_cent} and \ref{fig:fov_par} identify the sources in PACS (for central footprints) and SPIRE (for both footprints). \textit{Herschel} fluxes are presented in Table \ref{tab:cat1}.

There are 230/263 catalogue sources with a \textit{Spitzer} counterpart and 131/263 detected by \textit{WISE} (112 overlap). Of the \textit{Spitzer} sources, 146/230 have both PACS and SPIRE detections, while 36 are only detected in PACS and 48 by SPIRE alone. The mean grouping `multiplicity' of the priors for the PACS sources is $\sim$1.3, which indicates that matching PACS to IRAC is relative trivial. For the SPIRE sources this mean multiplicity rises to $\sim$3.4.

The 33/263 sources without a \textit{Spitzer} counterpart are almost exclusively located beyond the available IRAC and MIPS coverage, and are direct-detections in either PACS or SPIRE. Only two were direct detections in PACS, and both are also detected by SPIRE. One of those (HLSJ001401.3--302224 in A2744) lies just beyond the \textit{Spitzer} imaging and just inside the PACS coverage, but is well detected by WISE. The other (HLSJ041606.6--240528 in M0416) is located within the core of the cluster, in a region densely populated and extremely blended in IRAC (and also in WISE). This is the only case for which we were unable to extract IRAC photometry for a \textit{Herschel} source within the IRAC image bounds. Unfortunately, MIPS is unavailable for M0416.

The remaining 31 sources are SPIRE direct-detections in the parallel fields beyond current IRAC, MIPS or PACS coverage. 18/31 are detected by \textit{WISE}. Therefore, only 14/263 \textit{Herschel} sources (1 PACS + 13 SPIRE) lack near-infrared counterparts and SED information between 1--70~$\mu$m. Pre-empting the next section, it is worth noting here that despite the lack of \textit{Spitzer} or \textit{WISE} data, two of these 14 sources have VLA (and hence secure \textit{HST}) counterparts, and a further five have unambiguous \textit{HST} associations.

\section{Counterpart analysis}
\label{sec:cps}

Analysis of the intrinsic properties (e.g. IR luminosity, dust temperature) of \textit{Herschel} sources relies on a knowledge of the redshift. This is particularly true for background objects, as the magnification factor ($\mu$) is also a function of redshift behind the lensing mass.

We therefore attempt to provide a best-effort `value-added' catalogue for the sources detected by \textit{Herschel}, identifying the most likely optical counterpart, which in turn leads to redshift information.

\subsection{Counterpart identification}

\begin{figure*}
\centering
\includegraphics[width=175mm]{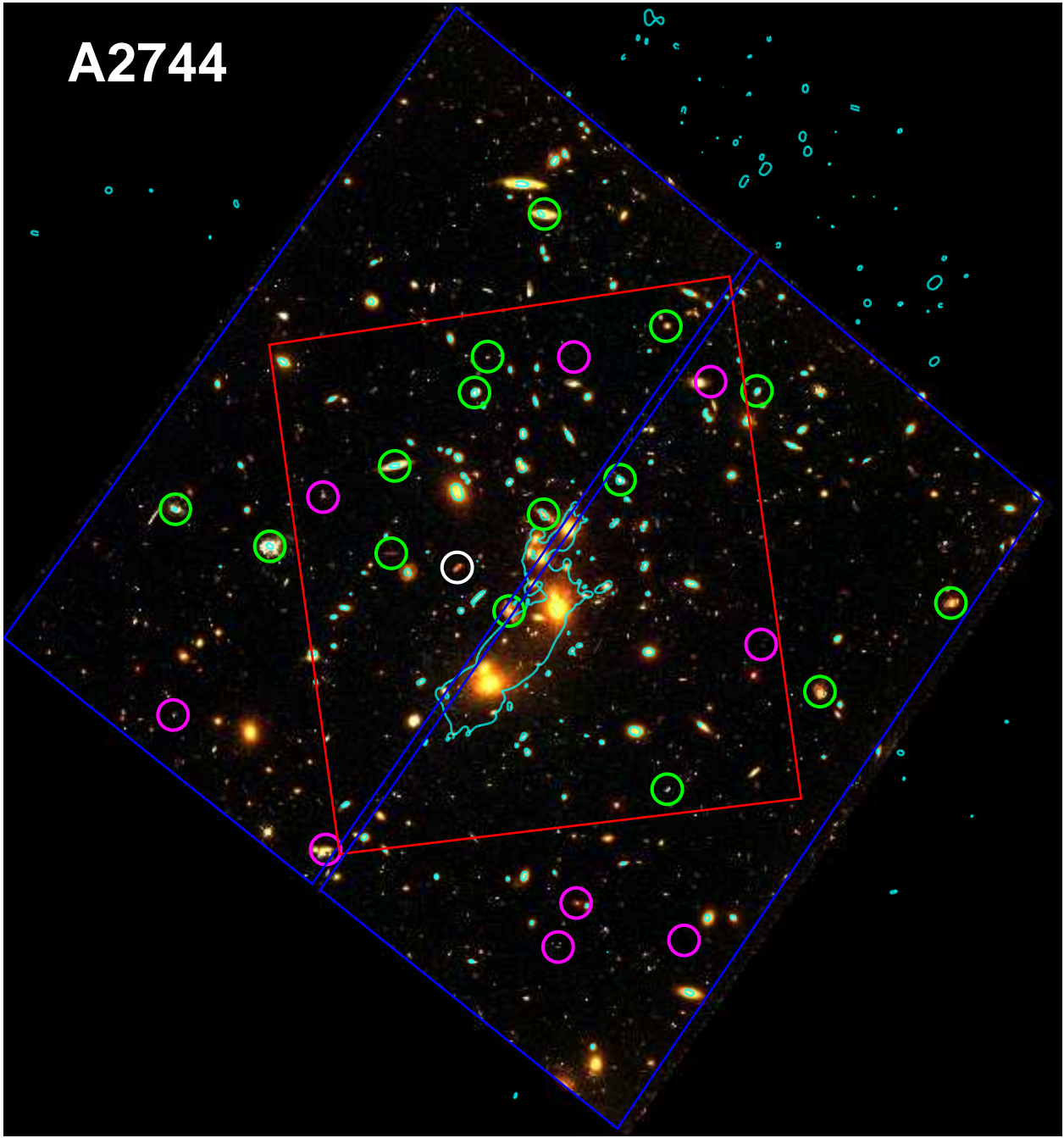}
\caption{\textit{HST} 3-colour image of the A2744 central field, with footprints as in Figure \ref{fig:fov_cent} and lensing critical lines for background sources at $z=1$ given in cyan (from the CATS models; see Section \ref{sec:mag}). \textit{Herschel}-detected sources are shown by circles (corresponding to the PACS PSF FWHM $\sim8$~arcsec) and show many counterparts are readily identifiable. Circle colours indicate the origin of the counterpart redshift: green$=$spectroscopic, white$=$CLASH optical photometric, magenta$=$IR-based estimate.}
\label{fig:a2744_hst}
\end{figure*}

\begin{figure*}
\centering
\includegraphics[width=175mm]{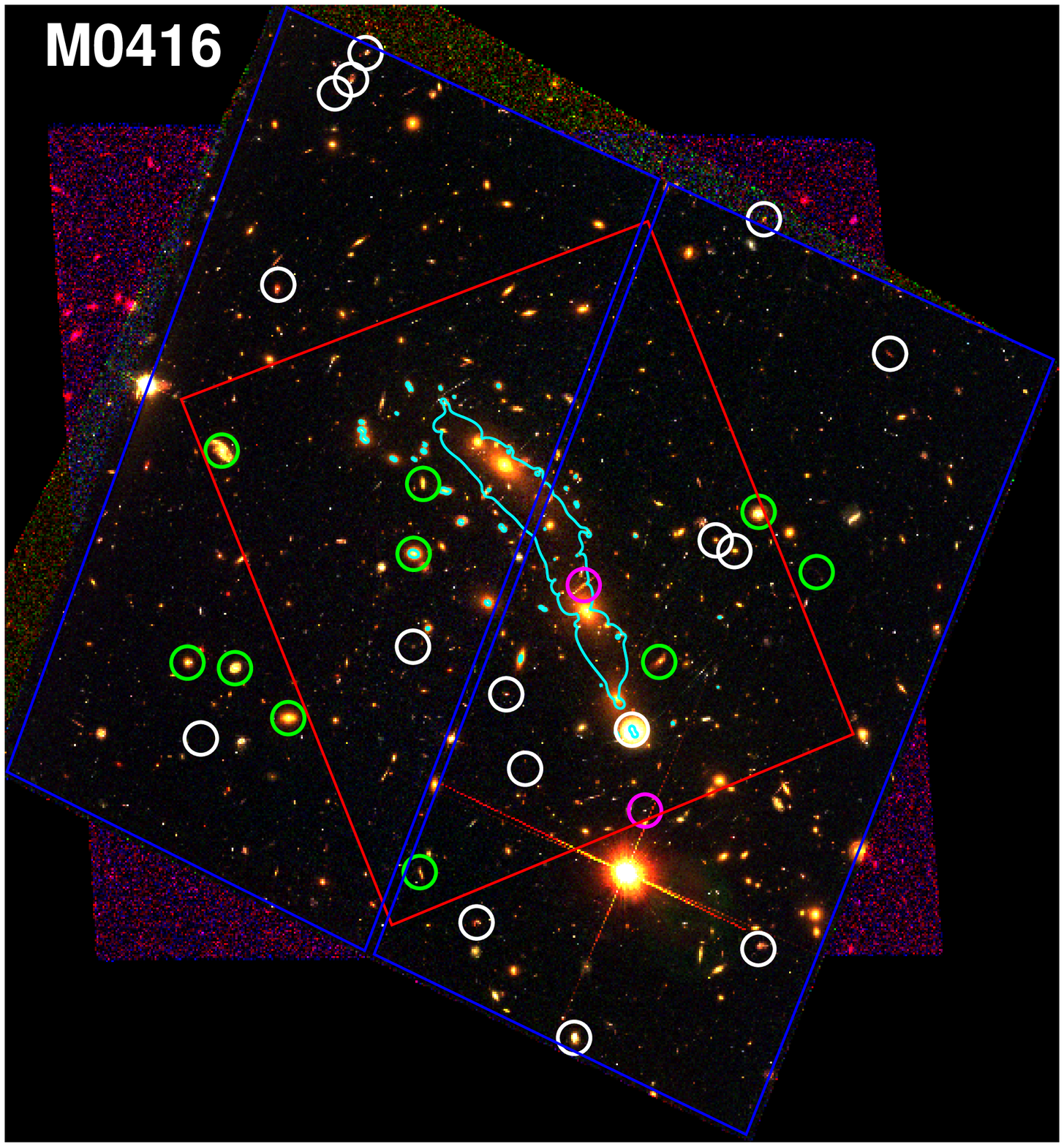}
\caption{\textit{HST} 3-colour image of the M0416 central field. Layout as in Figure \ref{fig:a2744_hst}.}
\label{fig:m0416_hst}
\end{figure*}

\begin{figure*}
\centering
\includegraphics[width=175mm]{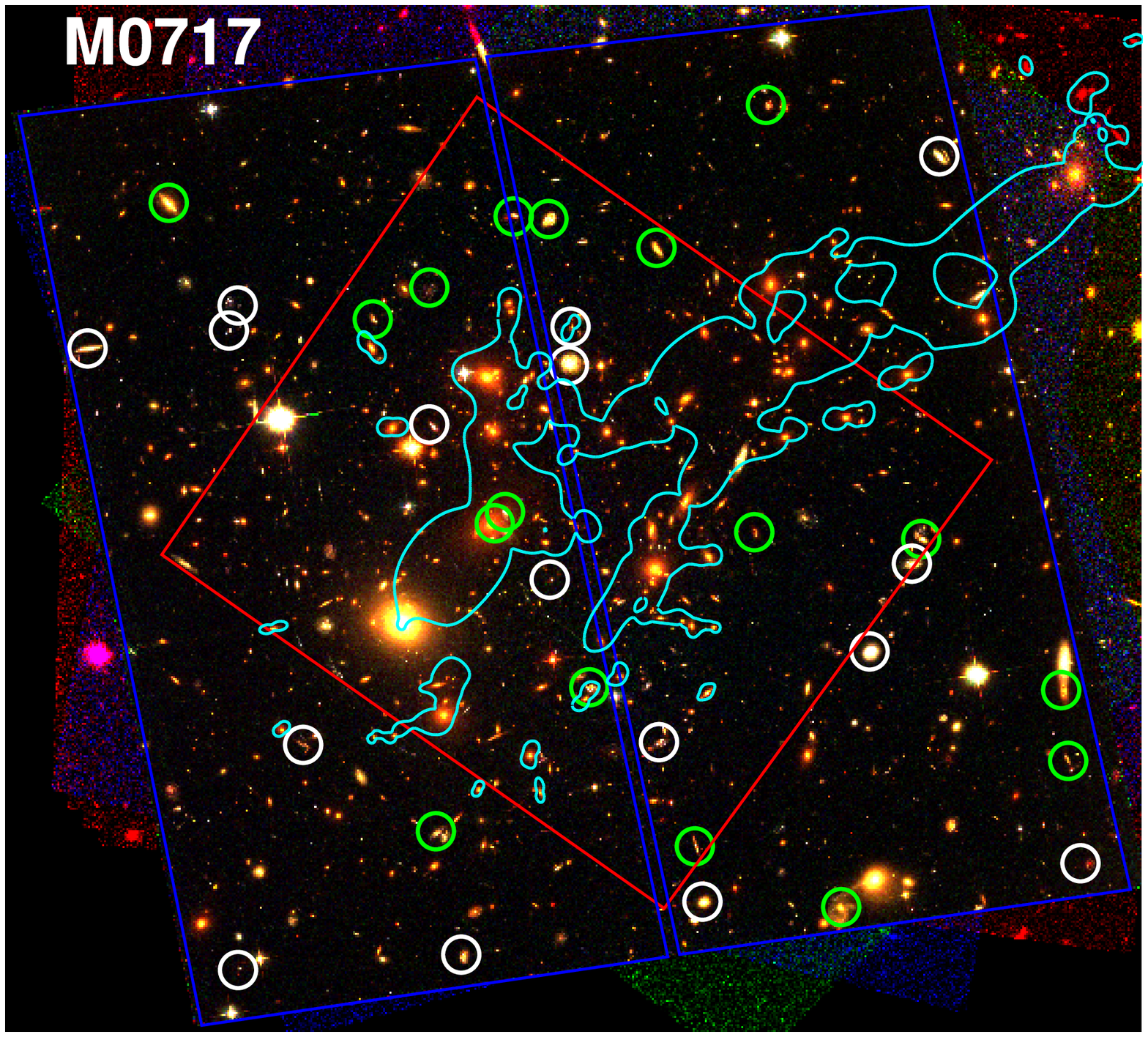}
\caption{\textit{HST} 3-colour image of the M0717 central field. Layout as in Figure \ref{fig:a2744_hst}.}
\label{fig:m0717_hst}
\end{figure*}

For each IR source we locate the best optical counterpart using a variety of techniques and all available data. The first-guess counterpart for many \textit{Herschel} sources was the \textit{HST} object at the position of the prior-based IRAC source. For the three clusters with complete HFF \textit{HST} images (A2744, M0416 and M0717), Figures \ref{fig:a2744_hst}--\ref{fig:m0717_hst} highlight that there is an obvious optical counterpart for many of the \textit{Herschel}-detected sources. We verify these by ensuring that the IR SED (i.e. dust peak and temperature) is reasonable when assuming that the IR source is at the optical redshift. We similarly eliminate other nearby optical sources as potential candidates.

Previous work on optical--FIR source matching has employed a similar approach to our PSF fitting \citep[e.g.][]{hwa10-75,elb11-119,dom14-2,pap15-129}. Working with the larger GOODS catalogues, \citeauthor{elb11-119} kept only fluxes that were considered reliable by a `clean index', somewhat similar to our `multiplicity' in that it encapsulates the number of prior sources within the \textit{Herschel} PSF FWHM. However, the small size of the HFF sample allowed us to individually examine each source, and find the best candidate counterpart for each. This is particularly important for sources in which the shortest-wavelength (i.e. highest resolution) \textit{Hershel} flux originated from a `grouped' \textit{Spitzer} counterpart; we do not want to just use the brightest IRAC source. Instead, we consult the imaging and SED to choose the best fit counterpart, as described in Section \ref{sec:priors}. When multiple potential counterparts are at a similar on-sky position but different redshift, the shape of the IR continuum (discussed in Section \ref{sec:irsed}) can be used as an effective discriminator.

Furthermore, we do not just use the IR SED and \textit{HST} imaging to identify the candidate counterpart. As mentioned in Section \ref{sec:vla}, high spatial resolution radio observations offer a valuable additional constraint on the origin of the IR emission, often allowing us to pinpoint the responsible \textit{HST} counterpart or even sub-component \citep{tho15-1874,gea15-502}. 91 \textit{Herschel} sources have an associated radio detection, from 187/263 located within available VLA maps. We use these to successfully validate our chosen \textit{HST}/IRAC counterpart, and find very few mismatches. We estimate that of the remaining 172 \textit{Herschel}-detected sources, the counterpart of $<$5\% would change given additional VLA observations (such as the upcoming HFF programme, PI: Murphy). Such future refinements will be applied to the online version of our ``value-added'' \textit{Herschel} catalogue as required.

We emphasise here the prior-based source extraction, ``by-eye'' multi-wavelength counterpart matching (including radio data) and optical--IR SED fitting were not executed sequentially or in isolation. The best counterpart for each individual \textit{Herschel} source was selected based on all of the available information simultaneously.

\subsection{Counterpart redshift}

In this paper, the counterparts are primarily used to yield a source redshift. More than 40\% (112/263) have spectroscopic redshifts, taken from various studies (as described in Section \ref{sec:speczs}). A further 89 have well-constrained photometric redshift estimates from the multi-band CLASH catalogues (see Section \ref{sec:photzs}). The remaining 62 sources have a redshift estimated from the best-fitting IR SED templates, which is described further in Section \ref{sec:irz}. This will be updated in the online ``value-added'' catalogue as spectroscopic surveys target further \textit{Herschel} counterparts. Table \ref{tab:redshifts} provides a full breakdown of the number of sources with each type of redshift, while Figures \ref{fig:a2744_hst}--\ref{fig:m0717_hst} also indicate their origin. We discuss the distribution of redshifts in Section \ref{sec:distributions}.

\begin{table}
\caption{Number of sources within each \textit{HST} footprint, binned by the origin of the counterpart redshift.}
\label{tab:redshifts}
\begin{tabular}{lrrrrrrrrrrr}
\hline
\multicolumn{1}{c}{Cluster}  & \multicolumn{4}{c}{Central} & \multicolumn{4}{c}{Parallel} & \multicolumn{1}{c}{\textbf{Total}} \\
& \multicolumn{1}{c}{S$^1$} & \multicolumn{1}{c}{P$^2$} & \multicolumn{1}{c}{IR$^3$} & \multicolumn{1}{c}{\textbf{Tot}} & \multicolumn{1}{c}{S$^1$} & \multicolumn{1}{c}{P$^2$} & \multicolumn{1}{c}{IR$^3$} & \multicolumn{1}{c}{\textbf{Tot}} & \\
\hline
A2744 & 15 & 1 & 9 & \textbf{25} & 3 & 0 & 10 & \textbf{13} & \textbf{38} \\
A370 & 26 & 0 & 14 & \textbf{40} & 2 & 0 & 9 & \textbf{11} & \textbf{51} \\
M0416 & 10 & 16 & 2 & \textbf{28} & 0 & 7 & 0 & \textbf{7} & \textbf{35} \\
M0717 & 17 & 16 & 0 & \textbf{33} & 3 & 10 & 0 & \textbf{13} & \textbf{46} \\
M1149 & 23 & 16 & 0 & \textbf{39} & 2 & 14 & 0 & \textbf{16} & \textbf{55} \\
AS1063 & 11 & 9 & 2 & \textbf{22} & 0 & 0 & 16 & \textbf{16} & \textbf{38} \\
\hline
\textbf{Total} & 102 & 58 & 27 & \textbf{187} & 10 & 31 & 35 & \textbf{76} & \textbf{263} \\
\hline
\end{tabular}
\\
\raggedright
$^1$ Spectroscopic redshift\\
$^2$ CLASH \textit{HST} or Subaru photometric redshift\\
$^3$ IR SED redshift estimate\\
\end{table}

\subsection{Magnification factors}
\label{sec:mag}

Sources within the foreground or associated with the cluster itself (typically taken as $z<z_{\rm cl}+0.1$) are assumed to have a zero magnification.

Recent years have seen a huge effort to constrain lensing models for the Frontier Fields. Within the central areas, we employ the CATS models \citep{jau12-3369, ric14-268, jau14-1549} derived using the publicly available \textsc{Lenstool} package \citep{jul09-1319}. Within CATS, the cluster mass is represented by a combination of two types of component: (1) known cluster galaxies, with spectroscopically confirmed redshifts and masses scaled from their luminosity, (2) one or more large-scale, smooth haloes. The models were refined by comparing predictions to the known sets of well-constrained multiple images in each cluster. The CATS lens models cover a $5.3\times6.1$~arcmin area.

We compare the CATS-derived magnifications with those from the ``light-traces-mass'' models based on the method of \citet{zit09-1985}. We find a good agreement between Zitrin and CATS, with a median difference in the magnification factor of $\Delta\mu\sim0.1$ (rms dispersion $\sim$0.5).

In the parallel fields, we are forced to use the low-resolution, wide-field initial solution from \textsc{SaWLens} \citep{mer09-681, mer11-333}, which is a non-parametric model based on strong and weak lensing. In the central regions, where a direct comparison is possible, CATS and \textsc{SaWLens} also agree well, with a median difference in magnification for our sources of $\Delta\mu<0.1$. In the outer regions, magnification factors are generally less well-constrained, but are also typically small ($\mu\approx1$).

\subsection{IR Spectral Energy Distributions (SEDs)}
\label{sec:irsed}

Derivation of the physical properties requires an IR model template fit to the dust emission continuum.

Template fitting is executed in the rest-frame, with observed (image plane) fluxes (ie uncorrected for magnification), as presented in the SED figures later in this section. IR properties, however, are given in the source plane: intrinsic values corrected for magnification. We assume that differential (wavelength-dependent) effects are negligible. All uncertainties for output properties are bootstrapped via 1000 Monte Carlo simulations based on the estimated errors in flux, redshift and template-to-template variation.

\subsubsection{IR redshift estimate}
\label{sec:irz}

For all sources, we first estimate an approximate redshift ($z_{\rm phot\_FIR}$) from a modified blackbody with a specific characteristic dust temperature ($T_{\rm dust}$). Rather than assuming the same temperature in every case, we attempt to soften the temperature--redshift degeneracy using the best-fitting \citet{rie09-556} template. The technique is possible because the templates exhibit a skewness which varies smoothly with temperature. Hence, the best fit skew gives a coarse temperature estimation.

\begin{figure}
\centering
\includegraphics[viewport=20mm 7mm 140mm 143mm,height=84mm,angle=270,clip]{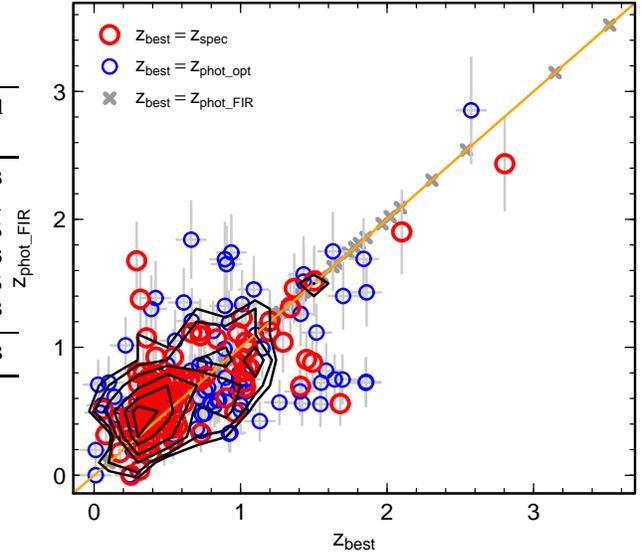}
\caption{Comparison of the FIR redshift estimate ($z_{\rm phot\_FIR}$) with spectroscopic (large red circles) and optical photometric (small blue circles) counterpart redshifts. Black contours visualise the joint distribution of both these populations. Generally, the FIR estimate is reasonable, showing a scatter of $\Delta z$$\sim$0.3. Grey crosses show the 61/263 sources for which we use $z_{\rm best} = z_{\rm phot\_FIR}$. This is solely to show the relative range of redshifts for sources without a good estimate, and they are of course not included in the scatter estimates.}
\label{fig:zfir_test}
\end{figure}

Figure \ref{fig:zfir_test} demonstrates the validity of this method, comparing $z_{\rm phot\_FIR}$ to available spectroscopic (or well-constrained optical photometric) redshifts. For many sources, the FIR-derived redshift is reasonably good, with a mean scatter in the offset of $\sim$0.3, equating to $\delta z_{\rm phot\_FIR}/(1+z_{\rm spec})\sim 0.40$. This is not dissimilar to the uncertainty estimated by previous attempts at IR-based redshifts, e.g. \citet{pea13-2753}. Several of the most obvious outliers result from poor dust peak sampling, which hinders quantification of the skewness. For instance, the two spectroscopic sources with over-estimated $z_{\rm phot\_FIR}$ are at low redshift ($z_{\rm best} \sim 0.3$) and are within a parallel field which lacks the PACS coverage to adequately constrain the continuum peak. Their temperature is estimated to be higher than it should, resulting in a correspondingly higher redshift as well.

For the 62 sources without a spectroscopic or well-constrained (CLASH) optical photometric redshift, we use $z_{\rm phot\_FIR}$ when deriving intrinsic properties. We visually inspect the SEDs of all these sources to ensure that the IR continuum templates at the specified redshift are a close fit to the observed NIR (e.g. IRAC) photometry. For sources outside \textit{Spitzer} coverage (such as the two sources highlighted above), the WISE photometry has insufficient precision to discriminate between different redshifts.

Sources for which $z_{\rm best}=z_{\rm phot\_FIR}$ are flagged in our value-added catalogue, and whilst $z_{\rm phot\_FIR}$ appears well-behaved, we would still caution against their use for critical applications (such as ALMA spectral set-up).

\subsubsection{Characteristic dust temperature}

The characteristic dust temperature ($T_{\rm dust}$) of the IR component is calculated via the best-fitting single-temperature modified blackbody. Dust heating by the Cosmic Microwave Background is assumed to be negligible, which is reasonable for sources at $z<4$ \citep{dac13-13}. Dust temperature is degenerate with the emissivity index $\beta$ \citep[e.g.][]{bla03-733}, and we assume $\beta=2.0$. Using $\beta=1.5$ would systematically increase $T_{\rm dust}$ by $\sim$10\%.

Dust temperature is also degenerate with redshift, and the two values are inextricably linked for sources with only a $z_{\rm phot\_FIR}$ ($T_{\rm dust}$ is used by the estimation technique). Therefore, for those 62 sources, we simply give the coarse $T_{\rm dust}$ estimate described in Section \ref{sec:irz}. 

\subsubsection{Star formation rate (SFR)}

The total IR luminosity ($L_{\rm IR}$) is derived from the best-fitting \citet{rie09-556} template to the \textit{Herschel} fluxes, allowing for an overall normalisation parameter. Following the standard definition of $L_{\rm IR}$, we integrate over the rest-frame wavelength range $\lambda=8-1000$~$\mu$m. The SFR estimate follows directly via the \citet{ken98-189} relation, modified to match a \citet{kro02-82}-like initial mass function as in \citet{rie09-556}.\footnote{SFR [$M_{\sun}$ yr$^{-1}$] $=$ 0.66$\times$SFR$_{\rm K98}$ $=$ 1.14$\times$10$^{-10}$ $L_{\rm IR}$ [L$_{\sun}$]} Note that we do not make a correction for unobscured SFR.

For all sources, we also calculate SFR via an identical integration of the best-fitting \citet{ber13-100} template to all available IR photometry. The SED figures throughout this paper show the best-fit template and derived SFR parameters from both \citeauthor{rie09-556} and \citeauthor{ber13-100}. We find a good agreement between the two derived SFRs: a median discrepancy of 0.08~dex and an rms scatter of 0.18~dex. The uncertainty introduced by the intrinsic shape of the template sets is dominated by the photometric error for all sources, while the redshift uncertainty is also significant for those without a spectroscopic redshift. For simplicity, the catalogues tabulate IR properties derived from the \citeauthor{rie09-556} templates only.

\subsubsection{AGN contamination}
\label{sec:agn}

A strong active galactic nucleus (AGN) may artificially boost the SFR by contributing flux in the mid-IR. For all SEDs with at least one flux measurement in the 20--200~$\mu$m wavelength range, we find the best-fitting sum of a \citeauthor{rie09-556} template and the mean low-luminosity AGN from \citet{mul11-1082}.

Very few sources (6) have a non-negligible ($>$10\%) contribution to $L_{\rm IR}$ from the AGN component. For those, we calculate $L_{\rm IR}$ and SFR from the star-forming dust component alone (ie integrate only the \citeauthor{rie09-556} contribution). For all other sources, we revert to the best-fitting \citet{rie09-556} template without the additional AGN component.

\subsection{`Value-added' IR properties catalogue}
\label{sec:fullcat}

\begin{figure*}
\centering
\includegraphics[viewport=20mm 0mm 187mm 205mm,width=52.5mm,angle=270,clip]{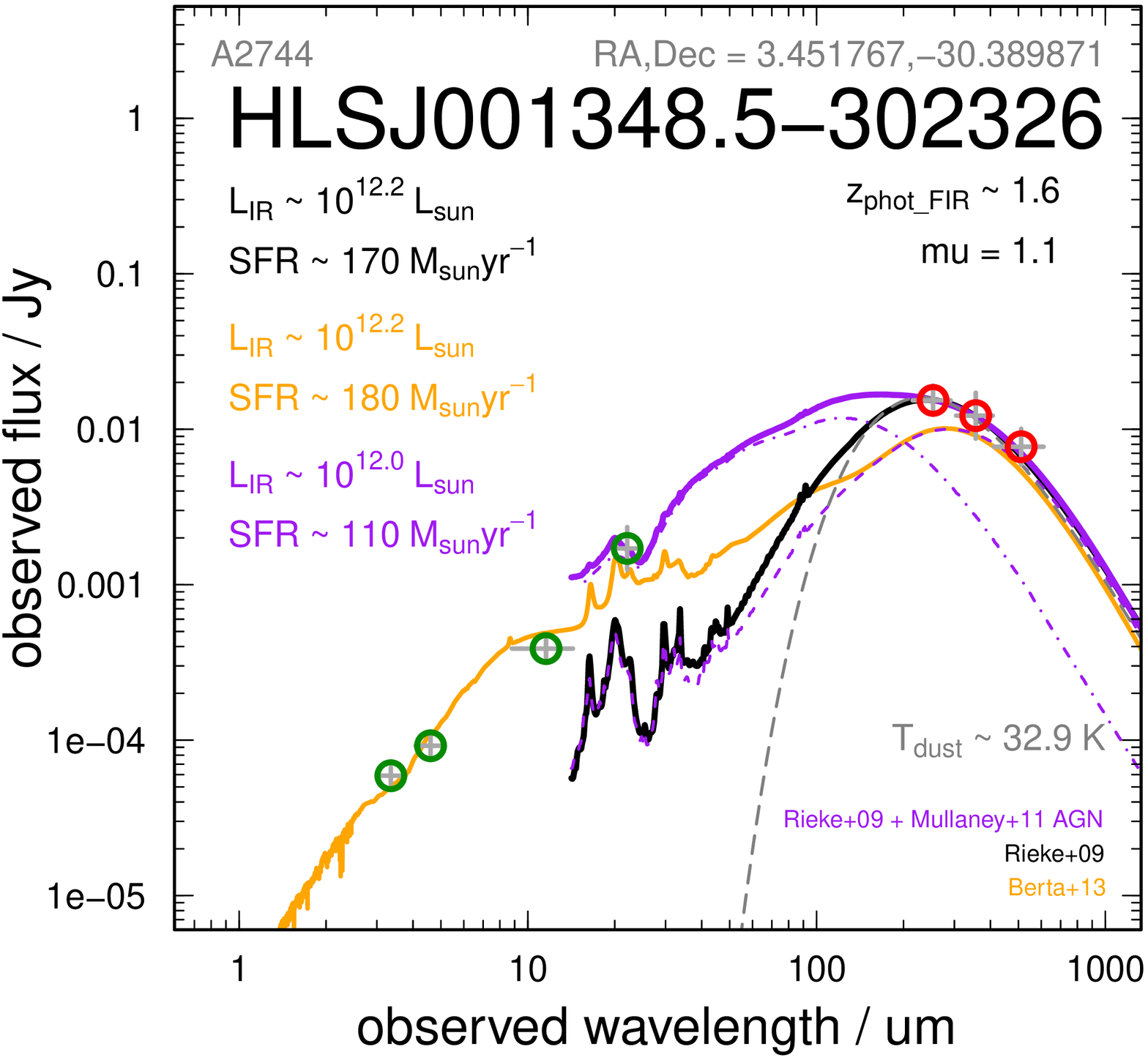}
\includegraphics[viewport=20mm 30mm 187mm 205mm,width=52.5mm,angle=270,clip]{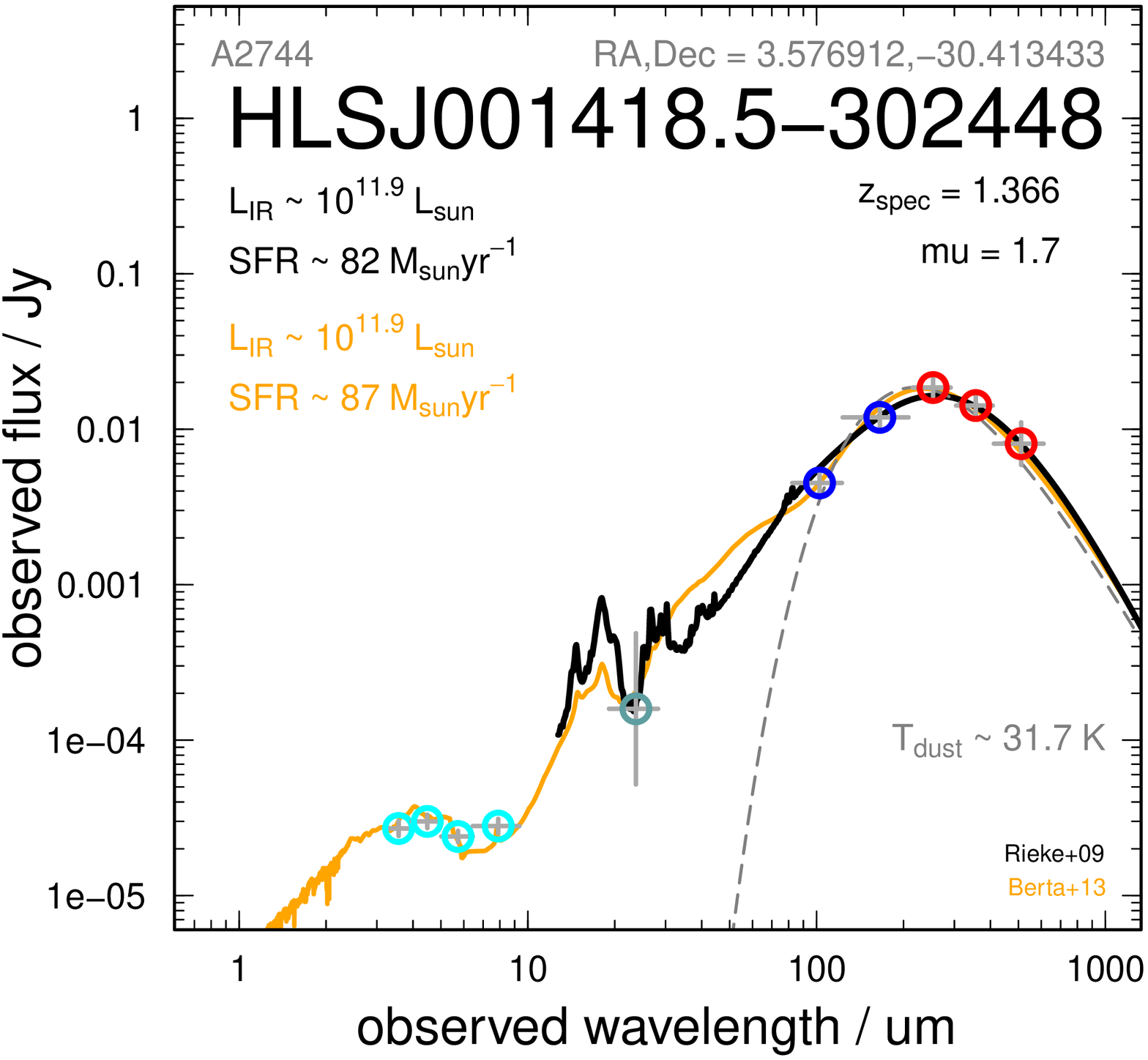}
\includegraphics[viewport=20mm 30mm 187mm 205mm,width=52.5mm,angle=270,clip]{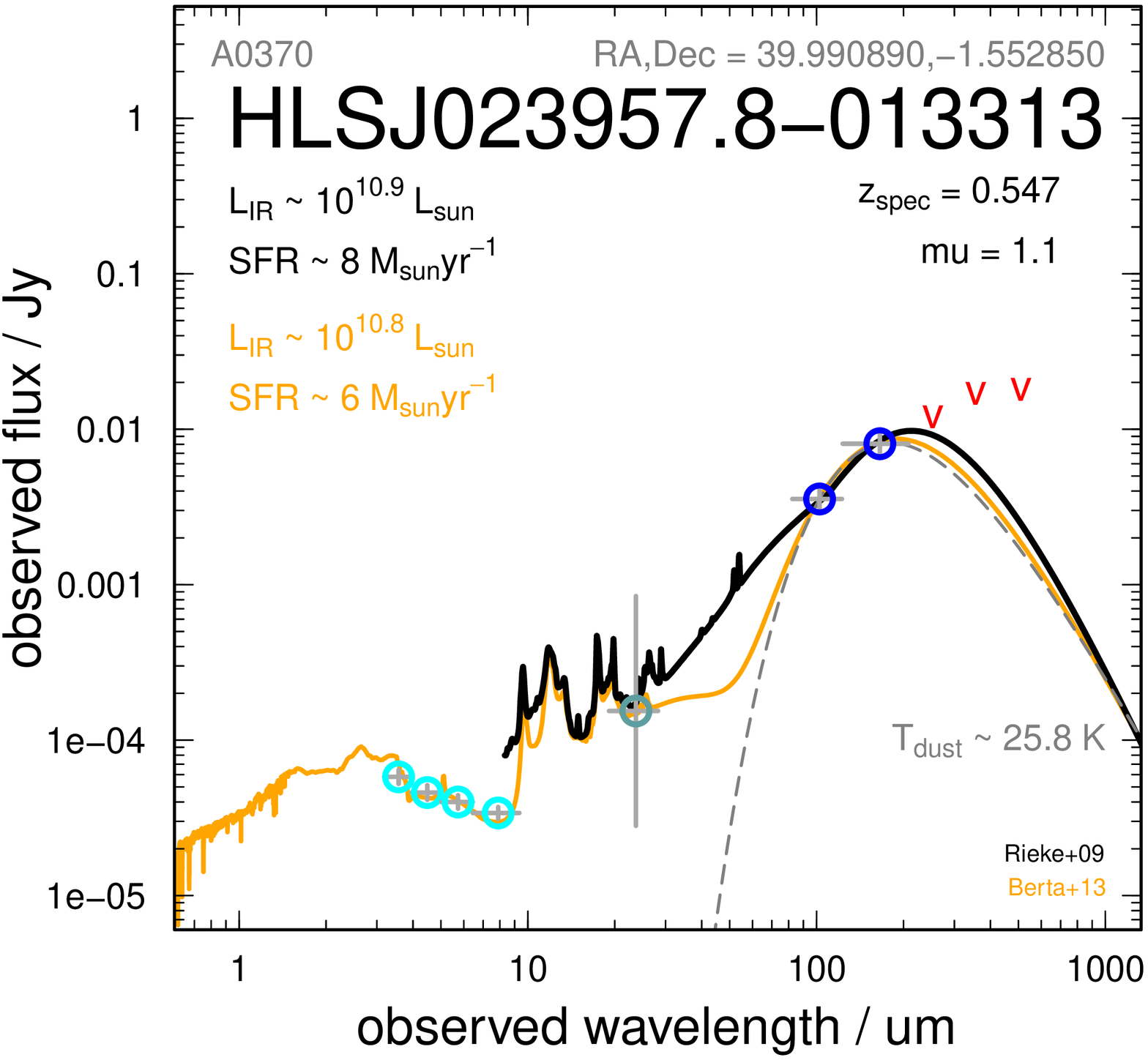}\\
\includegraphics[viewport=20mm 0mm 187mm 205mm,width=52.5mm,angle=270,clip]{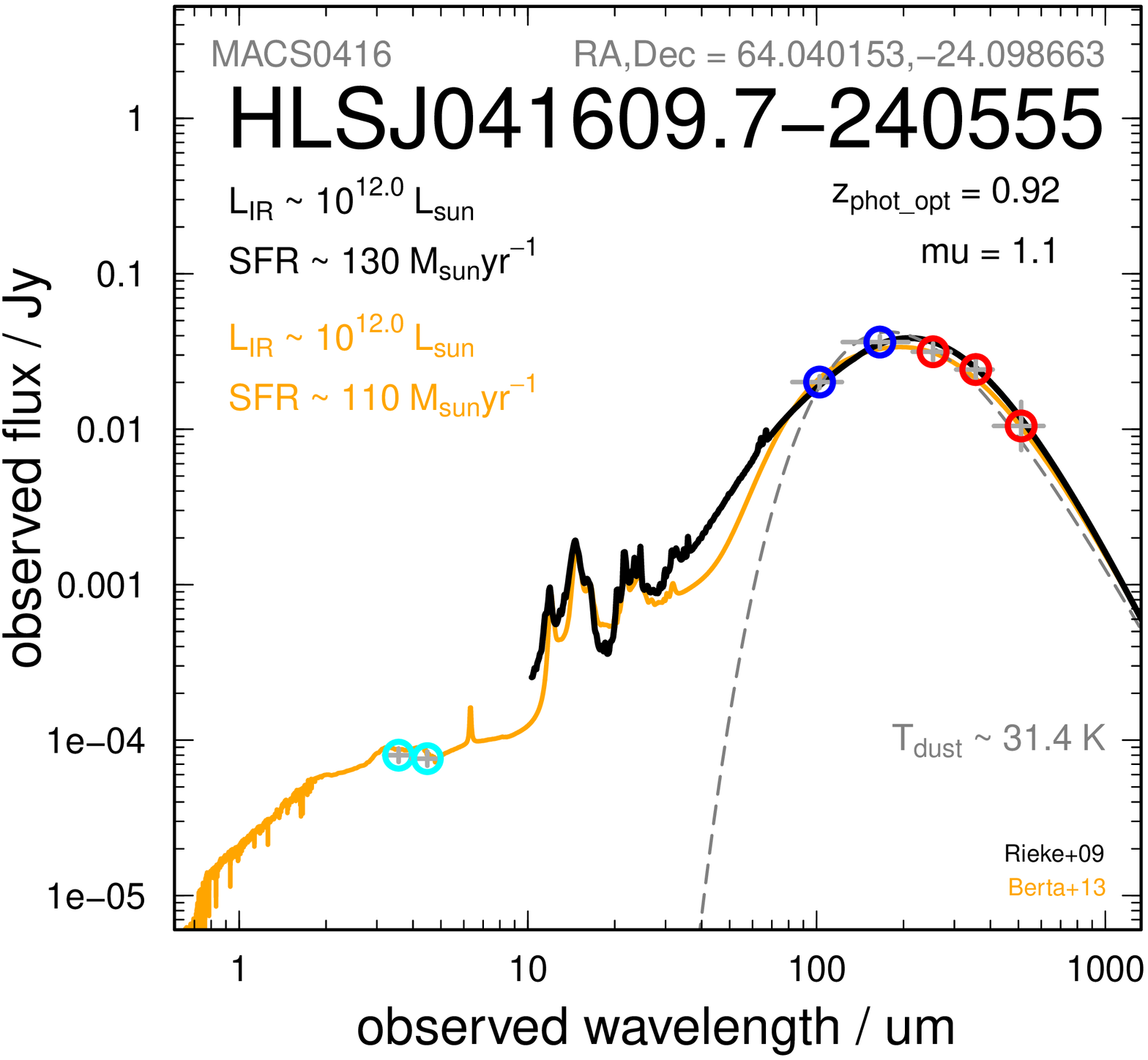}
\includegraphics[viewport=20mm 30mm 187mm 205mm,width=52.5mm,angle=270,clip]{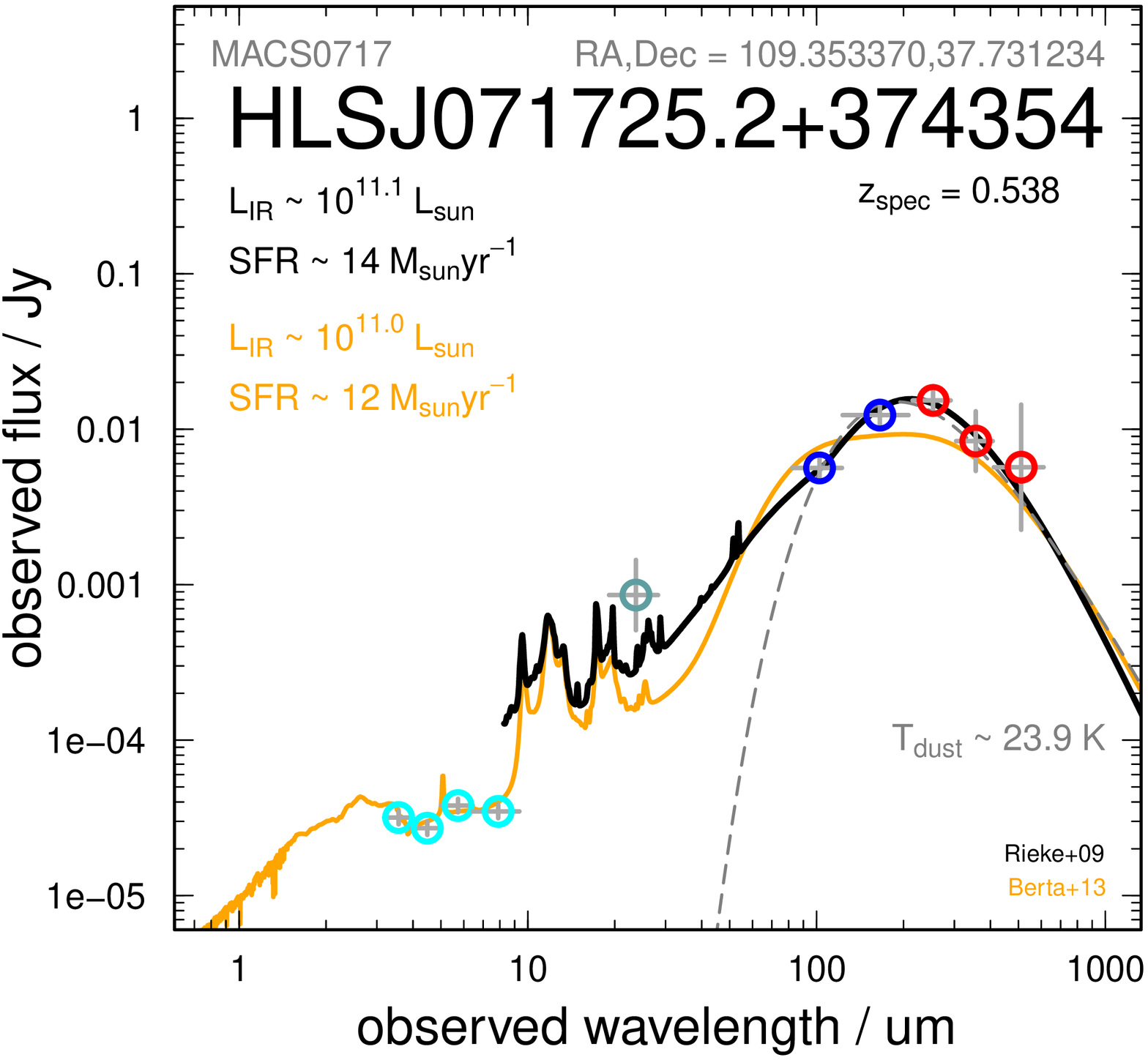}
\includegraphics[viewport=20mm 30mm 187mm 205mm,width=52.5mm,angle=270,clip]{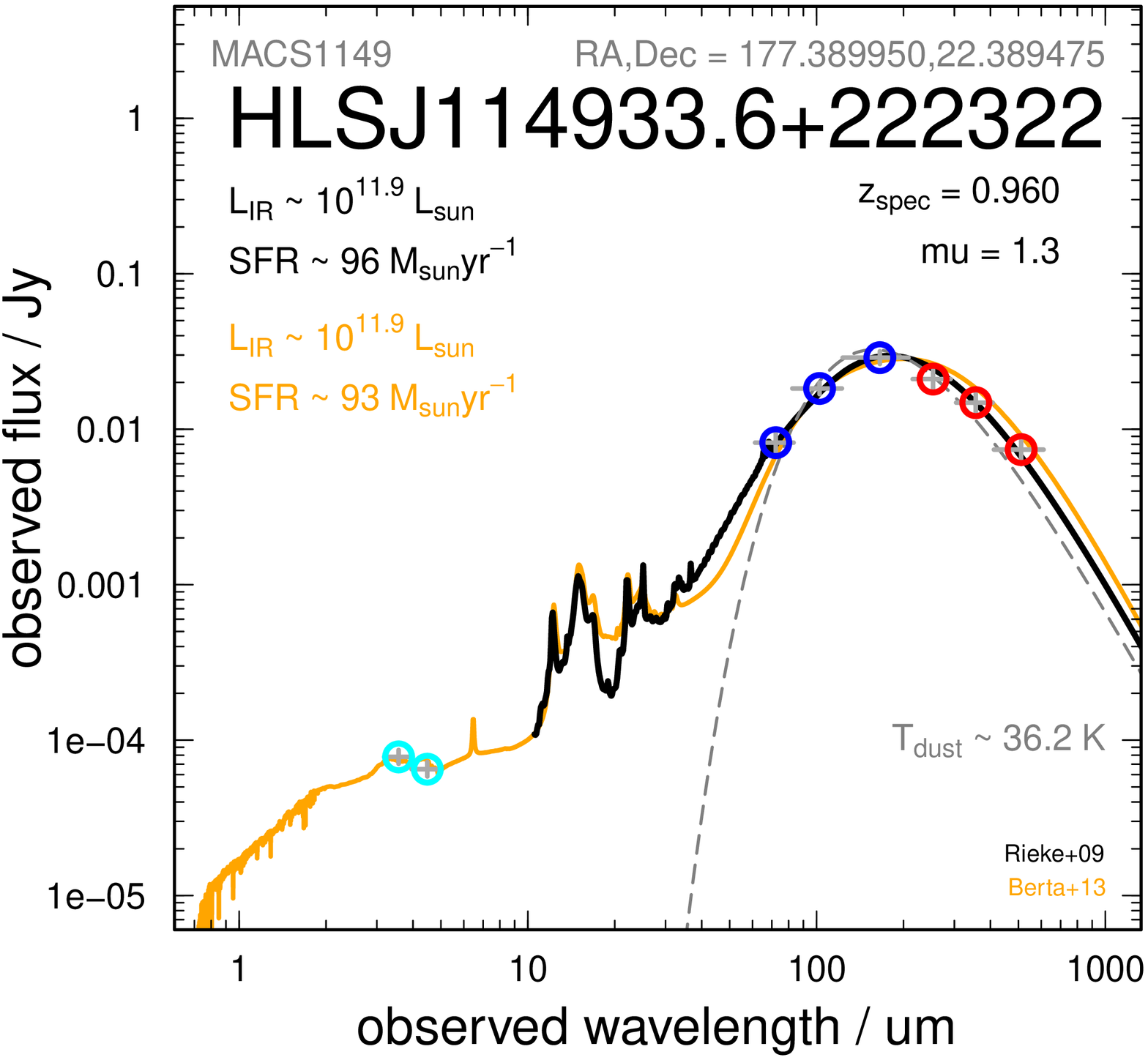}\\
\includegraphics[viewport=20mm 0mm 210mm 205mm,width=60mm,angle=270,clip]{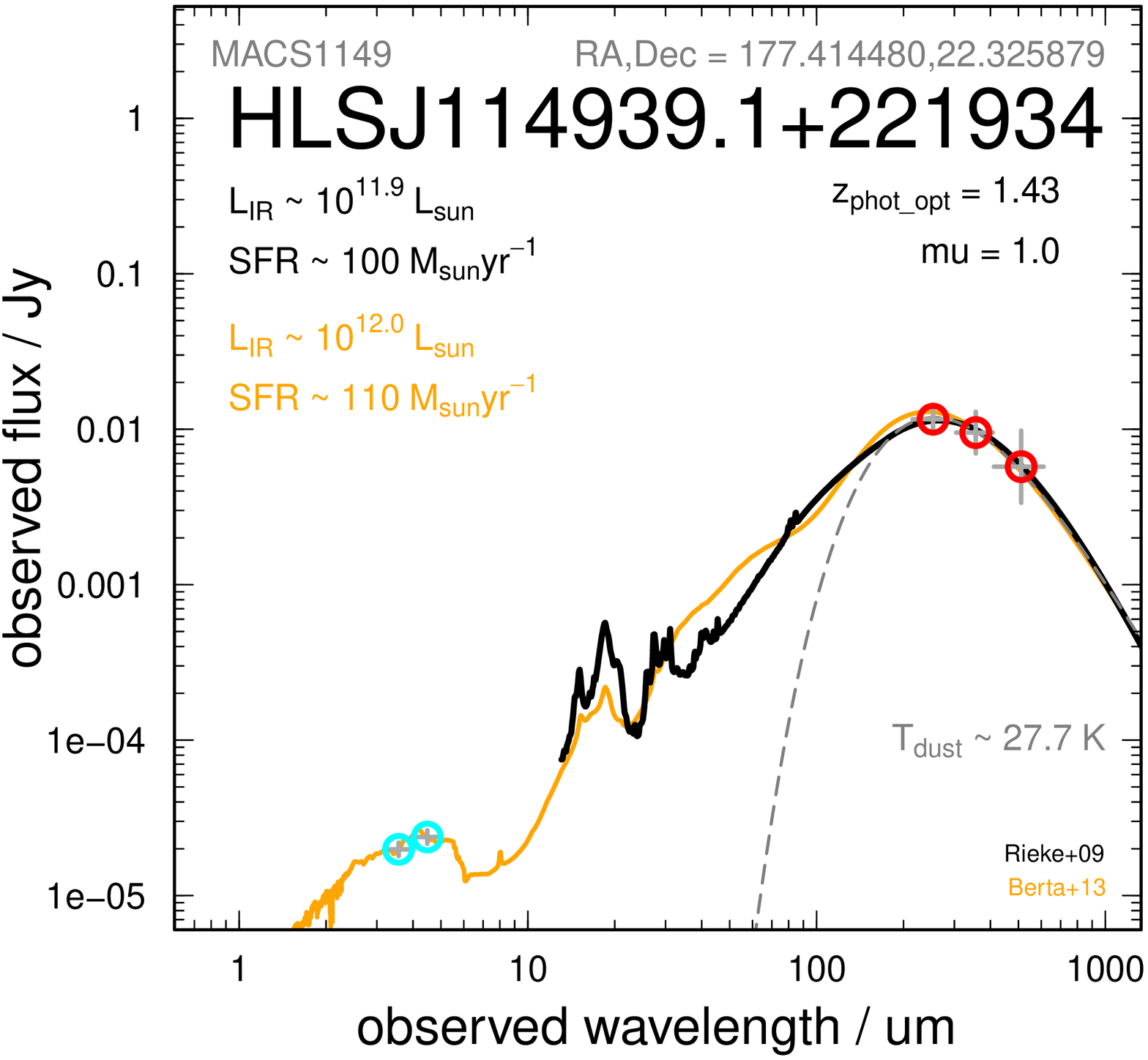}
\includegraphics[viewport=20mm 30mm 210mm 205mm,width=60mm,angle=270,clip]{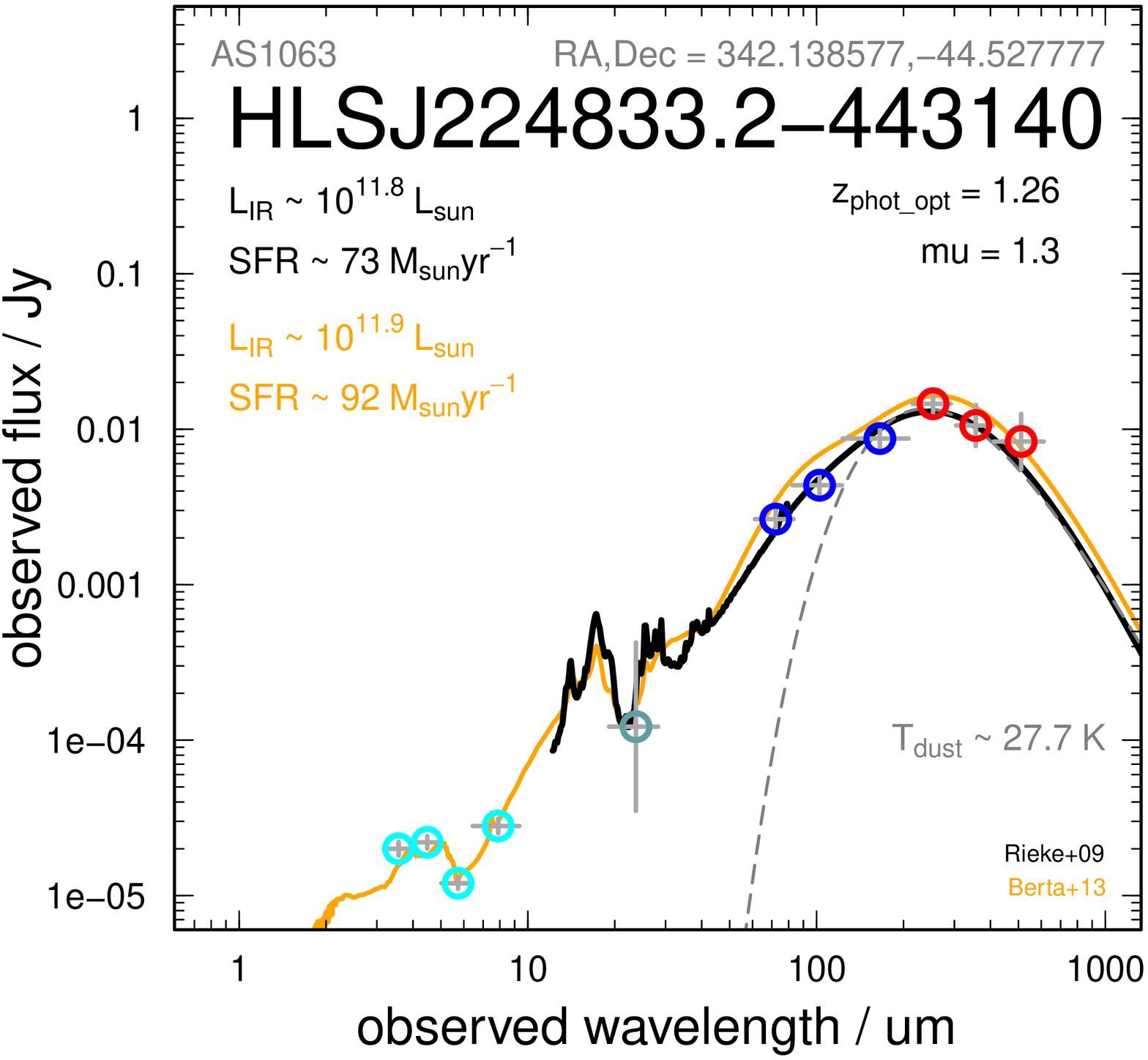}
\includegraphics[viewport=20mm 30mm 210mm 205mm,width=60mm,angle=270,clip]{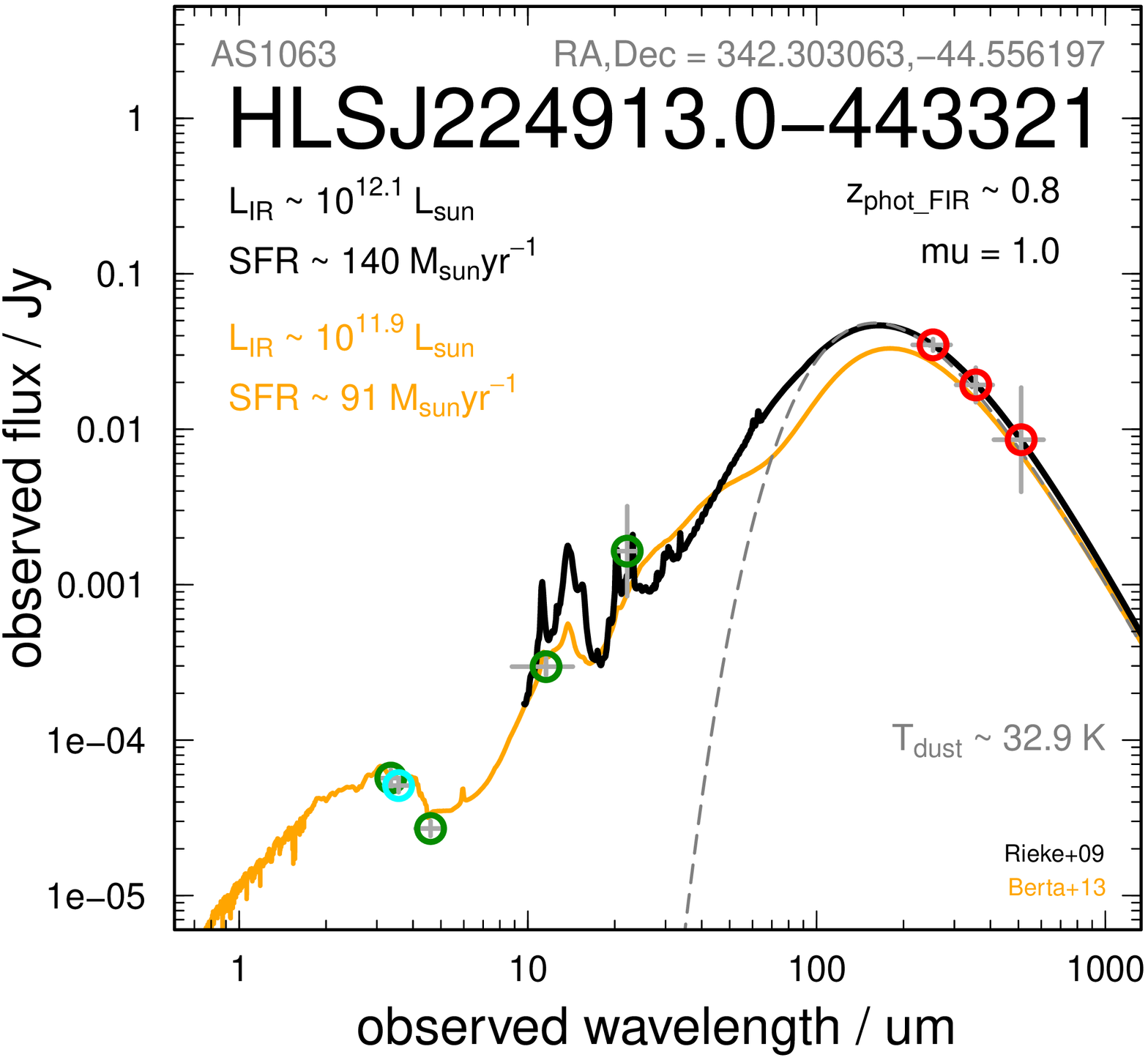}
\caption{Example IR SEDs illustrating the observed photometry and fits. In each panel, observed (uncorrected) fluxes are shown by open circles: IRAC$=$cyan, MIPS$=$grey, \textit{WISE}$=$green, PACS$=$blue, SPIRE$=$red. Non-detections (5$\,\sigma$ upper limits) are denoted by a `v'. Also plotted are the best-fitting modified-blackbody (grey dashes), \citet{rie09-556} template (black solid line) and \citet{ber13-100} template (orange solid line). Derived IR properties from each of the two template sets (in matching colours) take into account the magnification, $\mu$ (shown to the right). Redshifts are spectroscopic ($z_{\rm spec}$), optical photometric ($z_{\rm phot\_opt}$) or IR-derived ($z_{\rm phot\_FIR}$; see Section \ref{sec:irz}). For SEDs with a fitted AGN component, the two component model and star-formation IR properties are shown in purple.}
\label{fig:exseds}
\end{figure*}

In Figure \ref{fig:exseds} we provide a selection of SEDs to illustrate the observed data and the derivation of intrinsic properties. The examples originate from all 6 clusters (both central and parallel fields), are at varied redshifts (taken from spectroscopy, optical photometry and IR estimation) and exhibit a range of photometric completeness. An example of a source with a non-negligible AGN component is also included. Several of the most interesting SEDs are deliberately not shown in Figure \ref{fig:exseds} as they are discussed individually in Section \ref{sec:disc}.

The derived intrinsic properties for all 263 \textit{Herschel}--HFF sources are listed in Table \ref{tab:cat2}. We defer the derivation of stellar masses to a later paper, when fully homogenised optical--IR SEDs are available. We remind the reader that the online version of this catalogue will be updated as further spectroscopic redshifts become available.

\begin{table*}
\caption{Optical counterparts for \textit{Herschel}-detected sources within HFF. Derived properties account for magnification ($\mu$).}
\label{tab:cat2}
\begin{tabular}{lrrrcrrrrrc}
\hline
\multicolumn{1}{c}{ID$^1$} & \multicolumn{1}{c}{Field$^2$} & \multicolumn{5}{c}{Optical/NIR/radio counterpart} & \multicolumn{1}{c}{$L_{\rm IR}$$^6$} & \multicolumn{1}{c}{SFR$_{\rm IR}$$^6$} & \multicolumn{1}{c}{$T_{\rm dust}$$^7$} & \multicolumn{1}{c}{AGN$^8$} \\
 & & \multicolumn{1}{c}{RA$^3$} & \multicolumn{1}{c}{Dec$^3$} & \multicolumn{1}{l}{(ref)$^3$} & \multicolumn{1}{c}{$z$ (ref)$^4$} & \multicolumn{1}{c}{$\mu$ (ref)$^5$} & \multicolumn{1}{c}{log($L$/$L_{\sun}$)} & \multicolumn{1}{c}{$M_{\sun}$ yr$^{-1}$} & \multicolumn{1}{c}{K} & \\
\hline
HLSJ001412.7--302359 & A2744 C & 3.552819 & --30.399785 & (H) & \textbf{0.688} (1) & 1.21 (1) & 11.21 $\pm$ 0.09 & 18.3 $\pm$ 3.3 & 26.7 $\pm$ 1.6 & 0 \\
HLSJ001415.3--302423 & A2744 C & 3.563939 & --30.406277 & (H) & \textbf{0.653} (1) & 1.28 (1) & 11.17 $\pm$ 0.09 & 16.7 $\pm$ 3.0 & 25.9 $\pm$ 1.5 & 0 \\
HLSJ001416.7--302304 & A2744 C & 3.569295 & --30.384239 & (H) & \textbf{0.296} (1) & -- ~~--~ & 10.16 $\pm$ 0.12 & 1.6 $\pm$ 0.4 & 19.8 $\pm$ 1.8 & 0 \\
HLSJ001416.5--302410 & A2744 C & 3.568938 & --30.402831 & (H) & \textit{$\sim$0.5} (2) & 1.30 (1) & 10.88 $\pm$ 0.16 & 8.6 $\pm$ 3.2 & 27.7 $\pm$ 2.4 & 1 \\
HLSJ001418.5--302246 & A2744 C & 3.577030 & --30.379500 & (H) & \textbf{0.498} (3) & 1.40 (1) & 10.90 $\pm$ 0.09 & 9.1 $\pm$ 1.6 & 26.3 $\pm$ 1.4 & 0 \\
HLSJ001417.6--302301 & A2744 C & 3.573229 & --30.383567 & (H) & \textit{$\sim$0.9} (2) & 1.95 (1) & 11.97 $\pm$ 0.13 & 107.2 $\pm$ 31.4 & 23.0 $\pm$ 1.6 & 0 \\
HLSJ001418.5--302448 & A2744 C & 3.576912 & --30.413433 & (H) & \textbf{1.366} (4) & 1.68 (1) & 11.86 $\pm$ 0.08 & 81.5 $\pm$ 14.7 & 31.7 $\pm$ 1.5 & 0 \\
HLSJ001418.0--302529 & A2744 C & 3.575482 & --30.424512 & (I) & \textit{$\sim$3.5} (2) & 1.42 (1) & 12.64 $\pm$ 0.09 & 501.5 $\pm$ 105.5 & 27.7 $\pm$ 2.4 & 0 \\
... & ... & ... & ... & ... & ... & ... & ... & ... & ... & ... \\
\hline
\\
\end{tabular}
\\
\raggedright
 \textit{[Note: The full table is published in the electronic version of the paper. A portion is shown here for illustration.]}\\
$^1$ \textit{Herschel} ID as in Table \ref{tab:cat1}\\
$^2$ C=central region; P=parallel\\
$^3$ Counterpart RA and Dec, from the most accurate available position (V=VLA, H=HST, O=[ground-based] optical, I=IRAC, W=WISE, P=PACS, S=SPIRE)\\
$^4$ Counterpart \textbf{spectroscopic}, optical photometric or \textit{IR estimated} redshift from (1) \citet{owe11-27}, (2) $z_{\rm phot\_FIR}$, (3) \citet{cou98-188}, (4) \citet{raw14-196}, (5) \citet{wan15-29}, (6) \citet{bus02-787}, (7) \citet{bra09-947}, (8) \citet{wol12-2}, (9) \textit{HST} GRISM, (10) \citet{bam05-109}, (11) \citet{ivi98-583}, (12) \citet{sou88-19}, (13) \citet{hen87-473}, (14) \citet{sou99-70}, (15) Magellan/IMACS, (16) CLASH \textit{HST} photo-z, (17) CLASH Subraru photo-z, \citet{ume14-163}, (18) \citet{bal13-9}, (19) \citet{ebe14-21}, (20) LBT/MODS, (21) \citet{smi09-163}, (22) Walth et al. (in prep.), (23) \citet{gom12-79}, (24) \citet{kar15-11}\\
$^5$ $\mu$ from (1) CATS models \citep{jul09-1319, jau12-3369, ric14-268, jau14-1549} or (2) SaWLens wide-field method \citet{mer09-681, mer11-333}. `--' for foreground and cluster sources\\
$^6$ From the best-fitting \citet{rie09-556} template\\
$^7$ Characteristic dust temperature of the best-fitting single-temperature modified blackbody\\
$^8$ Flag for alternate IR SED fit using SF+AGN model, effectively correcting $L_{\rm IR}$ and SFR$_{\rm IR}$ for AGN contamination (see Section \ref{sec:agn}) \\
\end{table*}

\section{Discussion}
\label{sec:disc}

We now present a brief exploratory overview of the \textit{Herschel}-detected population.

\begin{figure*}
\centering
\includegraphics[viewport=20mm 12mm 120mm 118mm,height=61mm,angle=270,clip]{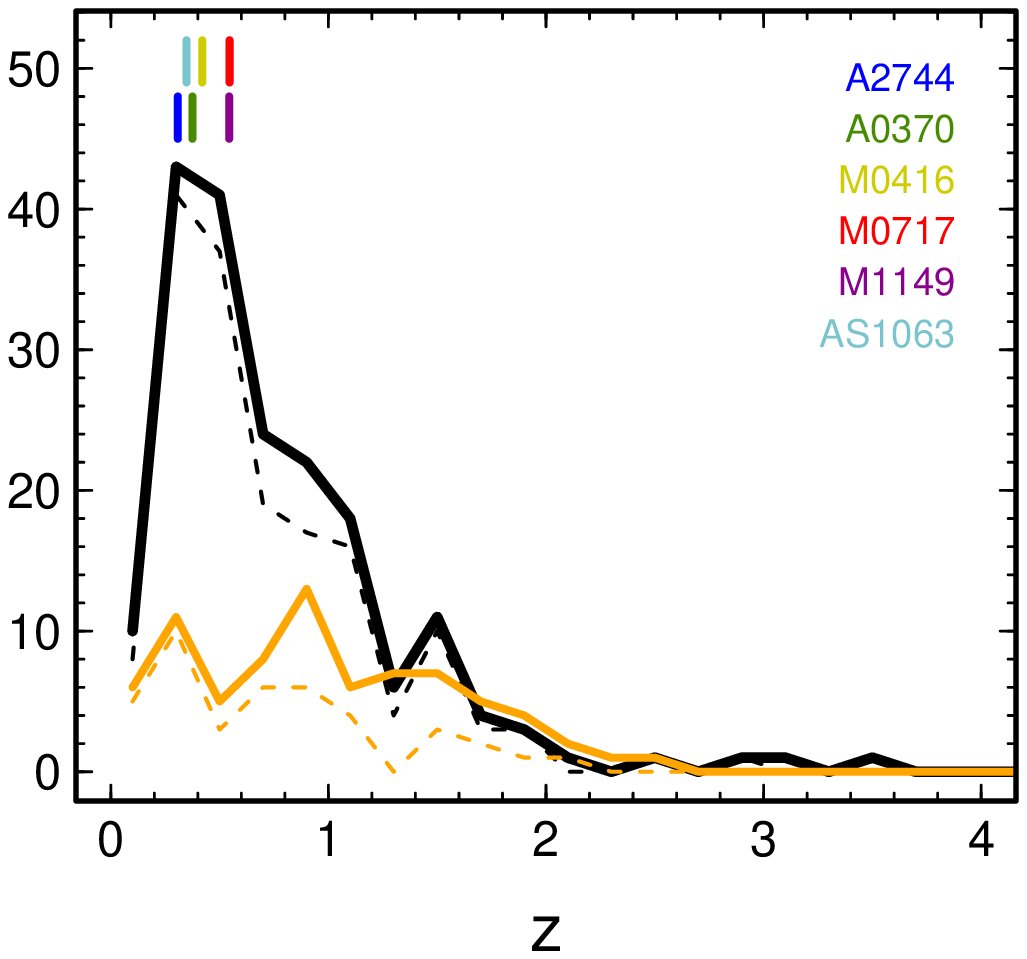}
\includegraphics[viewport=20mm 20mm 120mm 118mm,height=56.5mm,angle=270,clip]{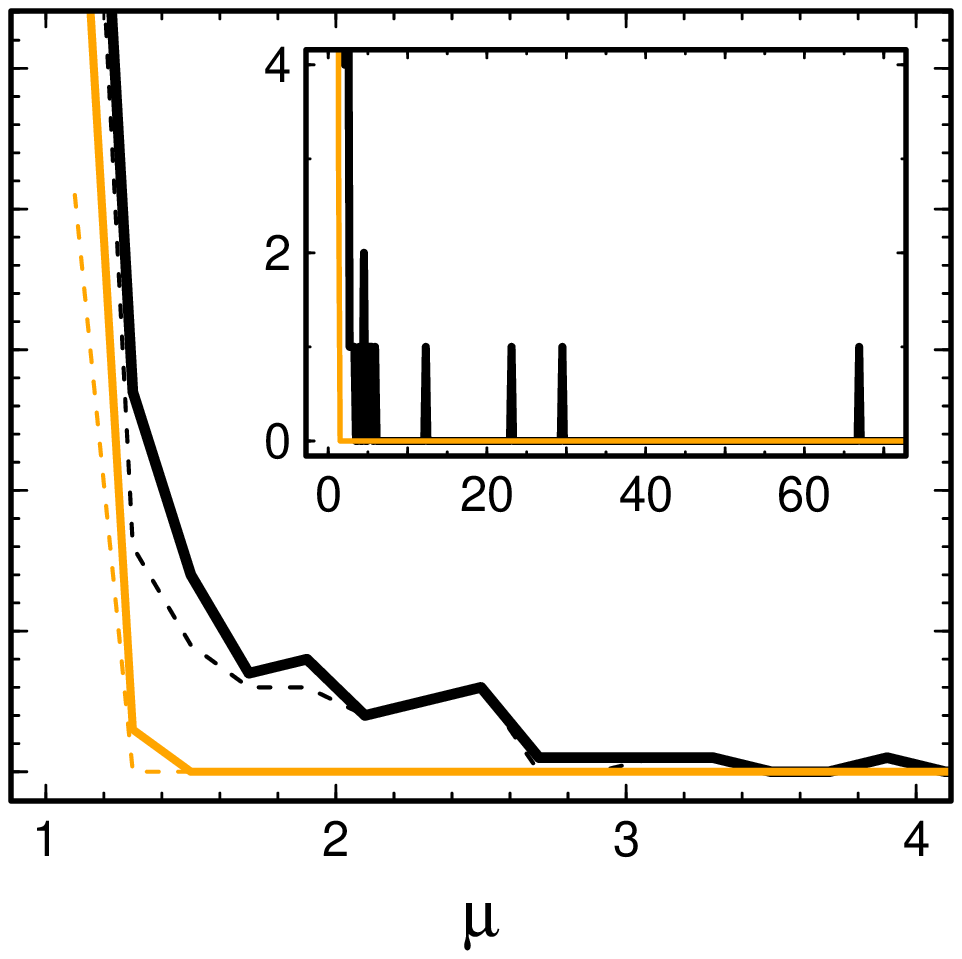}
\includegraphics[viewport=20mm 20mm 120mm 118mm,height=56.5mm,angle=270,clip]{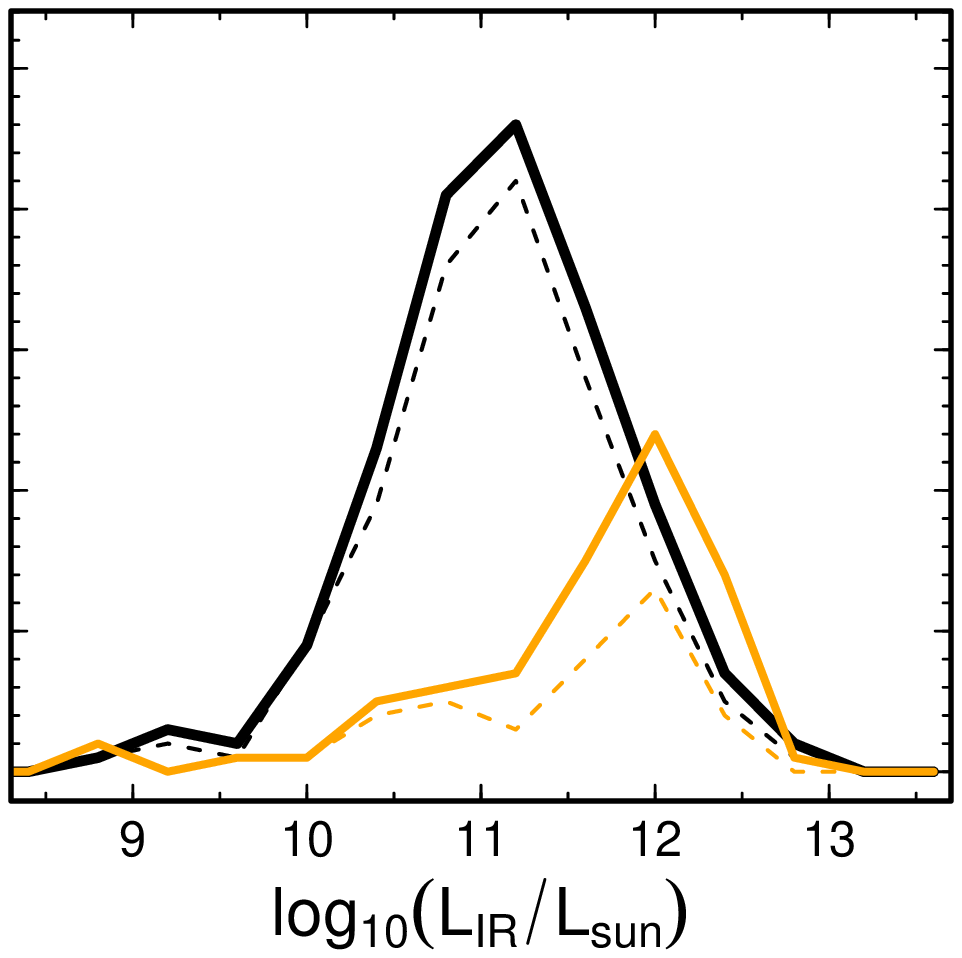}
\caption{Property distributions for the \textit{Herschel}-detected population, divided into central (black) and parallel (orange) regions, and either including (solid) or excluding (dashed) sources with only an IR-estimated redshift. \textit{Left panel:} redshift $z$, with cluster redshifts indicated by the coloured lines towards the top-left. \textit{Centre}: magnification factor $\mu$, with an inset panel showing $\mu>4$. \textit{Right:} total IR luminosity $L_{\rm IR}$, corrected for magnification.}
\label{fig:hists}
\end{figure*}

\subsection{IR property distributions}
\label{sec:distributions}

Figure \ref{fig:hists} presents the distribution of three properties: redshift $z$, magnification $\mu$ and IR luminosity $L_{\rm IR}$. The distributions remain unchanged regardless of the inclusion of the sources with only an IR-estimated redshift, supporting their general validity.

In the $z$ distribution, a broad peak at $z\sim0.4$ dominates the central population, indicating a substantial population of cluster members. Indeed, 49 \textit{Herschel} sources located in the central footprints have a cluster counterpart. In the parallel fields, there are only seven. Naively this suggests a prevalence of IR-bright cluster galaxies in the core, contradicting the notion that clusters effectively quench star-formation \citep[e.g.][]{fad08-9,hai15-101}. However, there are several important factors we need to take into account. First, star-forming galaxies at the cluster redshift typically exhibit a dust peak in the PACS bands, and so they are less likely to be detected and/or well-constrained in the parallel fields with SPIRE imaging only. Second, line of sight projection means that the central footprints actually probe the full range of cluster radii, and it is virtually impossible to say whether a cluster source coincident with the core is truly at a low cluster-centric radius. Third, the number density of all cluster members (SF+quiescent) is much higher in the cluster core so the number count of SF members is a worse tracer of activity than e.g. the star-forming fraction \citep{hai13-126}. In total, 56/263 (21\%) IR-bright sources are identified as cluster members.

A further 44/263 (17\%) of the \textit{Herschel} detections are in the cluster foreground, and are not discussed futher in this study.

The majority of \textit{Herschel}-detected sources in HFF are beyond the cluster redshift (163/263; 62\%). They have a median redshift of $z=1.0$. Two sources are at $z>3$, while a further seven lie at $2<z<3$. As mentioned in Section \ref{sec:fluxcat}, we may be bias against the faintest, highest-redshift sources, which would have an IR dust peak in the redder SPIRE bands. These source could be too faint for the bluer \textit{Herschel} bands (PACS and even SPIRE 250~$\mu$m) meaning that they are unlikely to be detected in enough IR bands to be included in our catalogue. Additional imaging at longer wavelengths (e.g. JCMT/SCUBA-2, ALMA) increases the likelihood of identifying these sources, as illustrated by e.g. \citet{boo13-1} and also by the four highest redshift sources in our catalogue, which all have existing (sub)-millimetre photometry beyond SPIRE. We return to those four sources in Section \ref{sec:highz}.

While the number of $z>1.2$ sources is roughly equal in central and parallel regions, the population of background galaxies at $z<1.2$ is significantly larger in the inner fields. This is reflected in the median redshift of background sources: $z=0.95$ in the centre and $z=1.2$ for the parallel fields. This is likely to result from a combination of the limitations of detecting $z<1$ sources without PACS (or \textit{Spitzer}) coverage and also the increased lensing power within the central regions for the `sweet spot' of magnification at $z\sim1$.

For the background population, the median $\mu=1.24$, rising to $\mu=1.43$ for the 109 behind the central footprints, as all sources within parallel field undergo $\mu<1.25$. The distributions in Figure \ref{fig:hists} clearly shows this difference, which is unsurprising as the critical lines of the cluster mass models all fall within those core regions. In total, 29 sources are more than doubled in brightness ($\mu>2$). 

The (magnification-corrected) $L_{\rm IR}$ distribution in the cores also peaks at lower intrinsic luminosity, which is entirely expected given what we know about the data, physics and galaxy population: 1) the magnification is stronger in the core, allowing us to see further down the luminosity function; 2) the lack of \textit{Spitzer} and PACS imaging in the parallel fields decreases the effective detection limit; 3) the smaller population of objects in the cluster outskirts limits the number of nearby, intrinsically faint objects seen without lensing. In total, the background sample of 162 sources includes 99 LIRGs ($10^{11} < L_{\rm IR} < 10^{12}$~L$_{\sun}$) and 23 sub-LIRGs ($L_{\rm IR} < 10^{11}$~L$_{\sun}$).

\begin{figure}
\centering
\includegraphics[viewport=20mm 0mm 115mm 118mm,height=84mm,angle=270,clip]{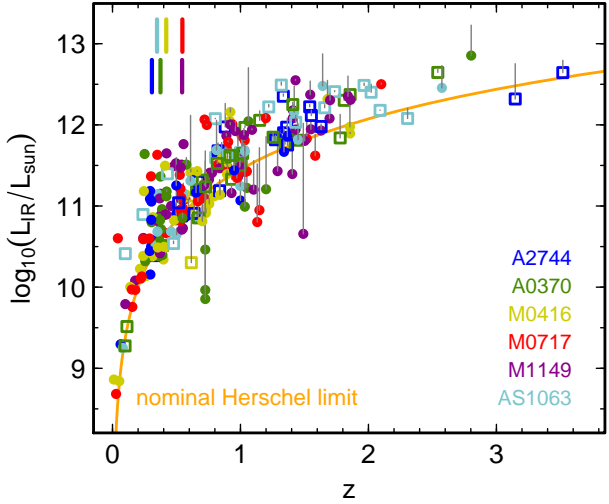}
\caption{Magnification-corrected IR luminosity ($L_{\rm IR}$) versus the counterpart redshift for \textit{Herschel}-detected sources in HFF. Closed circles are from the central regions and open squares from the parallel fields, colour-coded by the cluster (cluster redshifts are indicated by solid lines towards the upper-left). Grey vertical lines indicate the magnification effect for individual sources. The solid orange curve presents the nominal limit of the \textit{Herschel} observations without lensing.}
\label{fig:lir_z}
\end{figure}

Figure \ref{fig:lir_z} brings together all the previous information and effectively illustrates the benefits of lensing surveys by highlighting the most interesting sources. We can immediately see that the highest redshift sources ($z>2$) are rare, and typically are not boosted from significantly below the nominal detection limit. The highest magnification, and intrinsically faintest background sources tend to be located within the redshift range $0.6 < z <1.6$: more than ten sub-LIRGs, and a similar number of LIRGs, all boosted from below the nominal detection limit.

\begin{figure}
\centering
\includegraphics[viewport=20mm 0mm 145mm 170mm,height=84mm,angle=270,clip]{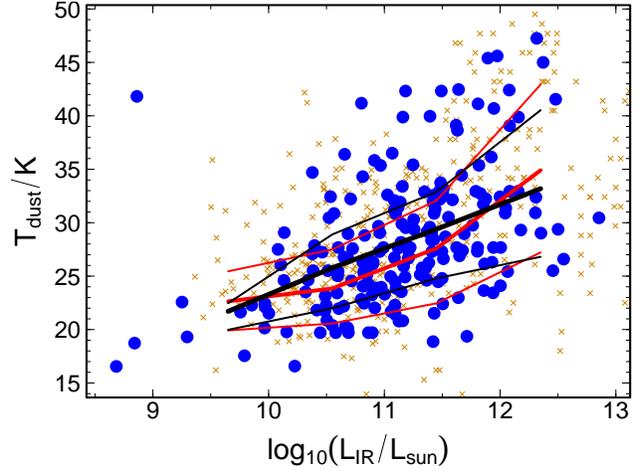}
\caption{Magnification-corrected IR luminosity ($L_{\rm IR}$) versus the characteristic dust temperature ($T_{\rm dust}$) for HFF sources (blue filled circles). Those with only $z_{\rm phot\_FIR}$ are excluded. Black lines show the population trend (and $\pm$20\% percentiles). Comparison samples from \citet{bla03-733} and \citet{hwa10-75} (corrected to match $\beta=2$) are shown by orange crosses, with the general trend displayed by red lines.}
\label{fig:lir_tbb}
\end{figure}

We compare the HFF population to larger samples by exploring the $L_{\rm IR}$--$T_{\rm dust}$ plane. Figure \ref{fig:lir_tbb} shows that sources from both HFF and previous studies, e.g. \citet{bla03-733} and \citet{hwa10-75} (corrected to match $\beta=2$), sample very similar parent populations. This emphasises the obvious point: lensing is unlikely to select atypical sources, but rather allows us to observe larger numbers of `normal' sources to a fainter limit.

An exploration of the morphology of the IR-bright population is beyond the scope of this paper. Although Table \ref{tab:cat2} offers a unique launchpad for a quantitative analysis of the optical counterparts (such as \citealt{man15-763}), we defer such an effort to future papers.

\subsection{Highest redshift sources}
\label{sec:highz}

\begin{figure*}
\centering
\includegraphics[viewport=20mm 30mm 210mm 205mm,width=65mm,angle=270,clip]{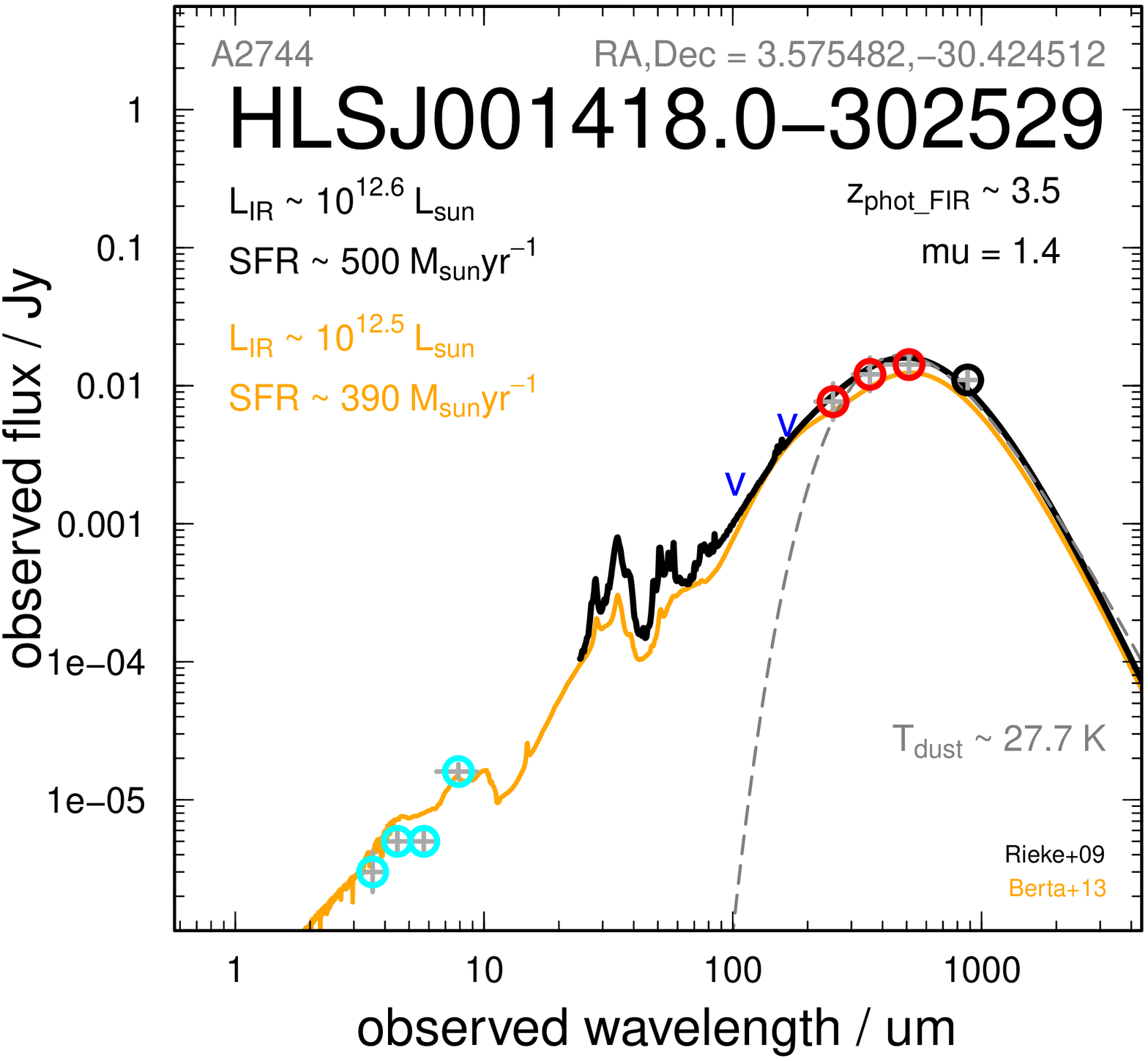}
\includegraphics[viewport=20mm 0mm 210mm 205mm,width=65mm,angle=270,clip]{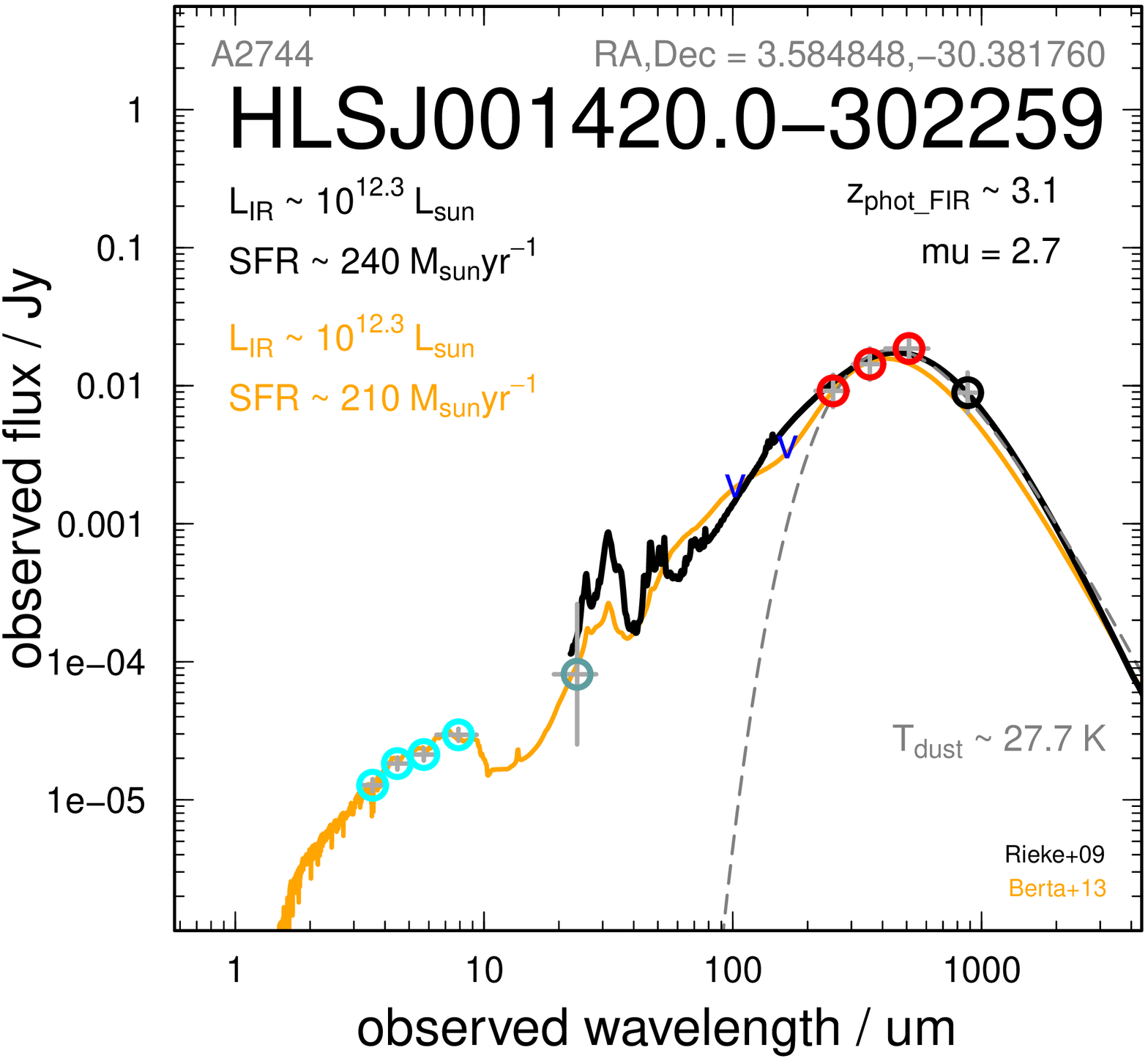}\\
\includegraphics[viewport=20mm 30mm 187mm 205mm,width=57mm,angle=270,clip]{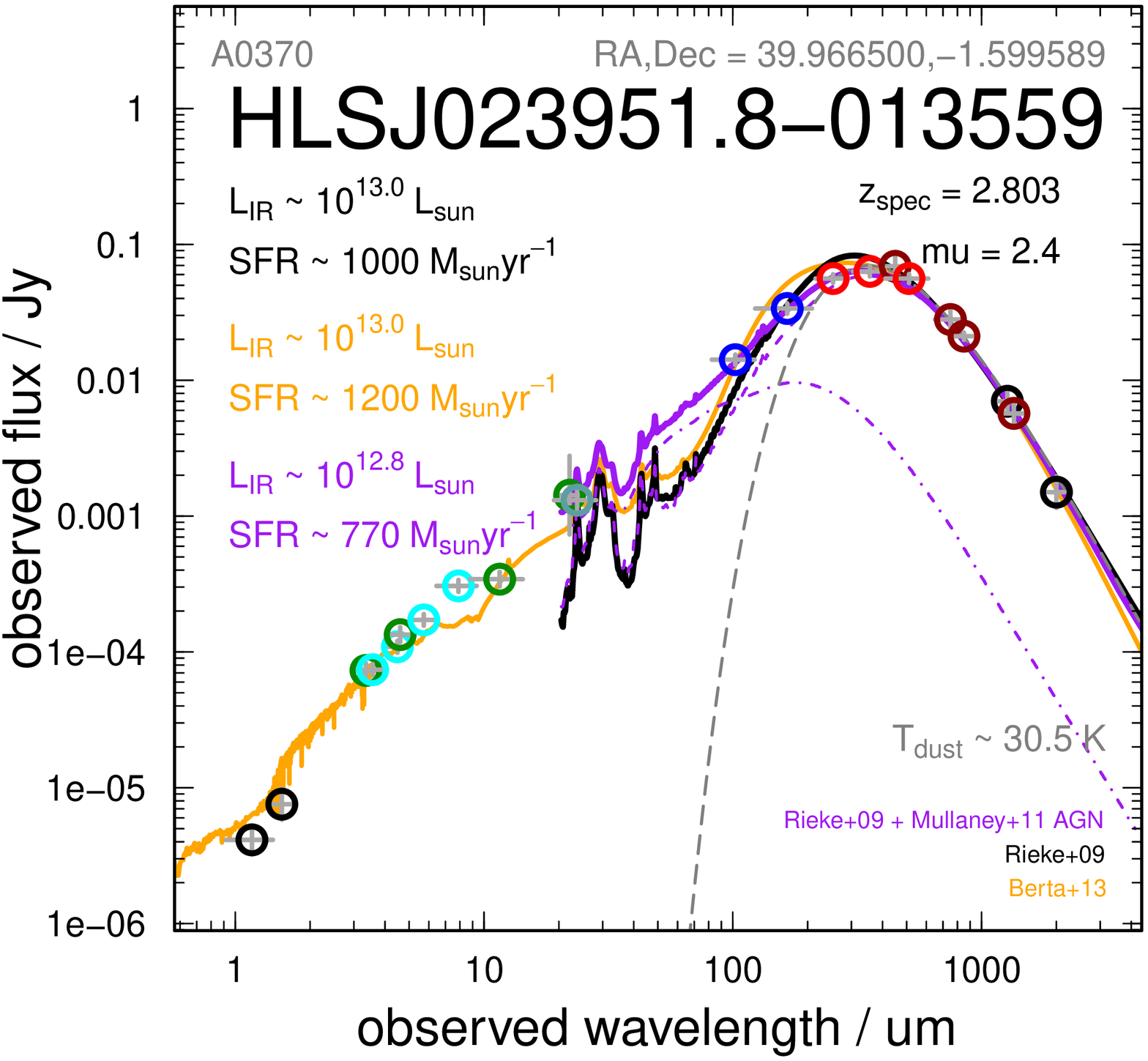}
\includegraphics[viewport=20mm 0mm 187mm 205mm,width=57mm,angle=270,clip]{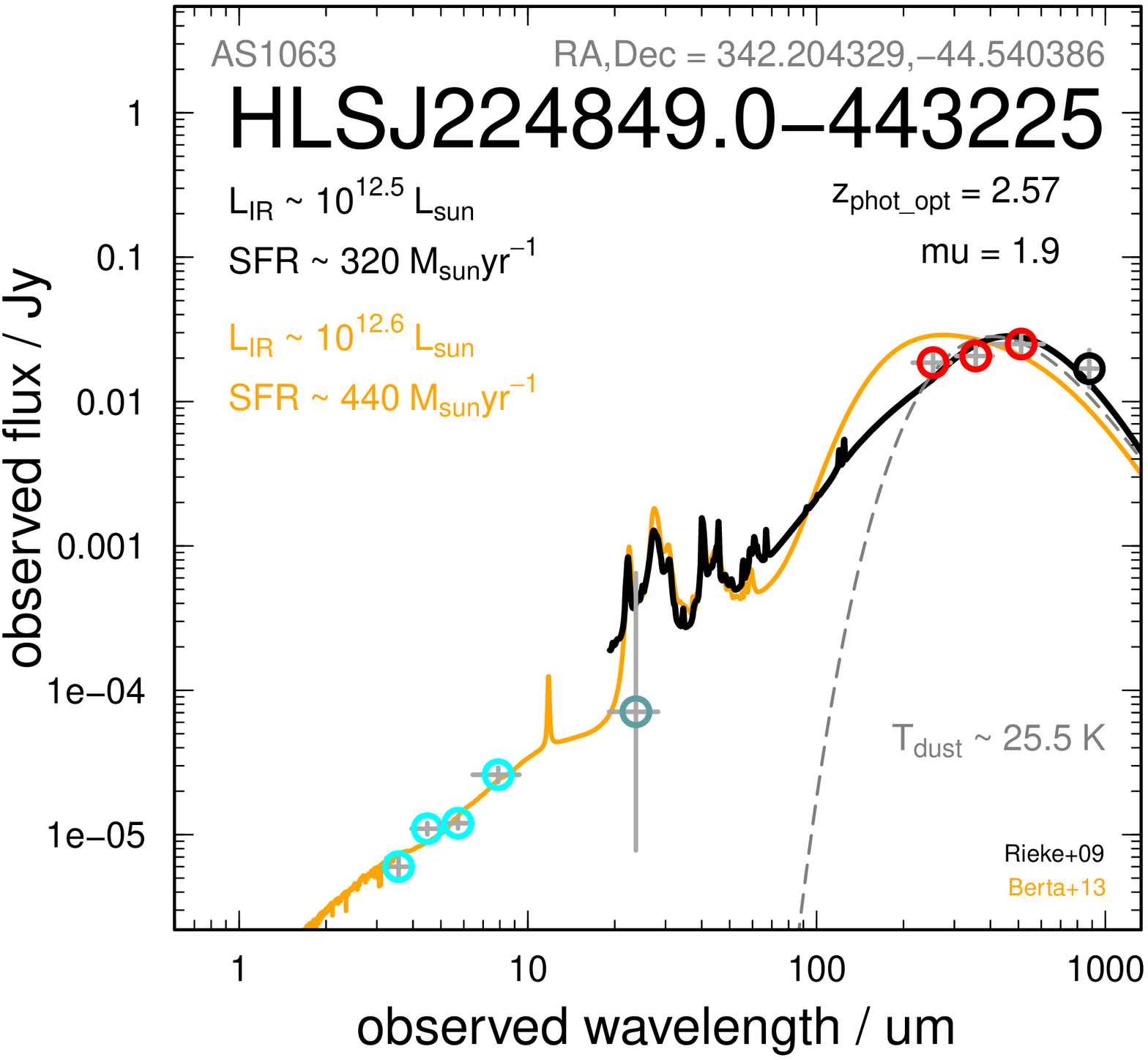}
\caption{IR SEDs for the four highest redshift, \textit{Herschel}-detected sources located in the HFF. Panel layout as in Figure \ref{fig:exseds}, with additional JCMT/SCUBA fluxes shown in brown and APEX/LABOCA, PdBI and IRAM-30m/GISMO continuum photometry in black.}
\label{fig:highz}
\end{figure*}

We now briefly describe a few of the most interesting \textit{Herschel}-detected sources within the HFF, starting with those at highest redshift. The four highest redshift sources are magnified by $\mu\sim1.5-3$, but are all intrinsically bright ULIRGs ($L_{\rm IR}>10^{12}$~L$_{\sun}$) magnified from near to, or above, the nominal \textit{Herschel} limit (Figure \ref{fig:highz}). In \textit{HST} imaging, they are small, faint, red and point-like, as expected for high redshift sources without very strong magnification.

The two $z>3$ sources, both located behind A2744, only have IR-derived redshift estimates. Fortunately, the $z_{\rm phot\_FIR}$ constraint is also aided by the 870~$\mu$m detection. Furthermore, \textit{Spitzer} fluxes (unused in the redshift estimate) provide a good independent validation of a $z>3$ continuum shape (upper panels, Figure \ref{fig:highz}). These two sources are to be explored in more detail by Boone et al. (in preparation).

The highest redshift source in the sample with a confirmed spectroscopic redshift is HLSJ023951.9--013559 at $z=2.803$ behind A370, which is better known as SMM02399--0136 \citep[e.g.][]{ivi98-583,ivi10-198}. We combine our \textit{Herschel} fluxes with (sub)-millimetre continuum observations from JCMT/SCUBA at 450, 750, 850~$\mu$m and 1.35~mm \citep{ivi98-583,cow02-2197}, PdBI 1.27~mm \citep{gen03-633} and IRAM-30m/GISMO 2~mm. From a pure \citeauthor{rie09-556} fit, we derive an estimate of the SFR$=$$1000\pm45 M_{\sun}$ yr$^{-1}$ ($L_{\rm IR}=10^{13.0}$~L$_{\sun}$) given $\mu=2.37\pm0.08$. This agrees favourably with the PEP/HerMES catalogues by \citet{mag12-155} and the earlier estimate by \citeauthor{ivi98-583}, accounting for different magnification and IMF assumptions. However, the source is a well-known quasar host and exhibits an obvious power-law continuum in the mid-IR (Figure \ref{fig:highz}, lower-left panel). Via the AGN fitting method described in Section \ref{sec:agn}, we derive an intrinsic SFR$=$$812\pm35 M_{\sun}$ yr$^{-1}$ ($L_{\rm IR}=10^{12.85\pm0.02}$~L$_{\sun}$) and a characteristic dust temperature of $30.5\pm0.7$~K.

The final source displayed in Figure \ref{fig:highz} is located behind the AS1063 central footprint. The optical counterpart has an \textit{HST}-derived photometric redshift from CLASH, which puts the source at $z=2.57$ with a magnification of $\mu=1.9$. We derive SFR$=$$320\pm38 M_{\sun}$ yr$^{-1}$ ($L_{\rm IR}=10^{12.46\pm0.05}$~L$_{\sun}$) and $T_{\rm dust}=25.5\pm1.7$~K. As with other IR-detected galaxies behind AS1063, this source will be explored further in Walth et al., in preparation.

\begin{figure*}
\centering
\begin{minipage}[c]{110mm}
\includegraphics[width=115mm]{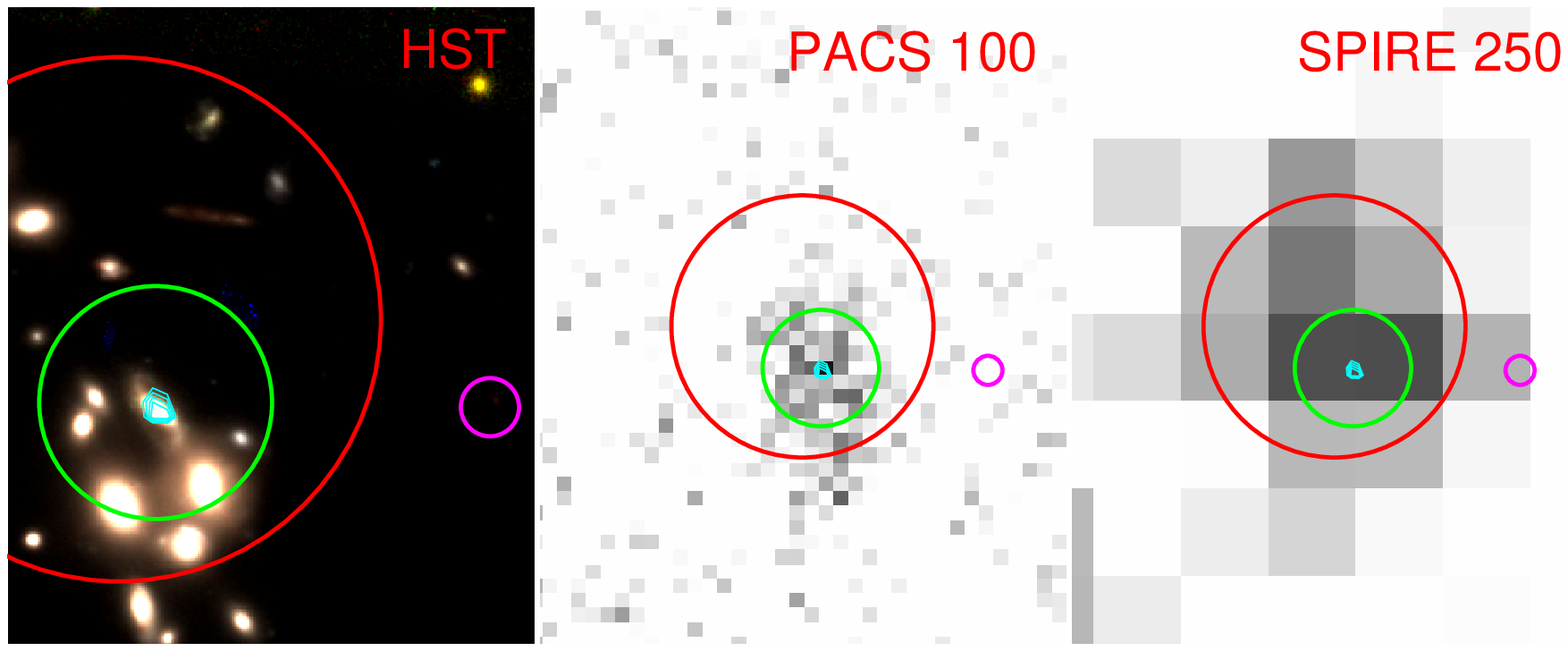}
\end{minipage}
\begin{minipage}[c]{65mm}
\includegraphics[viewport=20mm 0mm 210mm 205mm,width=65mm,angle=270,clip]{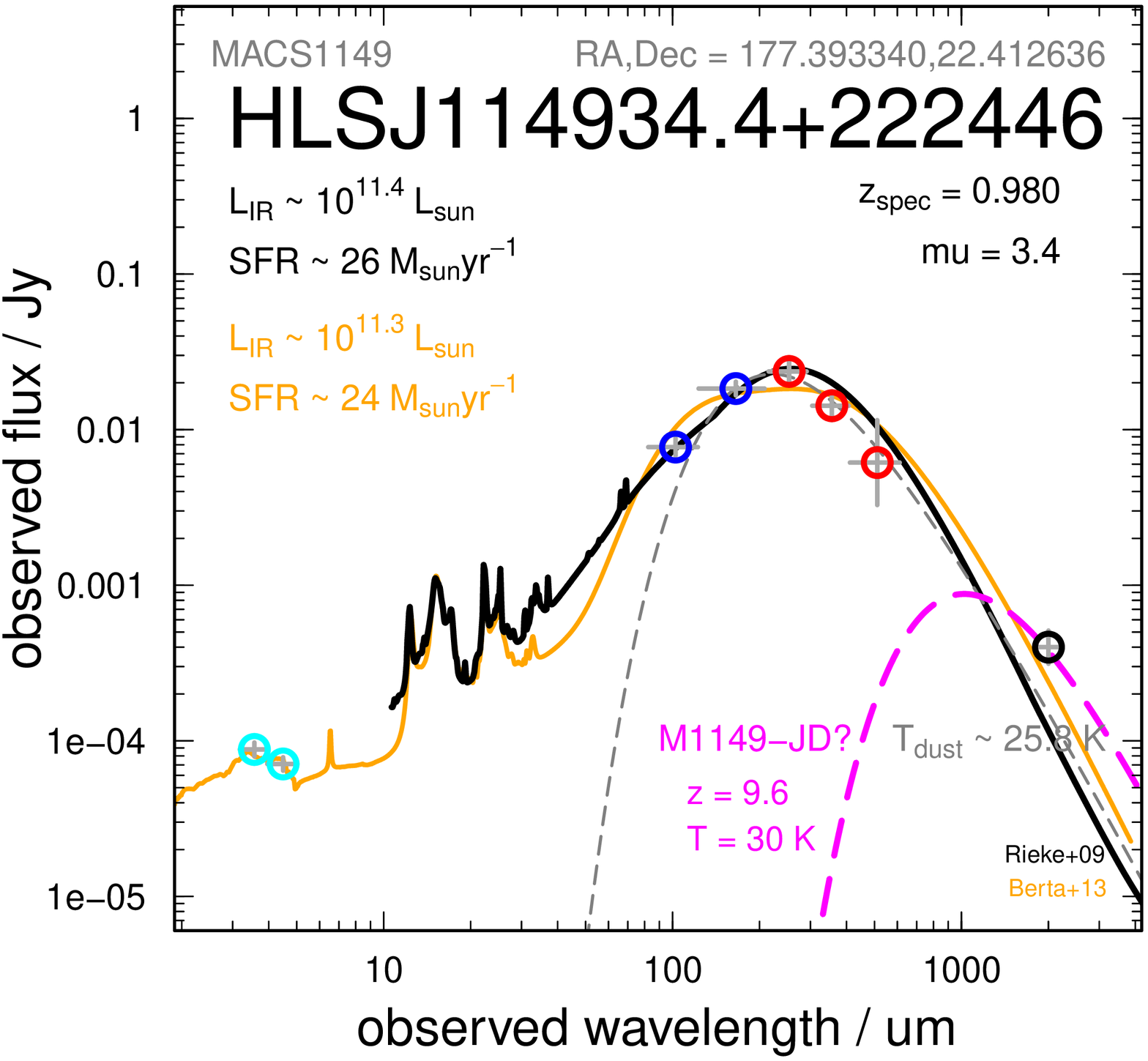}
\end{minipage}
\caption{\textit{Left panels:} The $z=9.6$ candidate M1149-JD \citep{dwe14-30}, marked by a 2~arcsec diameter magenta circle. Thumbnails show WFC3 1.05/1.25/1.60~$\mu$m, PACS 100~$\mu$m and SPIRE 250~$\mu$m. Green and red circles highlight the best fit PSF in PACS and SPIRE respectively (diameters indicate PSF FWHM). Both originate from the $z\sim1$ spiral galaxy, also detected by VLA (cyan contours). \textit{Right panel:} IR SED for the spiral galaxy (HLSJ114934.4+222446). Layout as in Figure \ref{fig:exseds}. A second dust SED (magenta dashes) shows that even if M1149-JD dominated the GISMO 2mm flux, it would not contribute significantly in the \textit{Herschel} bands (assuming $T_{\rm dust}=30$~K).}
\label{fig:m1149-jd}
\end{figure*}

One conspicuous absence from our list of high redshift sources in the HFF is the reported millimetre counterpart to the $z=9.6$ candidate M1149-JD \citep{dwe14-30}. We do find a \textit{Herschel} source (HLSJ114934.4+222446) within 10~arcsec of M1149-JD (Figure \ref{fig:m1149-jd}), but this is entirely consistent with a nearby magnified spiral galaxy at $z\sim1$ \citep{dwe15-sub,zav15-88}. The GISMO 2mm flux ($0.4\pm0.1$~mJy) appears to be 0.5~dex higher than expected for that spiral, so M1149-JD may be dominant at 2mm. However, any contribution by M1149-JD to the \textit{Herschel} flux would be undetected, as shown by the theoretical dust SED in Figure \ref{fig:m1149-jd}.

\subsection{Strongest magnification sources}
\label{sec:highmu}

We now discuss several sources with the highest magnification.  The 11 sources with $\mu>4$ are listed in Table \ref{tab:highmu}.

\begin{table*}
\caption{\textit{Herschel}-detected sources with the highest magnification ($\mu>4$).}
\label{tab:highmu}
\begin{tabular}{llrrrl}
\hline
\multicolumn{1}{c}{ID} & \multicolumn{1}{c}{Cluster} & \multicolumn{1}{c}{$z$$^1$} & \multicolumn{1}{c}{$\mu$$^2$} & \multicolumn{1}{c}{log($L_{\rm IR}/L_{\sun}$)$^3$} & \multicolumn{1}{c}{Notes}\\
\hline
HLSJ023953.0--013507 & \multirow{3}{*}{A0370} & \multirow{3}{*}{0.725} & \multirow{3}{*}{109 $\pm$ 16} & \multirow{3}{*}{10.09 $\pm$ 0.07} & \multirow{3}{6cm}{\textbf{Figure \ref{fig:a370_arc}.} Arc \citep[e.g.][]{sou88-19} catalogued as three \textit{Herschel} sources} \\
HLSJ023952.6--013506 \\
HLSJ023953.7--013503 \\
\hline
HLSJ114935.5+222350 & MACS1149 & \textbf{1.491} & 23.0 $\pm$ 7.0 & 10.66 $\pm$ 0.09 & \textbf{Figure \ref{fig:m1149_snhost}.} Multiple-image supernova host \citep[e.g.][]{smi09-163} \\
HLSJ071734.4+374432 & MACS0717 & 1.15 & 5.9 $\pm$ 0.8 & 10.95 $\pm$ 0.08 & \\
HLSJ023952.0--013348 & A0370 & \textbf{1.034} & 5.5 $\pm$ 0.4 & 10.99 $\pm$ 0.09 & \\
HLSJ071736.8+374507 & MACS0717 & 1.13 & 5.2 $\pm$ 0.7 & 10.80 $\pm$ 0.45 & \\
HLSJ224841.8--443157 & AS1063 & \textbf{0.610} & 5.1 $\pm$ 0.1 & 11.42 $\pm$ 0.08 & Large spiral (Walth et al., in prep) \\
HLSJ023956.6--013426 & A0370 & \textbf{1.062} & 4.6 $\pm$ 0.3 & 12.04 $\pm$ 0.09 & \textbf{Figure \ref{fig:a370_ear}.} Ring galaxy \citep[e.g.][]{sou99-70} \\
HLSJ023954.1--013532 & A0370 & \textbf{1.200} & 4.5 $\pm$ 0.2 & 11.21 $\pm$ 0.08 & \textbf{Figure \ref{fig:a370_rings}.} Interlinked rings \\
HLSJ041607.7--240432 & MACS0416 & \textit{$\sim$0.6} & 4.0 $\pm$ 0.1 & 10.30 $\pm$ 0.18 & \\
\hline
\end{tabular}
\\
\raggedright
$^1$ Counterpart redshift format as in Table \ref{tab:cat2}\\
$^2$ All magnification factors from the CATS models\\
$^3$ $L_{\rm IR}$ corrected for magnification
\end{table*}

\begin{figure*}
\centering
\begin{minipage}[c]{75mm}
\includegraphics[width=70mm]{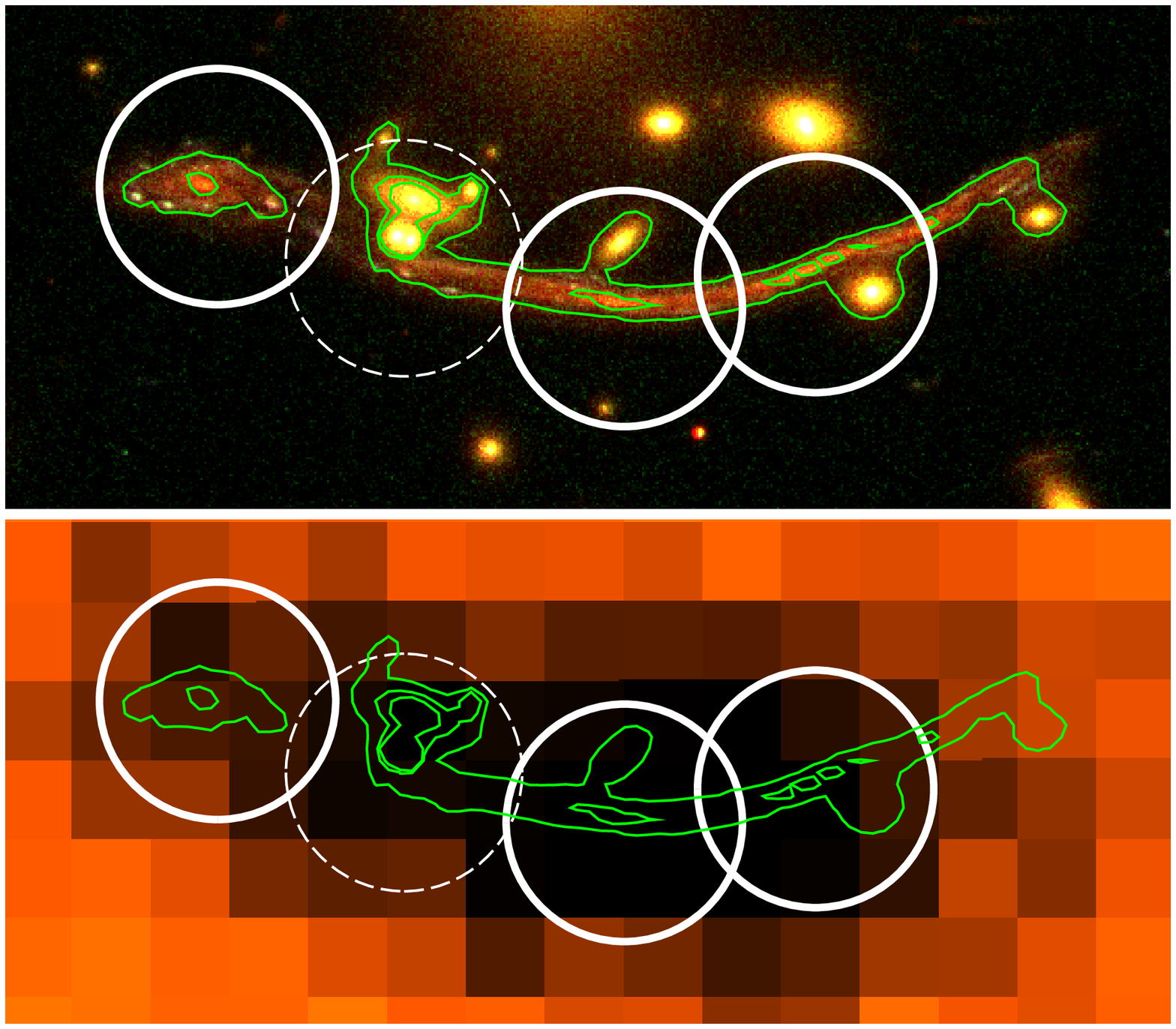}
\end{minipage}
\hspace{5mm}
\begin{minipage}[c]{65mm}
\includegraphics[viewport=20mm 0mm 210mm 205mm,width=65mm,angle=270,clip]{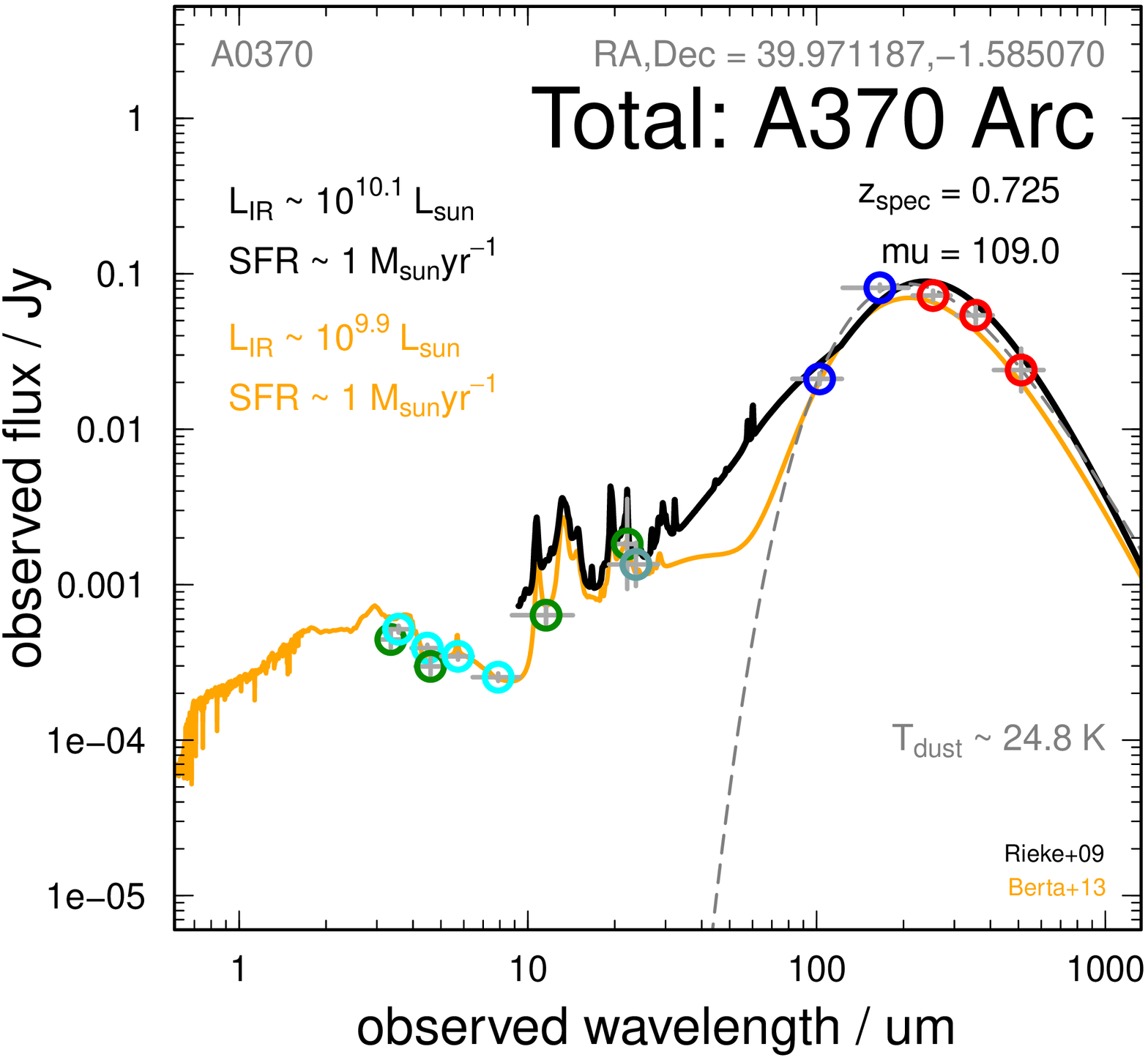}
\end{minipage}
\caption{The strongly lensed arc at $z=0.725$ behind A370. \textit{Left panels:} ACS 475/625/814~nm (\textit{upper}) and PACS 100~$\mu$m (\textit{lower}) maps of the arc (approximately outlined in green). The arc is catalogued as three separate \textit{Herschel} sources (see Table \ref{tab:highmu}) highlighted by solid white circles with 8~arcsec diameters (PACS PSF FWHM). A fourth IR source was only detected at 100~$\mu$m and is therefore cut from the catalogue (dashed circle). The western end of the optical arc is undetected by \textit{Herschel}. \textit{Right panel:} IR SED summed over the whole arc, assuming a magnification $\mu=109$. Panel layout as in Figure \ref{fig:exseds}.}
\label{fig:a370_arc}
\end{figure*}

The most spectacular of the highly magnified sources is undoubtably the strongly-lensed arc ($z=0.725$) in A370. This is well-known as the original spectroscopically-confirmed cluster-lensed arc \citep{sou88-19}, and the dusty star-forming component was first glimpsed using the \textit{Infrared Space Observatory} Camera (ISOCAM) by \citet{met03-791}. The $\sim$20~arcsec feature actually comprises five distorted images of the same background galaxy \citep{ric10-44}, as shown in the upper-left panel of Figure \ref{fig:a370_arc}.

The PSF fitting technique identifies four of the images as individual \textit{Herschel} sources, although one is only detected at 100~$\mu$m and is cut from the final catalogue. The western end of the optical arc is too faint for detection in any \textit{Herschel} band. The three catalogue sources are listed individually in the full Table \ref{tab:cat2}. The uppermost rows of Table \ref{tab:highmu} give the total intrinsic properties of the arc summed over all five images, with the total SED presented in Figure \ref{fig:a370_arc}. Assuming a magnification $\mu=109\pm16$, we derive a significantly sub-LIRG SFR$_{\rm total}=1.40\pm0.05$~$M_{\sun}$ yr$^{-1}$ ($L_{\rm IR} = 10^{10.1}$~L$_{\sun}$) and a $T_{\rm dust} = 24.8\pm0.4$~K. Reassuringly, each of the individual images also exhibit the same SED shape and $T_{\rm dust}$.

\begin{figure*}
\centering
\begin{minipage}[c]{75mm}
\includegraphics[width=60mm]{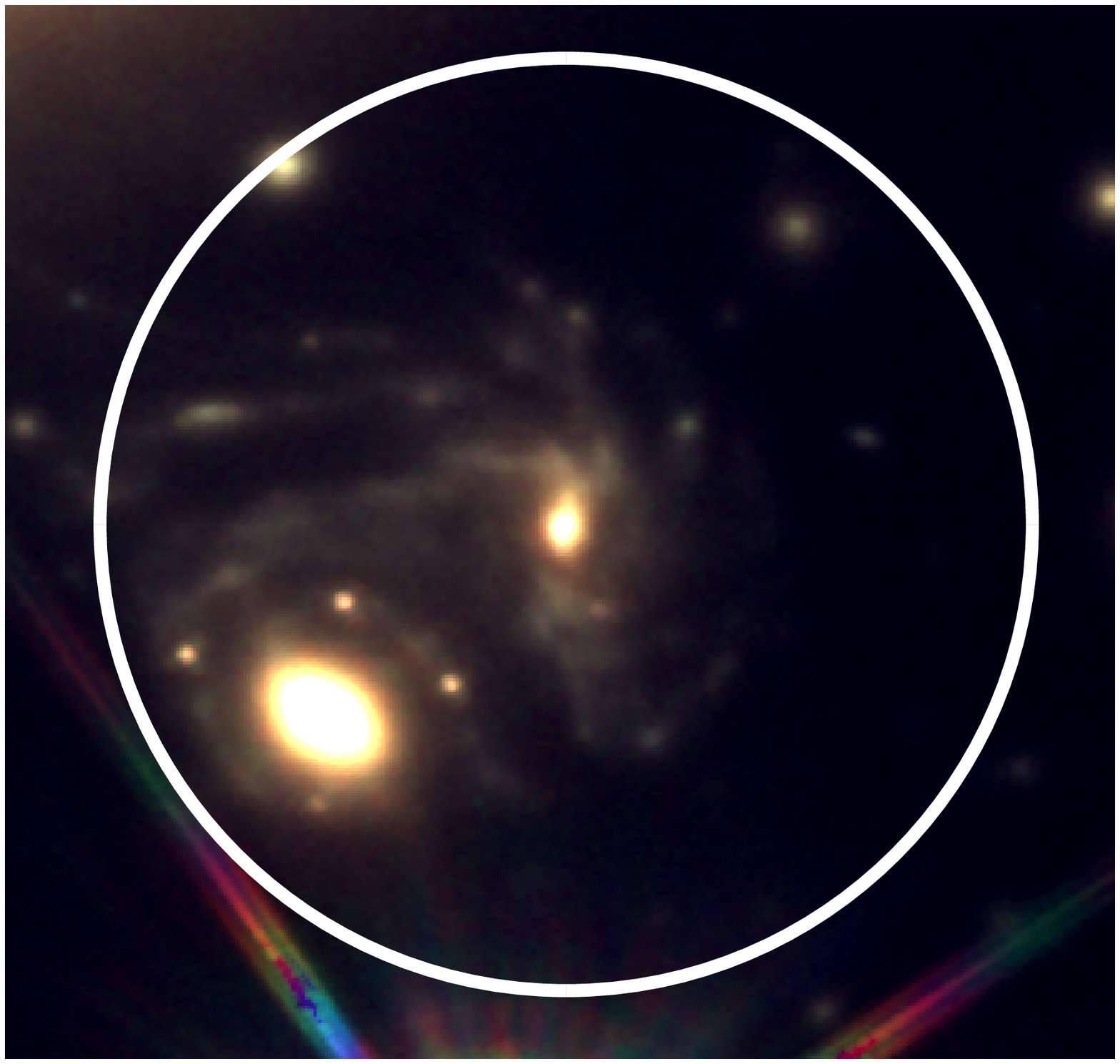}
\end{minipage}
\hspace{5mm}
\begin{minipage}[c]{65mm}
\includegraphics[viewport=20mm 0mm 210mm 205mm,width=65mm,angle=270,clip]{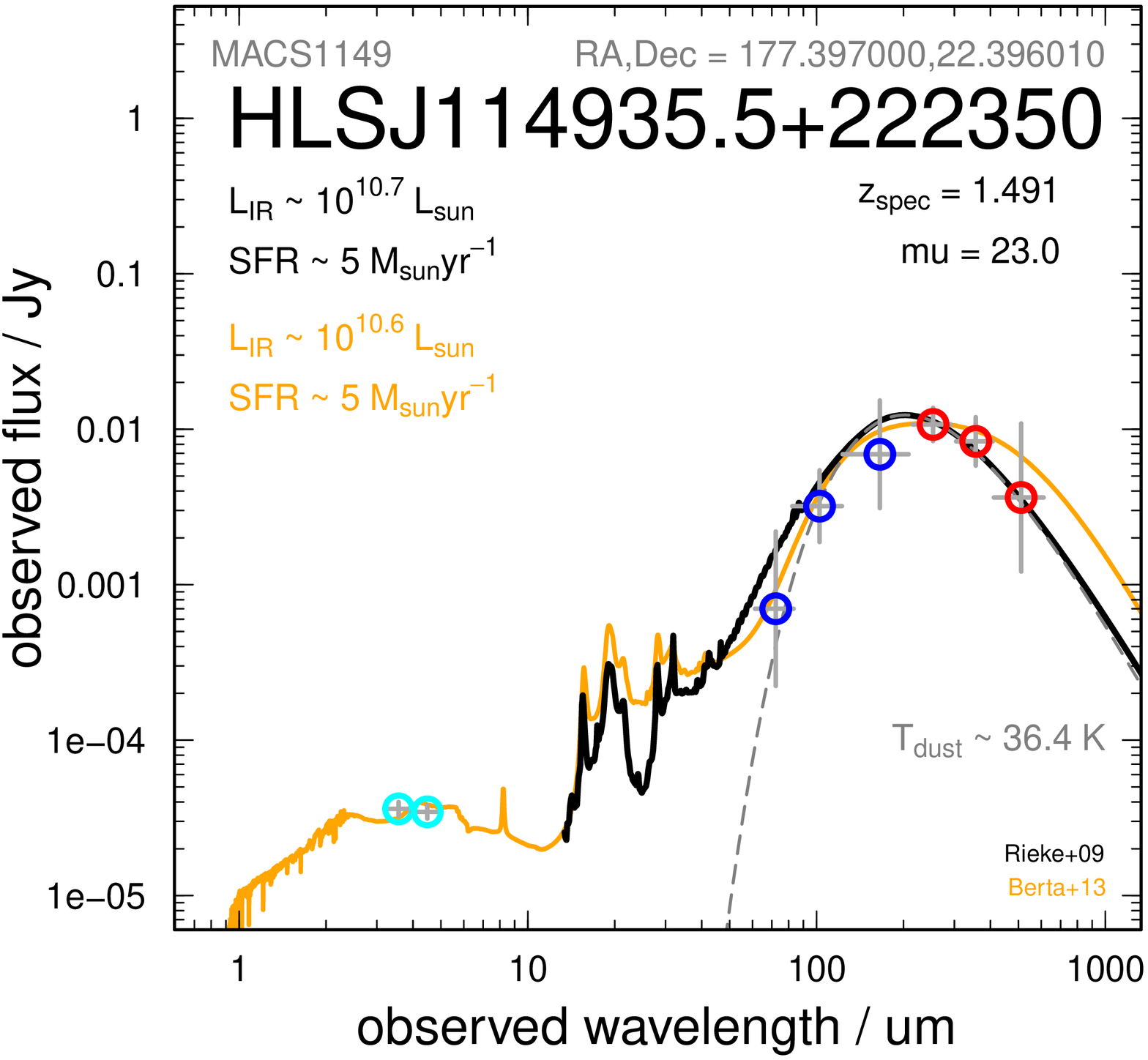}
\end{minipage}
\caption{The strongly lensed ($\mu\sim23\pm7$) supernova host at $z=1.49$ behind M1149 \citep[e.g.][]{smi09-163,sai13-2}. \textit{Left panel:} WFC3 1.05/1.25/1.60~$\mu$m image, including the supernova Einstein cross. The solid white circle indicates the location of the PACS detection, with an 8~arcsec diameter (PACS PSF FWHM). \textit{Right panel:} IR SED of this sub-LIRG (layout as in Figure \ref{fig:exseds}).}
\label{fig:m1149_snhost}
\end{figure*}

The second source in Table \ref{tab:highmu}, HLSJ114935.5+222350 behind M1149, is also widely known. The source corresponds to the highest magnification image ($\mu\sim23\pm7$) of a lensed galaxy at $z=1.49$, labelled A1.1 in \citet{smi09-163}. This source recently became the first known host of a multiply-imaged supernova, including a spectacular Einstein cross \citep{kel15-1123}. None of the other images are detected by \textit{Herschel}. Figure \ref{fig:m1149_snhost} presents the \textit{HST}/WFC3 image and IR SED, which shows an intrinsically faint (sub-LIRG) source, with a SFR=$5.1\pm0.9$~$M_{\sun}$ yr$^{-1}$ ($L_{\rm IR} = 10^{10.7\pm0.1}$~L$_{\sun}$) and a $T_{\rm dust} = 36\pm4$~K. Using Figure \ref{fig:lir_z}, we confirm that this is one of the faintest galaxies in the survey, more than an order of magnitude below the nominal detection limit of \textit{Herschel}.

HLSJ114935.5+222350 ($z = 1.49$) was studied in the compilation of 17 lensed sources by \citet{sai13-2}, named `J1149'. As the only overlap between the two samples, we briefly compare SEDs. The SPIRE photometry agrees perfectly ($<$5\%), which is well within the stated errors and should not be a surprise given the near identical data reduction and similar (PSF-fitting) flux derivation. Our PACS photometry in Table \ref{tab:cat1} is systematically higher (by $\sim40\%$), but the fluxes are very faint and very uncertain in both studies (between 50$-$100\% errors), so the $1-2$~mJy difference between our results and \citeauthor{sai13-2} can be explained by the noise. Our SED yields a higher SFR ($5.16\pm0.85$ compared to $3.4\pm2.8$~$M_{\sun}$ yr$^{-1}$), but both this property and $T_{\rm dust}$ agrees within the errors. We note that the extinction-corrected SFR derived from $V_{\rm 555}$ by \citet{smi09-163} agrees better with our higher IR estimate (SFR $\sim6$~$M_{\sun}$ yr$^{-1}$).

\begin{figure*}
\centering
\begin{minipage}[c]{75mm}
\includegraphics[viewport=0mm 25mm 185mm 210mm,width=60mm,clip]{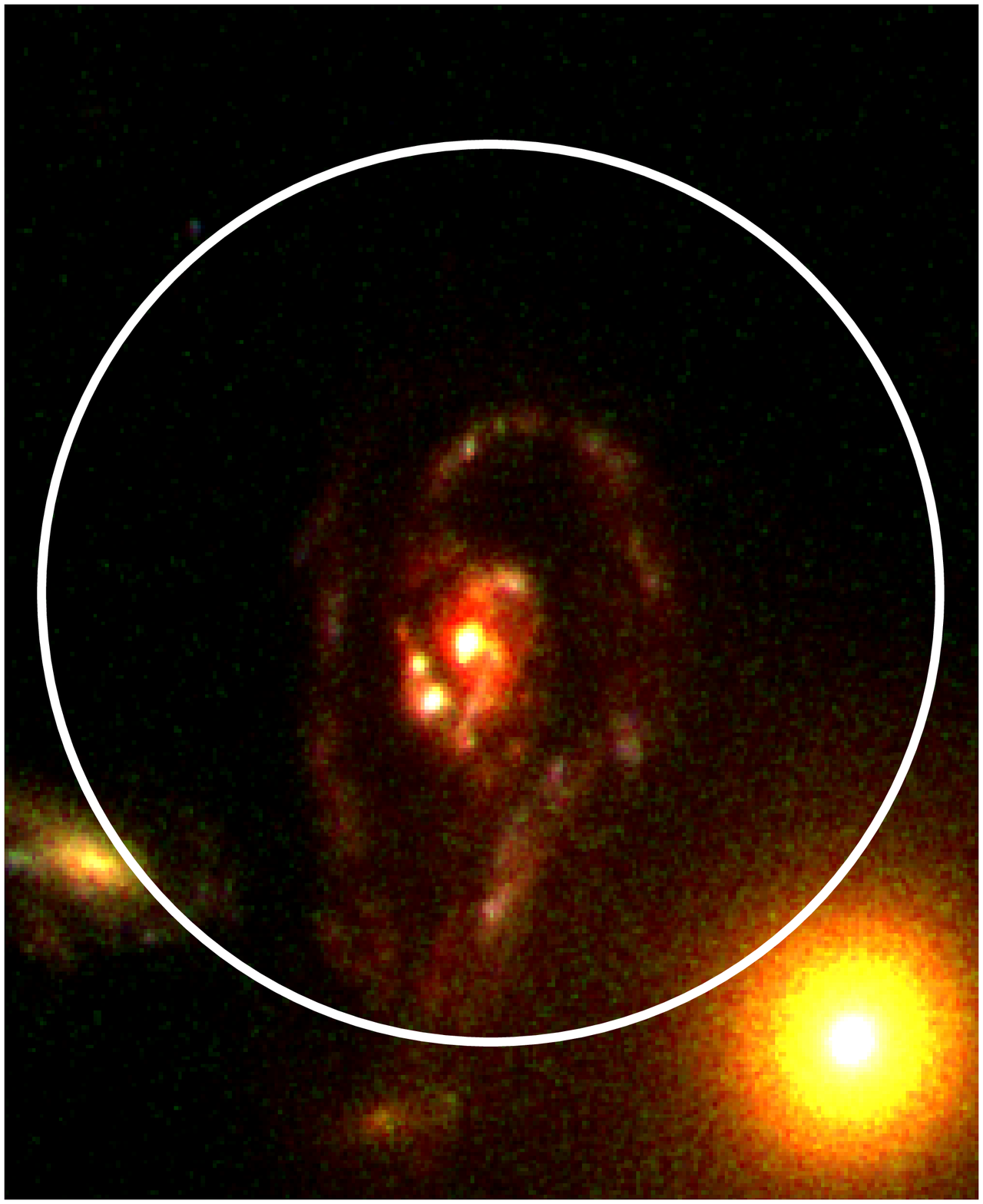}
\end{minipage}
\hspace{5mm}
\begin{minipage}[c]{65mm}
\includegraphics[viewport=20mm 0mm 210mm 205mm,width=65mm,angle=270,clip]{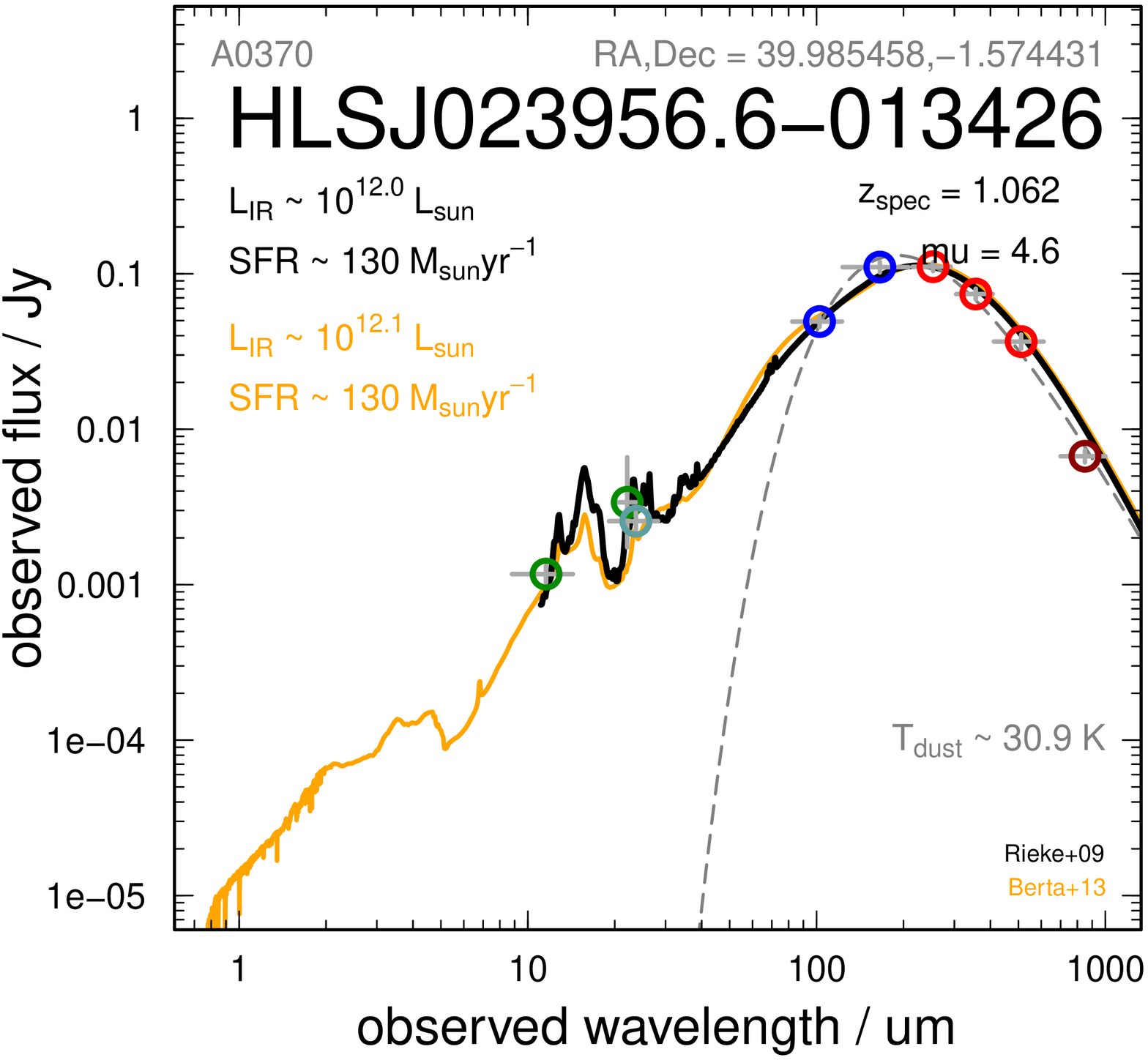}
\end{minipage}
\caption{The magnified ($\mu=4.6$) source HLSJ023956.6--013426 (SMMJ02399--0134; \citealt{sma98-21,sou99-70, mag12-155}) at $z=1.062$ behind A370. Layout as in Figure \ref{fig:m1149_snhost} (but with \textit{HST} 475/625/814~nm image).}
\label{fig:a370_ear}
\end{figure*}

Next we described the ring galaxy HLSJ023956.6--013426 at $z=1.062$ behind A370 ($\mu=4.6\pm0.3$), as shown in Figure \ref{fig:a370_ear}. The galaxy was extensively explored by \citet{sou99-70}, although previously detected as a SCUBA source \citep[][A370 L3]{sma98-21}. This distorted, ring-shaped source is one of the brightest \textit{Herschel} point sources within the HFF footprints, and we calculate an intrinsic SFR=$126\pm5$~$M_{\sun}$ yr$^{-1}$ ($L_{\rm IR} = 10^{12.04\pm0.01}$~L$_{\sun}$) and $T_{\rm dust} = 31\pm1$~K. \textit{Herschel} photometry (from the PEP/HerMES catalogue) has previously been published for this source \citep[][SMMJ02399--0134]{mag12-155} allowing us another opportunity to check our derived properties. The authors applied a lower $\mu=2.5$, but correcting for this, the derived properties are reassuringly similar: $L_{\rm IR} = 10^{12.04\pm0.05}$~L$_{\sun}$ and $T_{\rm dust} = 31\pm1$~K.

\begin{figure*}
\centering
\begin{minipage}[c]{75mm}
\includegraphics[width=60mm]{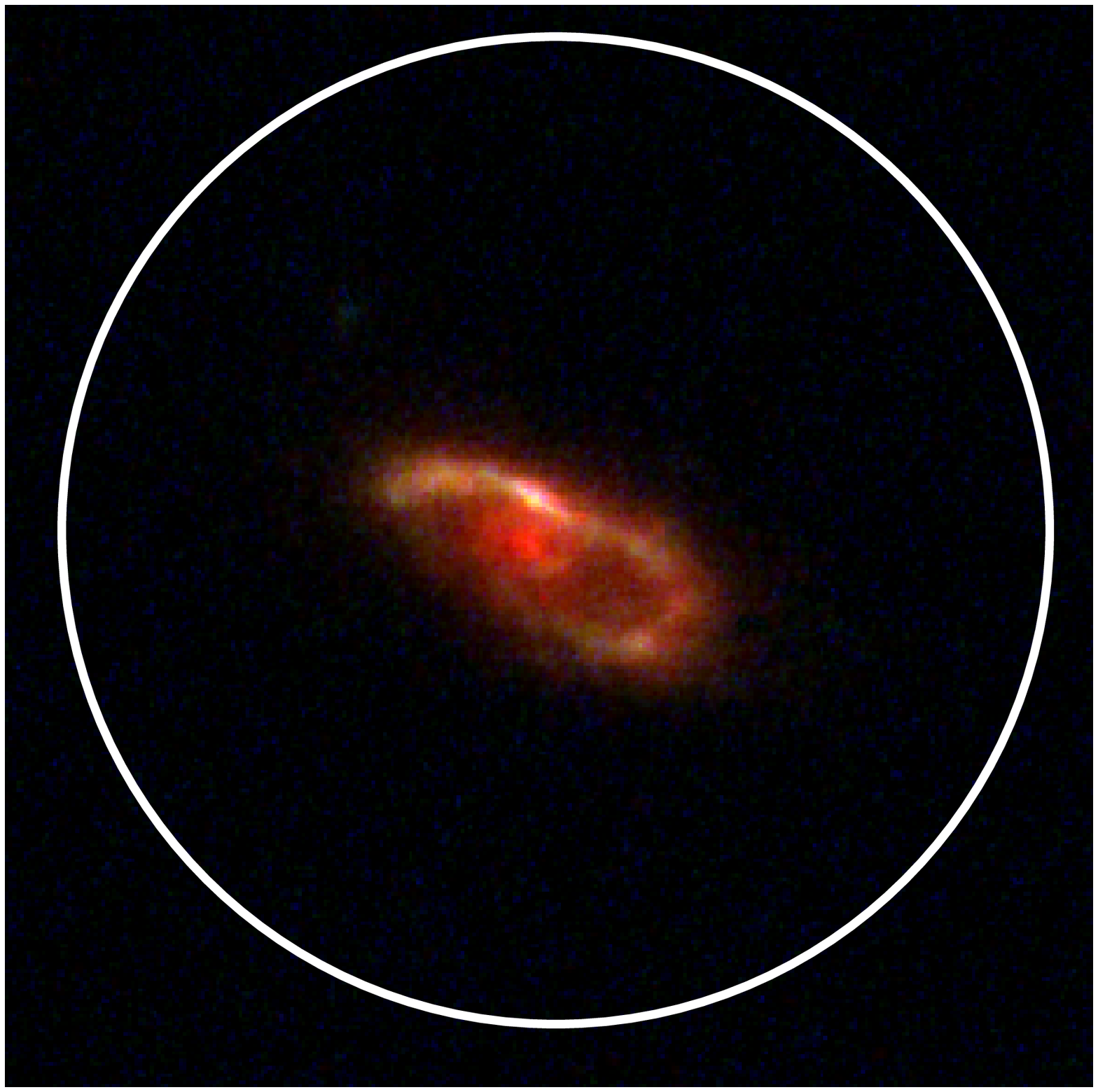}
\end{minipage}
\hspace{5mm}
\begin{minipage}[c]{65mm}
\includegraphics[viewport=20mm 0mm 210mm 205mm,width=65mm,angle=270,clip]{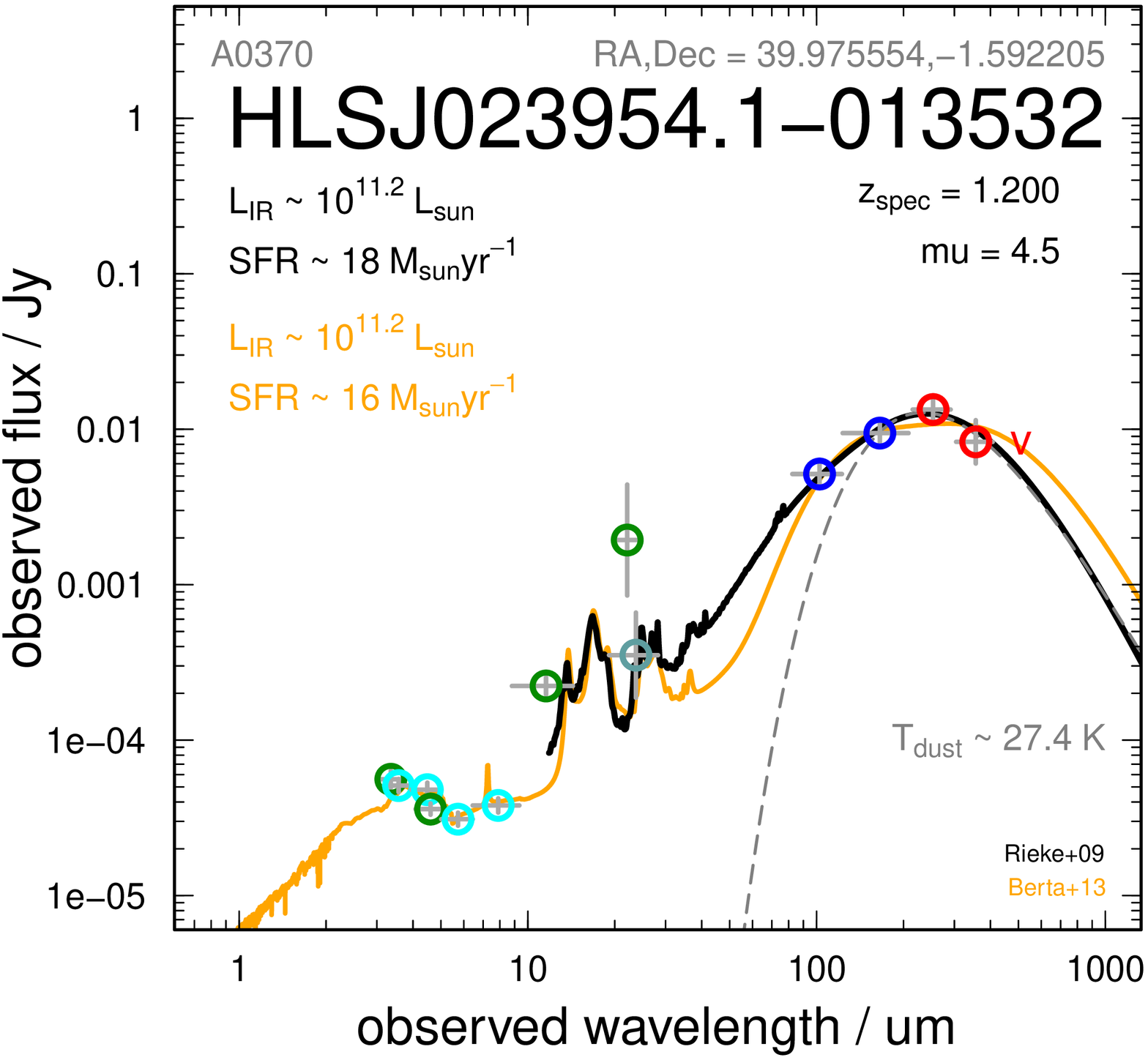}
\end{minipage}
\caption{A magnified ($\mu=4.5$) source at $z=1.2$ behind A370. Layout as in Figure \ref{fig:m1149_snhost} (but with \textit{HST} 0.625/0.814/1.60~$\mu$m image).}
\label{fig:a370_rings}
\end{figure*}

The next highest magnification source, HLSJ023954.1--013532 at $z=1.20$ (from \textit{HST} grism spectroscopy) behind A370 ($\mu=4.5$), also has an interesting morphology, as shown in Figure \ref{fig:a370_rings}. The source has a magnification-corrected SFR$=18\pm2$~$M_{\sun}$ yr$^{-1}$ ($L_{\rm IR} = 10^{11.21\pm0.04}$~L$_{\sun}$) and a typical $T_{\rm dust} = 27\pm2$~K. This is a prime example of a $z>1$ LIRG which has been magnified by cluster lensing to enable a very good constraint on the IR properties. Thus far, this galaxy has not been specifically targetted in any published observations.

The remaining strongly-lensed sources are a mixed bunch of distorted spiral or irregular morphologies, including a sub-LIRG spiral galaxy just behind AS1063 (described further in Walth et al. in preparation) and several LIRGs at $z\sim1$.

\section{Conclusions}
\label{sec:conc}

We present a complete census of the 263 \textit{Herschel}-detected sources within the \textit{HST} Frontier Fields, including 163 lensed sources located behind the clusters. Our primary aim is to provide a robust legacy catalogue of the \textit{Herschel} fluxes, which we combine with archival data from \textit{Spitzer} and \textit{WISE} to produce IR SEDs.

We optimally combine the IR photometry with data from \textit{HST}, VLA and ground-based observatories in order to identify optical counterparts and gain source redshifts. Hence we also provide a ``value-added" catalogue including magnification factor, intrinsic IR luminosity and characteristic dust temperature. We expect to update the derived properties for a few sources in the online catalogue, as 1) upcoming deeper/wider radio data helps to confirm our optical counterparts for sources without current VLA detections and 2) further spectroscopic observations secure the derived luminosity for the 151 sources without spectroscopic redshifts, particularly the 62 sources with only IR-based estimates. The catalogues will provide a useful reference for future multi-wavelength studies of the HFF.

There are nine \textit{Herschel} sources with counterparts at $z>2$ and a further nine IR-bright sources with $\mu>4$, six of which are sub-LIRG. Although we locate more than 20 background sub-LIRGs, it is clear that for an in-depth study of high-redshift, intrinsically faint sources, a larger cluster sample is required. Therefore, this work also serves as a preview of the full \textit{Herschel Lensing Survey} (HLS; \citealt{ega10-12}), covering 10$\times$ the cluster sample, and many thousands of \textit{Herschel}-detected sources \citep{ega-prep}.

\textit{Herschel} imaging, catalogues and source IR SEDs can be downloaded from the public flavour of the Rainbow Database.\footnote{https://rainbowx.fis.ucm.es}.

\section*{Acknowledgments}

TDR was supported by a European Space Agency (ESA) Research Fellowship at the European Space Astronomy Centre (ESAC), in Madrid, Spain. RJI acknowledges support from the ERC in the form of the Advanced Investigator Program, 321302, COSMICISM. IRS acknowledges support from STFC (ST/L00075X/1), the ERC Advanced Investigator programme DUSTYGAL 321334 and a Royal Society/Wolfson Merit Award.

Based on observations made with the NASA/ESA \textit{Hubble Space Telescope}, which is operated by the Association of Universities for Research in Astronomy, Inc., under NASA contract NAS 5-26555. These observations are associated with programmes \#11689, 13389, 13495, 13496, 13498, 13504. This publication also uses data products from the \textit{Wide-field Infrared Survey Explorer} (\textit{WISE}), which is a joint project of UCLA and JPL, Caltech, funded by NASA. This research made use of the NASA/IPAC Infrared Science Archive, which is operated by JPL, Caltech, under contract with NASA.

We made use of a private version of the Rainbow Cosmological Surveys Database, operated by the Universidad Complutense de Madrid (UCM), and PGP--G acknowledges support from Spanish Government MINECO grant AYA2012--31277.

This work also utilises gravitational lensing models produced by PIs Ebeling (CATS) and Merten \& Zitrin, funded as part of the \textit{HST} Frontier Fields program conducted by STScI. STScI is operated by the Association of Universities for Research in Astronomy, Inc. under NASA contract NAS 5-26555. The lens models were obtained from the Mikulski Archive for Space Telescopes (MAST).

\pagebreak

\input{cattable1.txt}

\pagebreak

\input{cattable2.txt}

\label{lastpage}
\end{document}

%% file: cattable1.txt
\begin{table*}
\caption{Observed PACS and SPIRE fluxes for \textit{Herschel}-detected sources within the \textit{HST} Frontier Fields.}
\begin{tabular}{lrrrrrrr}
\hline
\multicolumn{1}{c}{ID (ref)$^1$} & \multicolumn{1}{c}{Field$^2$} & \multicolumn{1}{c}{$S_{70}$} & \multicolumn{1}{c}{$S_{100}$} & \multicolumn{1}{c}{$S_{160}$} & \multicolumn{1}{c}{$S_{250}$} & \multicolumn{1}{c}{$S_{350}$} & \multicolumn{1}{c}{$S_{500}$} \\
 & & \multicolumn{1}{c}{mJy} & \multicolumn{1}{c}{mJy} & \multicolumn{1}{c}{mJy} & \multicolumn{1}{c}{mJy} & \multicolumn{1}{c}{mJy} & \multicolumn{1}{c}{mJy} \\
\hline
HLSJ001412.7--302359 (P) & A2744 C & -- & 5.2 $\pm$ 0.9 & 10.5 $\pm$ 1.8 & 13.1 $\pm$ 2.9 & 6.8 $\pm$ 3.0 & -- \\
HLSJ001415.3--302423 (P) & A2744 C & -- & 5.6 $\pm$ 0.8 & 10.2 $\pm$ 2.1 & 14.8 $\pm$ 2.8 & -- & -- \\
HLSJ001416.7--302304 (P) & A2744 C & -- & 2.5 $\pm$ 0.6 & 7.5 $\pm$ 1.3 & -- & -- & -- \\
HLSJ001416.5--302410 (P) & A2744 C & -- & 7.2 $\pm$ 0.9 & 11.8 $\pm$ 1.6 & 9.8 $\pm$ 5.5 & 8.4 $\pm$ 5.9 & 6.3 $\pm$ 9.2 \\
HLSJ001418.5--302246 (P) & A2744 C & -- & 7.6 $\pm$ 0.9 & 12.4 $\pm$ 1.6 & 12.6 $\pm$ 3.3 & 6.3 $\pm$ 2.8 & -- \\
HLSJ001417.6--302301 (P) & A2744 C & -- & 25.5 $\pm$ 2.1 & 62.4 $\pm$ 5.1 & 66.1 $\pm$ 5.2 & 47.5 $\pm$ 4.8 & 21.5 $\pm$ 4.8 \\
\hline
\end{tabular}
\end{table*}
\addtocounter{table}{-1}
\begin{table*}
\caption{\textit{Continued}}
\begin{tabular}{lrrrrrrr}
\hline
HLSJ023953.7--013503 (P) & A0370 C & -- & 7.9 $\pm$ 0.9 & 20.4 $\pm$ 2.0 & 16.1 $\pm$ 2.6 & 10.7 $\pm$ 5.1 & 4.0 $\pm$ 3.0 \\
HLSJ023954.1--013532 (P) & A0370 C & -- & 5.2 $\pm$ 0.8 & 9.4 $\pm$ 1.6 & 13.4 $\pm$ 2.6 & 8.3 $\pm$ 2.7 & -- \\
HLSJ023954.6--013427 (P) & A0370 C & -- & 7.4 $\pm$ 0.8 & 7.1 $\pm$ 1.4 & 3.8 $\pm$ 2.4 & -- & -- \\
HLSJ023954.8--013453 (P) & A0370 C & -- & 13.2 $\pm$ 1.1 & 15.7 $\pm$ 1.7 & 7.7 $\pm$ 4.3 & -- & -- \\
HLSJ023957.2--013259 (P) & A0370 C & -- & 3.7 $\pm$ 0.8 & 5.0 $\pm$ 1.0 & -- & -- & -- \\
HLSJ023957.8--013313 (P) & A0370 C & -- & 3.6 $\pm$ 0.6 & 8.1 $\pm$ 1.5 & -- & -- & -- \\
HLSJ023956.6--013426 (P) & A0370 C & -- & 49.3 $\pm$ 3.8 & 110.4 $\pm$ 8.6 & 111.0 $\pm$ 7.1 & 74.1 $\pm$ 5.8 & 36.6 $\pm$ 3.8 \\
HLSJ023957.5--013432 (P) & A0370 C & -- & 7.0 $\pm$ 0.8 & 11.2 $\pm$ 1.5 & -- & -- & -- \\
\hline
\end{tabular}
\end{table*}
\addtocounter{table}{-1}
\begin{table*}
\caption{\textit{Continued}}
\begin{tabular}{lrrrrrrr}
\hline
HLSJ071724.9+374410 (P) & M0717 C & -- & 4.4 $\pm$ 0.9 & 11.9 $\pm$ 1.6 & 8.2 $\pm$ 4.9 & -- & -- \\
HLSJ071727.7+374436 (P) & M0717 C & -- & 5.6 $\pm$ 0.7 & 6.3 $\pm$ 1.3 & -- & -- & -- \\
HLSJ071727.2+374606 (P) & M0717 C & -- & 9.6 $\pm$ 1.2 & 13.9 $\pm$ 1.7 & 9.6 $\pm$ 3.5 & -- & -- \\
HLSJ071727.5+374441 (P) & M0717 C & -- & 13.6 $\pm$ 1.2 & 26.8 $\pm$ 2.4 & 21.4 $\pm$ 3.2 & 8.3 $\pm$ 3.9 & -- \\
HLSJ071729.0+374320 (P) & M0717 C & -- & 7.2 $\pm$ 0.9 & 8.7 $\pm$ 1.5 & 11.7 $\pm$ 7.5 & -- & -- \\
HLSJ071728.5+374417 (P) & M0717 C & -- & 10.1 $\pm$ 1.0 & 9.6 $\pm$ 1.4 & 4.7 $\pm$ 3.3 & -- & -- \\
HLSJ071730.4+374617 (P) & M0717 C & -- & 7.6 $\pm$ 0.9 & 15.4 $\pm$ 1.7 & 12.5 $\pm$ 3.7 & -- & -- \\
HLSJ071730.7+374443 (P) & M0717 C & -- & 4.2 $\pm$ 2.6 & 15.0 $\pm$ 5.7 & 25.5 $\pm$ 5.1 & 20.1 $\pm$ 5.7 & 17.5 $\pm$ 14.8 \\
\hline
\end{tabular}
\end{table*}
\addtocounter{table}{-1}
\begin{table*}
\caption{\textit{Continued}}
\begin{tabular}{lrrrrrrr}
\hline
HLSJ114936.2+222417 (P) & M1149 C & -- & 2.3 $\pm$ 0.6 & 5.3 $\pm$ 1.3 & 9.5 $\pm$ 3.0 & 9.3 $\pm$ 3.0 & 6.7 $\pm$ 3.6 \\
HLSJ114936.1+222425 (P) & M1149 C & -- & 3.7 $\pm$ 0.7 & 7.9 $\pm$ 1.5 & 5.9 $\pm$ 2.5 & -- & -- \\
HLSJ114936.8+222254 (P) & M1149 C & -- & 10.3 $\pm$ 2.0 & 19.3 $\pm$ 2.5 & 19.2 $\pm$ 2.7 & 10.9 $\pm$ 4.1 & -- \\
HLSJ114937.3+222431 (P) & M1149 C & -- & 6.0 $\pm$ 0.8 & 12.1 $\pm$ 1.5 & 17.3 $\pm$ 2.7 & 8.6 $\pm$ 2.9 & 6.3 $\pm$ 3.2 \\
HLSJ114937.0+222542 (P) & M1149 C & -- & 3.0 $\pm$ 0.7 & 4.4 $\pm$ 1.3 & 4.9 $\pm$ 2.4 & -- & -- \\
HLSJ114938.1+222412 (P) & M1149 C & -- & 5.3 $\pm$ 0.7 & 9.6 $\pm$ 1.5 & 13.1 $\pm$ 3.0 & 8.4 $\pm$ 5.4 & -- \\
HLSJ114938.0+222428 (P) & M1149 C & -- & 8.2 $\pm$ 0.9 & 10.8 $\pm$ 1.5 & -- & -- & -- \\
HLSJ114938.8+222532 (P) & M1149 C & -- & 3.7 $\pm$ 0.8 & 4.4 $\pm$ 1.0 & -- & -- & -- \\
\hline
\end{tabular}
\end{table*}
\addtocounter{table}{-1}
\begin{table*}
\caption{\textit{Continued}}
\begin{tabular}{lrrrrrrr}
\hline
HLSJ224913.6--443238 (S) & AS1063 P & -- & -- & -- & 15.8 $\pm$ 2.6 & 6.4 $\pm$ 2.7 & -- \\
HLSJ224915.2--443220 (S) & AS1063 P & -- & -- & -- & 22.7 $\pm$ 2.6 & 20.9 $\pm$ 2.8 & 12.8 $\pm$ 2.8 \\
HLSJ224916.5--443042 (S) & AS1063 P & -- & -- & -- & 46.6 $\pm$ 3.9 & 31.5 $\pm$ 3.1 & 19.6 $\pm$ 3.7 \\
HLSJ224917.3--443119 (S) & AS1063 P & -- & -- & -- & 27.8 $\pm$ 3.0 & 17.7 $\pm$ 3.4 & 10.3 $\pm$ 2.9 \\
HLSJ224917.8--443256 (S) & AS1063 P & -- & -- & -- & 10.3 $\pm$ 3.0 & 6.1 $\pm$ 2.5 & -- \\
HLSJ224920.9--443337 (S) & AS1063 P & -- & -- & -- & 8.0 $\pm$ 2.9 & 9.7 $\pm$ 2.8 & 6.2 $\pm$ 1.8 \\
HLSJ224922.8--443235 (S) & AS1063 P & -- & -- & -- & 14.1 $\pm$ 2.4 & 7.8 $\pm$ 3.0 & 4.1 $\pm$ 6.5 \\
HLSJ224925.1--443252 (S) & AS1063 P & -- & -- & -- & 15.2 $\pm$ 3.0 & 11.8 $\pm$ 2.6 & 6.3 $\pm$ 2.6 \\
HLSJ224925.9--443332 (S) & AS1063 P & -- & -- & -- & 11.2 $\pm$ 2.5 & 4.4 $\pm$ 3.1 & -- \\
HLSJ224926.3--443223 (S) & AS1063 P & -- & -- & -- & 21.3 $\pm$ 2.6 & 21.1 $\pm$ 3.5 & 15.5 $\pm$ 3.3 \\
HLSJ224926.1--443305 (S) & AS1063 P & -- & -- & -- & 6.6 $\pm$ 2.4 & 8.7 $\pm$ 2.9 & 6.3 $\pm$ 2.6 \\
\hline
\\
\end{tabular}
\\
\raggedright
$^1$ \textit{Herschel} ID derived from PACS (P) or SPIRE (S) catalogue position\\
$^2$ C=central region; P=parallel\\
\end{table*}

%% file: cattable2.txt
\begin{table*}
\caption{Optical counterparts for \textit{Herschel}-detected sources within HFF. Derived properties account for magnification ($\mu$).}
\begin{tabular}{lrrrcrrrrrc}
\hline
\multicolumn{1}{c}{ID$^1$} & \multicolumn{1}{c}{Field$^2$} & \multicolumn{5}{c}{Optical/NIR/radio counterpart} & \multicolumn{1}{c}{$L_{\rm IR}$$^6$} & \multicolumn{1}{c}{SFR$_{\rm IR}$$^6$} & \multicolumn{1}{c}{$T_{\rm dust}$$^7$} & \multicolumn{1}{c}{AGN$^8$} \\
 & & \multicolumn{1}{c}{RA$^3$} & \multicolumn{1}{c}{Dec$^3$} & \multicolumn{1}{l}{(ref)$^3$} & \multicolumn{1}{c}{$z$ (ref)$^4$} & \multicolumn{1}{c}{$\mu$ (ref)$^5$} & \multicolumn{1}{c}{log($L$/$L_{\sun}$)} & \multicolumn{1}{c}{$M_{\sun}$ yr$^{-1}$} & \multicolumn{1}{c}{K} & \\
\hline
HLSJ001412.7--302359 & A2744 C & 3.552819 & --30.399785 & (H) & \textbf{0.688} (1) & 1.21 (1) & 11.21 $\pm$ 0.09 & 18.3 $\pm$ 3.3 & 26.7 $\pm$ 1.6 & 0 \\
HLSJ001415.3--302423 & A2744 C & 3.563939 & --30.406277 & (H) & \textbf{0.653} (1) & 1.28 (1) & 11.17 $\pm$ 0.09 & 16.7 $\pm$ 3.0 & 25.9 $\pm$ 1.5 & 0 \\
HLSJ001416.7--302304 & A2744 C & 3.569295 & --30.384239 & (H) & \textbf{0.296} (1) & -- ~~--~ & 10.16 $\pm$ 0.12 & 1.6 $\pm$ 0.4 & 19.8 $\pm$ 1.8 & 0 \\
HLSJ001416.5--302410 & A2744 C & 3.568938 & --30.402831 & (H) & \textit{$\sim$0.5} (2) & 1.30 (1) & 10.88 $\pm$ 0.16 & 8.6 $\pm$ 3.2 & 27.7 $\pm$ 2.4 & 1 \\
HLSJ001418.5--302246 & A2744 C & 3.577030 & --30.379500 & (H) & \textbf{0.498} (3) & 1.40 (1) & 10.90 $\pm$ 0.09 & 9.1 $\pm$ 1.6 & 26.3 $\pm$ 1.4 & 0 \\
HLSJ001417.6--302301 & A2744 C & 3.573229 & --30.383567 & (H) & \textit{$\sim$0.9} (2) & 1.95 (1) & 11.97 $\pm$ 0.13 & 107.2 $\pm$ 31.4 & 23.0 $\pm$ 1.6 & 0 \\
\hline
\end{tabular}
\end{table*}
\addtocounter{table}{-1}
\begin{table*}
\caption{\textit{Continued}}
\begin{tabular}{lrrrcrrrrrc}
\hline
HLSJ023953.7--013503 & A0370 C & 39.973640 & --1.584210 & (H) & \textbf{0.725} (12) & 12.38 (1) & 10.46 $\pm$ 0.08 & 3.3 $\pm$ 0.6 & 27.2 $\pm$ 2.4 & 0 \\
HLSJ023954.1--013532 & A0370 C & 39.975554 & --1.592205 & (H) & \textbf{1.200} (9) & 4.45 (1) & 11.21 $\pm$ 0.08 & 18.3 $\pm$ 3.3 & 27.4 $\pm$ 2.3 & 0 \\
HLSJ023954.6--013427 & A0370 C & 39.977550 & --1.574226 & (H) & \textbf{0.335} (9) & -- ~~--~ & 10.52 $\pm$ 0.09 & 3.8 $\pm$ 0.8 & 30.5 $\pm$ 2.7 & 0 \\
HLSJ023954.8--013453 & A0370 C & 39.978125 & --1.581442 & (V) & \textbf{0.387} (8) & -- ~~--~ & 10.96 $\pm$ 0.08 & 10.4 $\pm$ 1.9 & 29.1 $\pm$ 1.5 & 0 \\
HLSJ023957.2--013259 & A0370 C & 39.988083 & --1.549772 & (V) & \textbf{0.364} (8) & -- ~~--~ & 10.38 $\pm$ 0.10 & 2.7 $\pm$ 0.6 & 27.1 $\pm$ 3.2 & 0 \\
HLSJ023957.8--013313 & A0370 C & 39.990890 & --1.552850 & (O) & \textbf{0.547} (10) & 1.15 (1) & 10.85 $\pm$ 0.08 & 8.1 $\pm$ 1.5 & 25.8 $\pm$ 2.1 & 0 \\
HLSJ023956.6--013426 & A0370 C & 39.985458 & --1.574431 & (V) & \textbf{1.062} (14) & 4.57 (1) & 12.04 $\pm$ 0.10 & 125.8 $\pm$ 22.6 & 30.9 $\pm$ 0.6 & 0 \\
HLSJ023957.5--013432 & A0370 C & 39.989580 & --1.575560 & (H) & \textbf{0.256} (10) & -- ~~--~ & 10.32 $\pm$ 0.11 & 2.4 $\pm$ 0.6 & 23.5 $\pm$ 1.5 & 0 \\
\hline
\end{tabular}
\end{table*}
\addtocounter{table}{-1}
\begin{table*}
\caption{\textit{Continued}}
\begin{tabular}{lrrrcrrrrrc}
\hline
HLSJ071724.9+374410 & M0717 C & 109.353910 & +37.735573 & (V) & \textbf{0.536} (19) & -- ~~--~ & 11.09 $\pm$ 0.09 & 14.1 $\pm$ 2.9 & 24.5 $\pm$ 1.6 & 0 \\
HLSJ071727.7+374436 & M0717 C & 109.365486 & +37.743331 & (H) & 0.22 (16) & -- ~~--~ & 10.11 $\pm$ 0.23 & 1.4 $\pm$ 0.8 & 26.1 $\pm$ 2.8 & 0 \\
HLSJ071727.2+374606 & M0717 C & 109.363354 & +37.768311 & (H) & 0.15 (16) & -- ~~--~ & 9.97 $\pm$ 0.33 & 1.1 $\pm$ 0.8 & 22.1 $\pm$ 1.6 & 0 \\
HLSJ071727.5+374441 & M0717 C & 109.364760 & +37.744836 & (V) & \textbf{0.526} (19) & -- ~~--~ & 11.38 $\pm$ 0.08 & 27.5 $\pm$ 5.0 & 26.2 $\pm$ 1.0 & 0 \\
HLSJ071729.0+374320 & M0717 C & 109.370972 & +37.722271 & (H) & \textbf{0.229} (19) & -- ~~--~ & 10.00 $\pm$ 0.14 & 1.1 $\pm$ 0.4 & 24.6 $\pm$ 2.2 & 1 \\
HLSJ071728.5+374417 & M0717 C & 109.368772 & +37.737946 & (H) & 0.01 (16) & -- ~~--~ & 7.66 $\pm$ 9.22 & 0.0 $\pm$ 0.1 & 23.2 $\pm$ 2.2 & 0 \\
HLSJ071730.4+374617 & M0717 C & 109.376800 & +37.771464 & (V) & \textbf{0.899} (20) & 1.91 (1) & 11.42 $\pm$ 0.09 & 30.0 $\pm$ 5.4 & 32.6 $\pm$ 1.6 & 0 \\
HLSJ071730.7+374443 & M0717 C & 109.377710 & +37.745258 & (H) & \textbf{1.010} (20) & 2.54 (1) & 11.50 $\pm$ 0.13 & 35.9 $\pm$ 10.7 & 24.8 $\pm$ 2.5 & 0 \\
\hline
\end{tabular}
\end{table*}
\addtocounter{table}{-1}
\begin{table*}
\caption{\textit{Continued}}
\begin{tabular}{lrrrcrrrrrc}
\hline
HLSJ114936.2+222417 & M1149 C & 177.400957 & +22.404700 & (H) & 1.10 (16) & 2.27 (1) & 11.20 $\pm$ 0.09 & 18.1 $\pm$ 3.3 & 27.6 $\pm$ 2.4 & 0 \\
HLSJ114936.1+222425 & M1149 C & 177.400377 & +22.406727 & (H) & 1.41 (16) & 2.44 (1) & 11.39 $\pm$ 0.09 & 28.2 $\pm$ 5.2 & 40.0 $\pm$ 13.9 & 0 \\
HLSJ114936.8+222254 & M1149 C & 177.403620 & +22.381486 & (V) & \textbf{1.409} (20) & 1.39 (1) & 12.08 $\pm$ 0.08 & 137.4 $\pm$ 24.7 & 39.1 $\pm$ 2.2 & 0 \\
HLSJ114937.3+222431 & M1149 C & 177.405460 & +22.408510 & (H) & \textbf{1.020} (9) & 1.41 (1) & 11.62 $\pm$ 0.08 & 47.9 $\pm$ 8.6 & 27.7 $\pm$ 2.4 & 0 \\
HLSJ114937.0+222542 & M1149 C & 177.404060 & +22.428214 & (V) & 0.55 (17) & 1.01 (1) & 10.73 $\pm$ 0.13 & 6.1 $\pm$ 1.8 & 27.6 $\pm$ 3.1 & 0 \\
HLSJ114938.1+222412 & M1149 C & 177.408770 & +22.403077 & (V) & 0.99 (16) & 1.51 (1) & 11.48 $\pm$ 0.08 & 34.3 $\pm$ 6.2 & 32.2 $\pm$ 2.0 & 0 \\
HLSJ114938.0+222428 & M1149 C & 177.408320 & +22.407729 & (V) & \textbf{0.533} (19) & -- ~~--~ & 11.10 $\pm$ 0.09 & 14.2 $\pm$ 2.6 & 30.9 $\pm$ 2.2 & 0 \\
HLSJ114938.8+222532 & M1149 C & 177.411703 & +22.425523 & (H) & 0.54 (16) & -- ~~--~ & 10.71 $\pm$ 0.12 & 5.8 $\pm$ 1.6 & 32.2 $\pm$ 4.5 & 0 \\
\hline
\end{tabular}
\end{table*}
\addtocounter{table}{-1}
\begin{table*}
\caption{\textit{Continued}}
\begin{tabular}{lrrrcrrrrrc}
\hline
HLSJ224913.6--443238 & AS1063 P & 342.307337 & --44.544783 & (I) & \textit{$\sim$0.2} (2) & -- ~~--~ & 10.89 $\pm$ 0.41 & 8.9 $\pm$ 8.4 & 32.9 $\pm$ 41.4 & 0 \\
HLSJ224915.2--443220 & AS1063 P & 342.314834 & --44.538237 & (I) & \textit{$\sim$1.7} (2) & 1.03 (3) & 12.41 $\pm$ 0.10 & 291.7 $\pm$ 64.9 & 27.7 $\pm$ 2.7 & 0 \\
HLSJ224916.5--443042 & AS1063 P & 342.320281 & --44.512853 & (W) & \textit{$\sim$1.3} (2) & 1.09 (3) & 12.49 $\pm$ 0.11 & 351.5 $\pm$ 92.5 & 32.9 $\pm$ 2.8 & 0 \\
HLSJ224917.3--443119 & AS1063 P & 342.322117 & --44.520606 & (W) & \textit{$\sim$1.2} (2) & 1.07 (3) & 12.22 $\pm$ 0.15 & 189.1 $\pm$ 64.7 & 32.9 $\pm$ 3.6 & 0 \\
HLSJ224917.8--443256 & AS1063 P & 342.324002 & --44.548805 & (S) & \textit{$\sim$0.9} (2) & -- ~~--~ & 11.63 $\pm$ 0.31 & 48.1 $\pm$ 34.7 & 32.9 $\pm$ 28.2 & 0 \\
HLSJ224920.9--443337 & AS1063 P & 342.336299 & --44.560820 & (I) & \textit{$\sim$2.1} (2) & 1.01 (3) & 12.18 $\pm$ 0.13 & 171.7 $\pm$ 51.1 & 27.7 $\pm$ 3.8 & 0 \\
HLSJ224922.8--443235 & AS1063 P & 342.344357 & --44.543522 & (I) & \textit{$\sim$0.8} (2) & 1.06 (3) & 10.88 $\pm$ 0.53 & 8.6 $\pm$ 10.5 & 32.9 $\pm$ 9.9 & 1 \\
HLSJ224925.1--443252 & AS1063 P & 342.353444 & --44.549562 & (S) & \textit{$\sim$1.4} (2) & 1.21 (3) & 12.03 $\pm$ 0.16 & 122.5 $\pm$ 44.5 & 32.9 $\pm$ 4.5 & 0 \\
HLSJ224925.9--443332 & AS1063 P & 342.356469 & --44.559126 & (I) & \textit{$\sim$0.1} (2) & -- ~~--~ & 10.41 $\pm$ 0.33 & 2.9 $\pm$ 2.2 & 42.3 $\pm$ 44.0 & 0 \\
HLSJ224926.3--443223 & AS1063 P & 342.358424 & --44.539673 & (W) & \textit{$\sim$2.0} (2) & 1.07 (3) & 12.49 $\pm$ 0.09 & 348.1 $\pm$ 66.4 & 27.7 $\pm$ 2.6 & 0 \\
HLSJ224926.1--443305 & AS1063 P & 342.361422 & --44.552207 & (I) & \textit{$\sim$2.3} (2) & 1.34 (3) & 12.08 $\pm$ 0.13 & 136.3 $\pm$ 40.1 & 27.7 $\pm$ 4.0 & 0 \\
\hline
\\
\end{tabular}
\\
\raggedright
$^1$ \textit{Herschel} ID as in Table \ref{tab:cat1}\\
$^2$ C=central region; P=parallel\\
$^3$ Counterpart RA and Dec, from the most accurate available position (V=VLA, H=HST, O=[ground-based] optical, I=IRAC, W=WISE, P=PACS, S=SPIRE)\\
$^4$ Counterpart \textbf{spectroscopic}, optical photometric or \textit{IR estimated} redshift from (1) \citet{owe11-27}, (2) $z_{\rm phot\_FIR}$, (3) \citet{cou98-188}, (4) \citet{raw14-196}, (5) \citet{wan15-arxiv}, (6) \citet{bus02-787}, (7) \citet{bra09-947}, (8) \citet{wol12-2}, (9) \textit{HST} GRISM, (10) \citet{bam05-109}, (11) \citet{ivi98-583}, (12) \citet{sou88-19}, (13) \citet{hen87-473}, (14) \citet{sou99-70}, (15) Magellan/IMACS, (16) CLASH \textit{HST} photo-z, (17) CLASH Subraru photo-z, \citet{ume14-163}, (18) \citet{bal13-9}, (19) \citet{ebe14-21}, (20) LBT/MODS, (21) \citet{smi09-163}, (22) Walth et al. (in prep.), (23) \citet{gom12-79}, (24) \citet{kar15-11}\\
$^5$ $\mu$ from (1) CATS models \citep{jul09-1319, jau12-3369, ric14-268, jau14-1549} or (2) SaWLens wide-field method \citet{mer09-681, mer11-333}. `--' for foreground and cluster sources\\
$^6$ From the best-fitting \citet{rie09-556} template\\
$^7$ Characteristic dust temperature of the best-fitting single-temperature modified blackbody\\
$^8$ Flag for alternate IR SED fit using SF+AGN model, effectively correcting $L_{\rm IR}$ and SFR$_{\rm IR}$ for AGN contamination (see Section \ref{sec:agn}) \\
\end{table*}